\documentclass[11pt,a4paper]{article}
\pdfoutput=1
\usepackage{jheppub}
\usepackage{graphicx,psfrag,color}
\usepackage{bm}
\usepackage{mathbbol,verbatim}
\usepackage{slashed}
\usepackage{graphics}
\usepackage{color,ulem}
\allowdisplaybreaks

\usepackage{tikz}
\usetikzlibrary{trees}
\usetikzlibrary{decorations.pathmorphing}
\usetikzlibrary{decorations.markings}
\usetikzlibrary{decorations, decorations.markings, decorations.pathmorphing, arrows, graphs, shapes.geometric, snakes}
\usetikzlibrary{arrows}

\tikzset{
photon/.style={decorate, decoration={snake}, draw=red},
dark/.style={draw=gray, postaction={decorate},
        decoration={markings,mark=at position .55 with {\arrow[draw=gray]{>}}}},
antidark/.style={draw=gray, postaction={decorate},
        decoration={markings,mark=at position .55 with {\arrow[draw=gray]{<}}}},
electron/.style={draw=violet, postaction={decorate},
        decoration={markings,mark=at position .55 with {\arrow[draw=violet]{>}}}},
neutrino/.style={draw,color=violet,thick, postaction={decorate} },
neutrinolight/.style={draw=blue, postaction={decorate} },
quark/.style={draw=blue, postaction={decorate},
        decoration={markings,mark=at position .55 with {\arrow[draw=blue]{>}}}},
antiquark/.style={draw=blue, postaction={decorate},
        decoration={markings,mark=at position .55 with {\arrow[draw=blue]{<}}}},
heavyquark/.style={draw=purple, postaction={decorate},
        decoration={markings,mark=at position .55 with {\arrow[draw=purple]{>}}}},
antiheavyquark/.style={draw=purple, postaction={decorate},
        decoration={markings,mark=at position .55 with {\arrow[draw=purple]{<}}}},
        gluon/.style={decorate, draw=or,
        decoration={coil,amplitude=2pt, segment length=3pt}},
gluon/.style={decorate, draw=or,
        decoration={coil,amplitude=2pt, segment length=3pt}},
ZZ/.style={decorate, decoration={snake,amplitude=1.5pt, segment length=5pt}, draw=greeen},
left,
  }

\definecolor{greeen}{rgb}{0.03,0.84,0.13}
\definecolor{test}{rgb}{0.03,0.74,0.33}
\definecolor{viol}{rgb}{0.44,0,0.94}
\definecolor{or}{rgb}{0.95,0.65,0}

\begin{document}

\begin{flushright}
ULB-TH/17-20
\end{flushright}

\title{Leptogenesis Constraints on $B-L$ breaking Higgs Boson in TeV Scale Seesaw Models}

\author[a]{P. S. Bhupal Dev,}
\author[b]{Rabindra N. Mohapatra,}
\author[a,c]{Yongchao Zhang}
\affiliation[a]{Department of Physics and McDonnell Center for the Space Sciences,  Washington University, St. Louis, MO 63130, USA}
\affiliation[b]{Maryland Center for Fundamental Physics, Department of Physics, University of Maryland, College Park, MD 20742, USA}
\affiliation[c]{Service de Physique Th\'{e}orique, Universit\'{e} Libre de Bruxelles, Boulevard du Triomphe, CP225, 1050 Brussels, Belgium}

\emailAdd{bdev@wustl.edu}
\emailAdd{rmohapat@umd.edu}
\emailAdd{yongchao.zhang@physics.wustl.edu}


\abstract{In the type-I seesaw mechanism for neutrino masses, there exists a $B-L$ symmetry, whose breaking leads to the lepton number violating mass of the heavy Majorana neutrinos. This would imply the existence of a new neutral scalar associated with the $B-L$ symmetry breaking, analogous to the Higgs boson of the Standard Model. If in such models, the heavy neutrino decays are also responsible for the observed baryon asymmetry of the universe via the leptogenesis mechanism, the new seesaw scalar interactions with the heavy neutrinos will induce additional dilution terms for the heavy neutrino and lepton number densities. We make a detailed study of this dilution effect on the lepton asymmetry in three generic classes of seesaw models with TeV-scale $B-L$ symmetry breaking, namely, in an effective theory framework and in scenarios with global or local $U(1)_{B-L}$ symmetry. We find that requiring successful leptogenesis imposes stringent constraints on the mass and couplings of the new scalar in all three cases, especially when it is lighter than the heavy neutrinos. We also discuss the implications of these new constraints and prospects of testing leptogenesis in presence of seesaw scalars at colliders. }


\maketitle

\section{Introduction}
The type-I seesaw mechanism~\cite{seesaw1,seesaw2,seesaw3,seesaw4,seesaw5} provides a simple way to understand the smallness of light neutrino masses in terms of the heavy right-handed neutrino (RHN) Majorana masses and their mixing with light neutrinos.  Two pertinent questions arise here: (i) What is the seesaw (or the RHN mass) scale? (ii) Is there a symmetry (global or local) associated with the origin of the RHN masses that go into the seesaw formula? As far as the first question is concerned, in a phenomenological bottom-up  approach, the seesaw scale could be anywhere between eV and $10^{14}$ GeV~\cite{Mohapatra:2005wg, Drewes:2013gca}. However, due to observational bias (the `streetlight effect'), we are particularly interested in the TeV scale seesaw models, which can be effectively probed in a plethora of current and future experiments at both energy~\cite{Deppisch:2015qwa, Cai:2017mow} and intensity~\cite{deGouvea:2013zba, Alekhin:2015byh} frontiers.
Low-scale seesaw can also be theoretically motivated from Higgs naturalness perspective~\cite{Vissani:1997ys, Clarke:2015gwa, Bambhaniya:2016rbb}.
As far as any associated symmetry behind the seesaw mechanism is concerned, $B-L$ symmetry is an obvious and natural possibility~\cite{Marshak:1979fm, Mohapatra:1980qe}, whose breaking leads to the heavy Majorana masses in the seesaw formula.  Once this point of view is adopted, regardless of whether the symmetry is global or local, there should exist a Higgs field that breaks the symmetry and the real part of that complex Higgs field will be an important new particle intimately linked with the seesaw mechanism. Therefore,  a detailed study of its properties is likely to provide another window to the physics behind the neutrino masses.
The collider signatures of this $B-L$ breaking scalar have been studied in some recent papers~\cite{Maiezza:2015lza, Nemevsek:2016enw, Dev:2016vle, Dev:2017dui, Dev:2017ozg}. In this paper, we study some cosmological implications of this new scalar for the dynamical generation of matter-antimatter asymmetry in the early universe.

An important consequence of the type-I seesaw is the possibility that it could provide a way to understand the matter-antimatter asymmetry of the universe via leptogenesis~\cite{Fukugita:1986hr, Davidson:2008bu, Blanchet:2012bk, Fong:2013wr}. For TeV scale seesaw, it is necessary to use the resonant leptogenesis mechanism~\cite{Pilaftsis:1997dr, Pilaftsis:1997jf, Pilaftsis:2003gt} (see Ref.~\cite{Dev:2017wwc} for a recent review) in order to avoid the Davidson-Ibarra bound~\cite{Davidson:2002qv}. An important aspect of leptogenesis is the dilution and wash-out effect induced by different particles and their interactions in the theory. In particular, the strong (exponential) dependence of the lepton asymmetry on the wash-out effect  can be used to falsify high-scale leptogenesis at the LHC~\cite{Deppisch:2013jxa}, as well as in low-energy experiments~\cite{Deppisch:2015yqa}. As for probing TeV scale leptogenesis, it depends on the details of the model under consideration~\cite{Chun:2017spz}. In particular, the same dilution and wash-out effects can be used to put stringent constraints on the model parameters, if we want to have successful leptogenesis in these models. For example, in the left-right (LR) symmetric realization~\cite{Pati:1974yy, Mohapatra:1974gc, Senjanovic:1975rk} of seesaw, dilution and washout of the lepton asymmetry due to the $W_R$-mediated $\Delta L=1$  interactions~\cite{Frere:2008ct, Dev:2014iva, Dev:2015vra, Dhuria:2015cfa}  provides a {\it lower} bound of about 10 TeV on the mass of $W_R$ boson~\cite{Dev:2015vra}. Similar (but weaker) lower bounds have also been derived on the mass of the extra $Z'$ boson  that couples to RHNs as well as to the SM quarks and leptons in $U(1)$ models~\cite{Blanchet:2009bu, Blanchet:2010kw, Iso:2010mv, Okada:2012fs, Heeck:2016oda}. 

In this paper, we consider generic seesaw models with $B-L$  symmetry and derive leptogenesis constraints on the additional Higgs scalar $H$ responsible for $B-L$ breaking from its dilution and washout effect on the lepton asymmetry.\footnote{{More generic discussions of the effects of spontaneous breaking of baryon or lepton number on leptogenesis can be found in~\cite{Shuve:2017jgj}, where it was pointed out under what conditions the new annihilation modes are important.}} We consider three generic classes of theories: (i) where a scalar or pseudo-scalar has a generic coupling to the RHN in an effective field theory (EFT) framework; (ii) where $B-L$ is a global $U(1)$ symmetry (the so-called singlet Majoron model~\cite{Chikashige:1980ui}); and (iii) where $B-L$ is a local gauge symmetry. In all these cases, our constraints are independent of any details of leptogenesis, except for some of the key parameters such as the effective neutrino mass $\widetilde{m}$ and the CP asymmetry $\varepsilon_{\rm CP}$. 
We find that the allowed scalar mass regions are correlated with the mass of the RHNs for fixed $B-L$ breaking scale (or the effective couplings to RHNs). In particular, the dilution of lepton asymmetry due to the seesaw scalar can be {\it directly} tested at the (high-luminosity) LHC and future hadron colliders. The RHNs can be pair produced in the Higgs portal through the mixing of seesaw scalar with the SM Higgs, and then decays into same-sign charged leptons plus jets as a result of the Majorana nature of RHN, i.e. $pp \to h/H \to NN \to \ell^\pm \ell^\pm + 4j$. Therefore, if the heavy RHNs are discovered at the LHC or future higher energy/luminosity colliders~\cite{Deppisch:2015qwa, Cai:2017mow}, this will narrow the parameter region to search for the seesaw Higgs boson and its properties, which would provide a direct test of leptogenesis and existence of $B-L$ symmetry in this scenario.

The rest of this paper is organized as follows: In Sec.~\ref{sec:EFT}, we consider generic couplings of a scalar or a pseudo-scalar to the RHNs in an EFT framework and solve Boltzmann equations for the RHN number density and the lepton asymmetry in their presence to find viable regions allowed by the observed baryon asymmetry. The global $U(1)_{B-L}$ model is considered in Sec.~\ref{sec:global}, where the (almost) massless Majoron $J$ plays an important role in the dilution process as it directly couples to the RHNs, in addition to the scalar interactions. The local $U(1)_{B-L}$ case is investigated in Sec.~\ref{sec:local}, where we have a heavy $Z_R$ boson in addition to the seesaw scalar, and the gauge interactions and self-trilinear interactions of the scalar are both involved in the Boltzmann equations. The collider implications of leptogenesis are presented in Sec.~\ref{sec:collider}.
We summarize in Sec.~\ref{sec:conclusion}. The reduced cross sections and $Z_R$ boson couplings to SM fermions are given respectively in Appendix~\ref{sec:appendixA} and \ref{sec:appendixB}.

\section{Effective theory with a CP-even/odd scalar}
\label{sec:EFT}
We begin the discussion by considering an effective ``low energy'' theory where the RHN $N$ has a Yukawa coupling to a new (pseudo)scalar $H$ ($A$). It also has a Dirac Yukawa coupling to the SM neutrinos $\nu$, which is responsible for both type-I seesaw as well as the generation of lepton asymmetry via leptogenesis. The effective theory could be considered as the low energy simplified version of a more fundamental ultraviolet (UV)-completion at higher energy scale, such as the specific $U(1)_{B-L}$ models discussed in Sec.~\ref{sec:global} and \ref{sec:local}, yet some of the key features of dilution of lepton asymmetry could already be seen and easily understood in the EFT framework, c.f. the Feynman diagram in Fig.~\ref{fig:diagrams} and the plots in Figs.~\ref{fig:example}-\ref{fig:EFT3}. We analyze in this section the generic dilution effect induced by these couplings and obtain bounds on the (pseudo)scalar mass and its coupling to RHNs.

\subsection{Effective couplings}
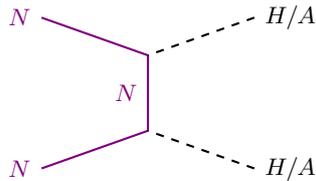
\begin{figure}[t!]
  \centering
  \begin{tikzpicture}[]
  \draw[neutrino,thick] (-1.4,1)node[left]{{\footnotesize$N$}} -- (0,0.5);
  \draw[neutrino] (-1.4,-1)node[left]{{\footnotesize$N$}} -- (0,-0.5);
  \draw[neutrino,thick](0,0.5)--(0,0)node[left]{{\footnotesize$N$}}--(0,-0.5);
  \draw[dashed,thick](1.4,1)node[right]{{\footnotesize$H/A$}}--(0,0.5);
  \draw[dashed,thick](1.4,-1)node[right]{{\footnotesize$H/A$}}--(0,-0.5);
  \end{tikzpicture}
  \caption{Feynman diagram for the scattering $N N \to HH,\,AA$ induced by the Yukawa couplings $f_{H,A}$ in Eq.~\eqref{eqn:LYukawa}. }
  \label{fig:diagrams}
\end{figure}

The lowest-order effective couplings of a CP-even (odd) scalar $H$ ($A$) to the RHN can be parameterized as\footnote{Here the Yukawa couplings $f_H$ and $f_A$ could be the same or different, depending on the specific model details. For example, in the global $U(1)$ model in Sec.~\ref{sec:global}, they are equal in the {\it original} Lagrangian; however, the CP-even scalar $H$ could mix with the SM Higgs but the CP-odd scalar does not, and as a result the couplings $f_{H,\,A}$ are not equal any more, with $f_H$ rescaled by the mixing angle $\cos\theta$. In addition, if the CP-odd scalar $A$ mixes with the SM Higgs, as discussed in Sec.~\ref{sec:collider}, it will no longer be a pure CP-odd state. However, the SM Higgs is well-determined to be CP-even like, with the mixing with a CP-odd scalar smaller than $\lesssim 0.2$ at the 95\% confidence level (CL)~\cite{Aad:2015tna, Aad:2016nal, Khachatryan:2016tnr}. }
\begin{eqnarray}
\label{eqn:LYukawa}
{\cal L}_Y^{\rm eff} & \ = \ & - \frac12 f_H \overline{N}^c H N
 - \frac{i}{2} f_A \overline{N}^c \gamma^5 A N +Y \overline{L}\phi N~+~{\rm H.c.}\, ,  
\end{eqnarray}
where $\phi$ denotes the SM Higgs doublet, $L$ the SM lepton doublet and we have suppressed the flavor indices for brevity. 
In most of the realistic seesaw models, there might be more than one physical scalar, or even multiplets, that couple to the RHN; see for instance the global $U(1)_{B-L}$ model in Sec.~\ref{sec:global} and the local one in Sec.~\ref{sec:local}  (in the local case the CP-odd component is eaten by the heavy $Z_R$ boson). Then the triple scalar couplings get involved in the dilution of RHNs through the processes $NN \to S_i \to S_j S_k$ ($i,\,j,\,k$ being scalar indices, see e.g. the diagrams in Fig.~\ref{fig:diagrams2}) and play an important role in leptogenesis, particularly when the Yukawa coupling $f$ is comparatively smaller. To capture the most important consequence of the presence of a (light) scalar/pseudoscalar in the type-I seesaw leptogenesis, we neglect the model-dependent scalar interactions in this section, and consider only the $t$-channel process
$NN \to HH/AA$,
mediated by the Yukawa coupling $f$ in Eq.~\eqref{eqn:LYukawa},\footnote{We consider the coupling of either $H$ or $A$ in Eq.~(\ref{eqn:LYukawa}) and not both of them simultaneously. So in what follows, we sometimes denote the coupling simply as $f$ (without the subscript), whose meaning should be clear from the context.} as shown in Fig.~\ref{fig:diagrams}.

It should be noted that even if the new scalar develops a non-vanishing vacuum expectation value (VEV), which is expected to be higher than the electroweak (EW) VEV at $v_{\rm EW} \simeq 174$ GeV, before the EW phase transition the heavy scalar does not mix with the SM Higgs in the early universe. Therefore, in addition to the RHN mass $m_N$ (and the leptogenesis-relevant quantities such as the effective neutrino mass $\widetilde{m}$), there are only two free parameters in the effective theory, i.e. the scalar mass $m_H$ ($m_A$) and the effective Yukawa coupling $f_{H(A)}$ in Eq.~(\ref{eqn:LYukawa}).


\subsection{Dilution of the lepton asymmetry}
The reduced cross section $\hat{\sigma} (NN \to HH/AA)$ can be found in Appendix~\ref{sec:appendixA} [cf.~Eqs.~(\ref{eqn:sigmaNNHH}) and (\ref{eqn:sigmaNNAA})], which are dictated by the scalar masses $m_{H,\,A}$ and the Yukawa coupling $f$. The relevant Boltzmann equations, which govern the evolution of the RHN number density and the lepton asymmetry, are then given by
\begin{eqnarray}
\label{eqn:Boltzmann}
\frac{n_\gamma H_N}{z} \frac{{\rm d} \eta_N}{{\rm d}z} \ &=& \
- \left[ \left( \frac{\eta_N}{\eta_N^{\rm eq}} \right)^2 - 1 \right] \, 2 \gamma_{HH,\,AA}
- \left( \frac{\eta_N}{\eta_N^{\rm eq}} - 1 \right) \left[ \gamma_D + \gamma_{s} + 2\gamma_{t} \right] \,, \\
\label{eqn:Boltzmann2}
\frac{n_\gamma H_N}{z} \frac{{\rm d} \eta_{\Delta L}}{{\rm d}z} \ &=& \
\gamma_D \left[ \varepsilon_{\rm CP} \left( \frac{\eta_N}{\eta_N^{\rm eq}} - 1 \right) - \frac23 \eta_{\Delta L} \right]
- \frac23 \eta_{\Delta L} \left[ \frac{\eta_N}{\eta_N^{\rm eq}} \gamma_{s} + 2\gamma_{t} \right] \,,
\end{eqnarray}
where $z\equiv m_N/T$ is a dimensionless parameter, $H_N\equiv H(z=1)\simeq 17m_N^2/M_{\rm Pl}$ is the Hubble expansion rate at temperature $T=m_N$ (with $M_{\rm Pl} =1.2\times 10^{19}$ GeV being the Planck mass), $n_\gamma=2T^3\zeta(3)/\pi^2$  is the number density of photons, and $\eta_N\equiv n_N/n_\gamma$ is the normalized number density of RHN (similarly $\eta_{\Delta L}=(n_L-n_{\bar{L}})/n_\gamma$ for the lepton asymmetry).  The $\gamma$'s are the various thermalized interaction rates: $\gamma_D$ for the RHN decay $N \to L\phi$,  and $\gamma_{s} = \gamma_{\phi s} + \gamma_{Vs}$ and $\gamma_{t} = \gamma_{\phi t} + \gamma_{V t}$ the standard $\Delta L = 1$ scattering processes as in Refs.~\cite{Giudice:2003jh, Pilaftsis:2003gt}  with the subscripts $s,t$ denoting respectively the $s$ and $t$-channel exchange of the SM Higgs doublet $\phi$ or the SM gauge bosons $V = W_{i},\,B$ (with $i=1,2,3$) before EW symmetry breaking. 
{Here the integration over different momenta has already been
performed, assuming implicitly kinetic equilibrium.}
The new scattering processes (cf. Fig.~\ref{fig:diagrams}) in our model correspond to the scattering rates $\gamma_{HH,\,AA}$, with the prefactor of 2 for the reduction of RHN by unit of two~\cite{Heeck:2016oda}.
The  thermal corrections to the SM particles are included in the calculation~\cite{Giudice:2003jh, Dev:2014laa}. If the Yukawa coupling $f$ is sizable and $m_N \gtrsim m_{H,\,A}$, or in other words, as long as the $\gamma_{HH,\,AA}$ term is comparable or larger than other terms on the right-hand-side (RHS) of Eq.~(\ref{eqn:Boltzmann}), this extra process could significantly dilute the RHN number density before the  sphaleron decoupling temperature $T_c\simeq 131.7$ GeV~\cite{DOnofrio:2014rug}, thus potentially  making type-I seesaw freeze-out leptogenesis ineffective, as we show below.\footnote{{In the limit of $m_N \ll m_{H,\,A}$ the inverse decay process $HH, AA \to NN$  might also impact the final lepton asymmetry, e.g. in models of nonthermal leptogenesis~\cite{Lazarides:1991wu, Giudice:1999fb, Asaka:1999yd} or leptogenesis via (light) sterile neutrino oscillations~\cite{Akhmedov:1998qx, Asaka:2005pn, Drewes:2017zyw}. A full study of this option goes beyond the scope of this paper, and is deferred to a future publication. }}

In general, the dilution effect depends on the RHN mass $m_N$, the scalar mass $m_{H,\,A}$, the effective Yukawa coupling $f_{H,A}$ and other model parameters such as the effective neutrino mass $\widetilde{m}\equiv v^2 (Y^\dag Y)_{11}/m_N$ (or Dirac Yukawa coupling) and the CP asymmetry $\varepsilon_{\rm CP}$, as well as the CP property of the extra scalar. Since in this paper we are mostly concerned with the role of the new scalar  in the lepton asymmetry generation in RHN decay $N \to L\phi$, we fix $\widetilde{m}\simeq \sqrt{\Delta m^2_{\rm atm}} \simeq 50 \, {\rm meV}$ throughout, without any significant tuning or cancellation in the type-I seesaw formula for light neutrino masses: $m_\nu \simeq -v^2 Y m_N^{-1} Y^{\sf T}$. Also, for the sake of simplicity, we have assumed implicitly that the scalar $H/A$ stays in equilibrium with the RHN through the Yukawa interaction and does not have any portals to talk (in)directly to the SM particles, otherwise we would have the extra process like $NN \to H/A \to$ SM particles contributing to the dilution. 
No specific flavor structure is considered, as we are focusing on the role of scalars in leptogenesis. To be concrete, one can assume there are only two RHNs (say $N_{e,\,\mu}$) above the EW scale with a universal coupling $f$, while the third one ($N_\tau$) decouples from the ``low-scale'' leptogenesis. A large CP-asymmetry $\varepsilon_{\rm CP}$ can then be generated by the resonant enhancement mechanism, if $\Delta m_N \simeq \Gamma_N/2 \ll m_N$, where $\Gamma_N$ is the average RHN decay width~\cite{Dev:2017wwc}.

\begin{figure}[t!]
  \centering
  \includegraphics[width=0.5\textwidth]{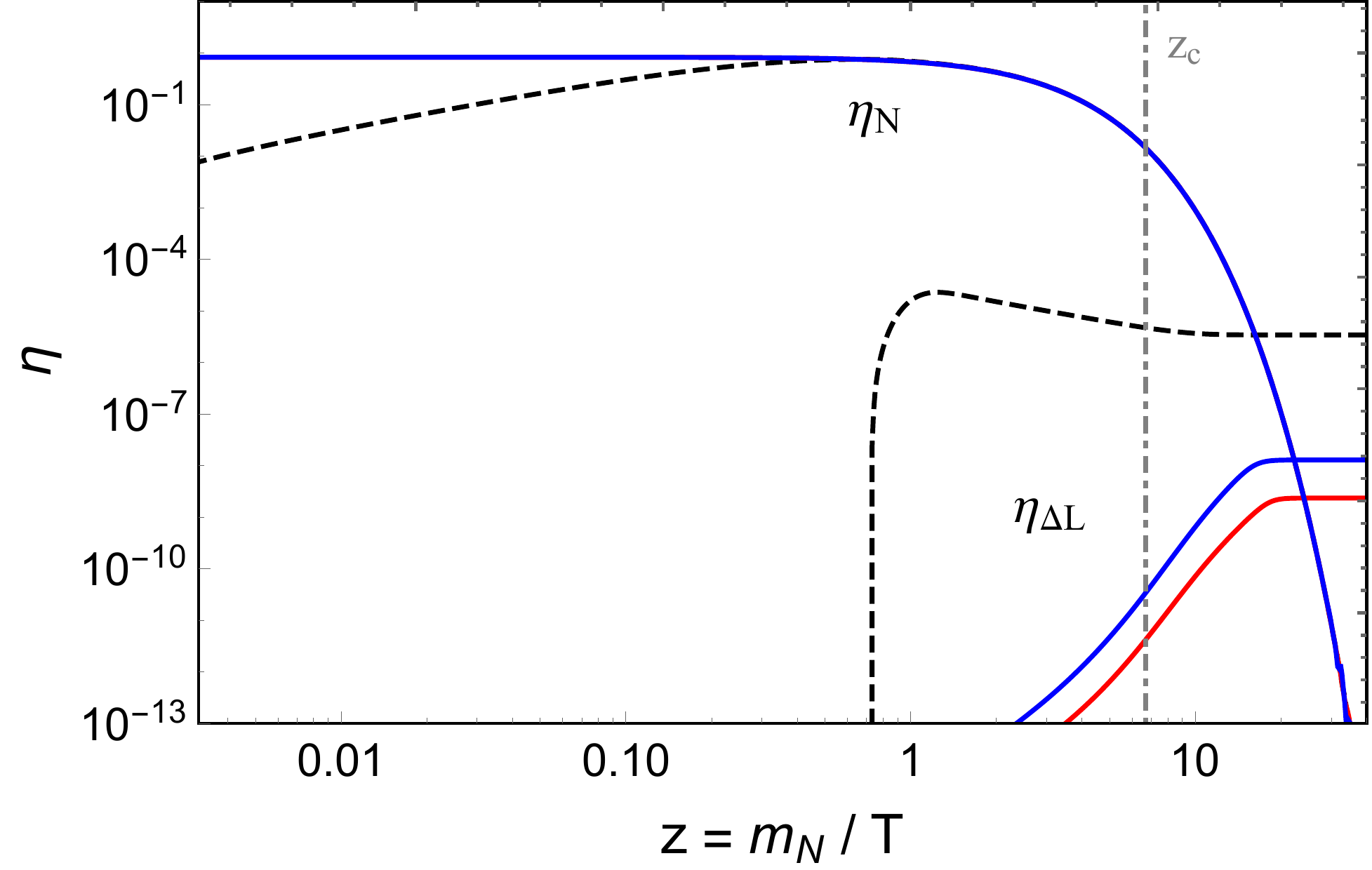}
  \caption{Evolution of the RHN and lepton asymmetry number densities, in presence of a CP-even scalar $H$ (red, solid) or CP-odd scalar $A$ (blue, solid) coupled to the RHN, compared to the standard case without them (dashed). Here we have set the effective neutrino mass $\widetilde{m} = 50$ meV, the RHN mass $m_N = 1$ TeV, the scalar mass $m_{H,\,A} = 500$ GeV, the effective Yukawa coupling $f = 1$ for both $H$ and $A$, and the CP-asymmetry $\varepsilon_{\rm CP} = 10^{-3}$. The red and blue lines for $\eta_N$  overlap with each other. The vertical dot-dashed line corresponds to the sphaleron decoupling temperature $z_c\equiv m_N/T_c$, beyond which the EW sphalerons become ineffective.}
  \label{fig:example}
\end{figure}

The evolution of the RHN and lepton asymmetry number densities for a benchmark scenario are presented in Fig.~\ref{fig:example}, where we set the CP-asymmetry $\varepsilon_{\rm CP} = 10^{-3}$ and the Yukawa coupling $f = 1$. The scalar $H$ ($A$) has a mass $m_{H(A)} = 500$ GeV, lighter than the RHN mass $m_{N} = 1$ TeV, such that the dilution process $NN \to HH\, (AA)$ is kinematically allowed in the parameter space of interest. It is clear from Fig.~\ref{fig:example} that in presence of the new scalar interactions with the RHN, the  lepton asymmetry decreases from its original value (lower dashed curve). {We should also mention here that we solve the Boltzmann equations with the thermal initial conditions $\eta_N=\eta_{\Delta L}=0$, which explains why in the absence of new scalar interactions, the $\eta_N$ value (upper dashed curve) traces its equilibrium distribution (almost coincident with the upper solid curve) at a lower temperature (or higher $z$), as can be seen from Eq.~\eqref{eqn:Boltzmann}. Apart from these additional new features, Fig.~\ref{fig:example} presents the standard picture of freeze-out leptogenesis, namely, the Boltzmann suppression of $\eta_N$ and thermal freeze-out of $\eta_{\Delta L}$ for $z\gg 1$. }

\begin{figure}[t!]
  \centering
  \includegraphics[width=0.5\textwidth]{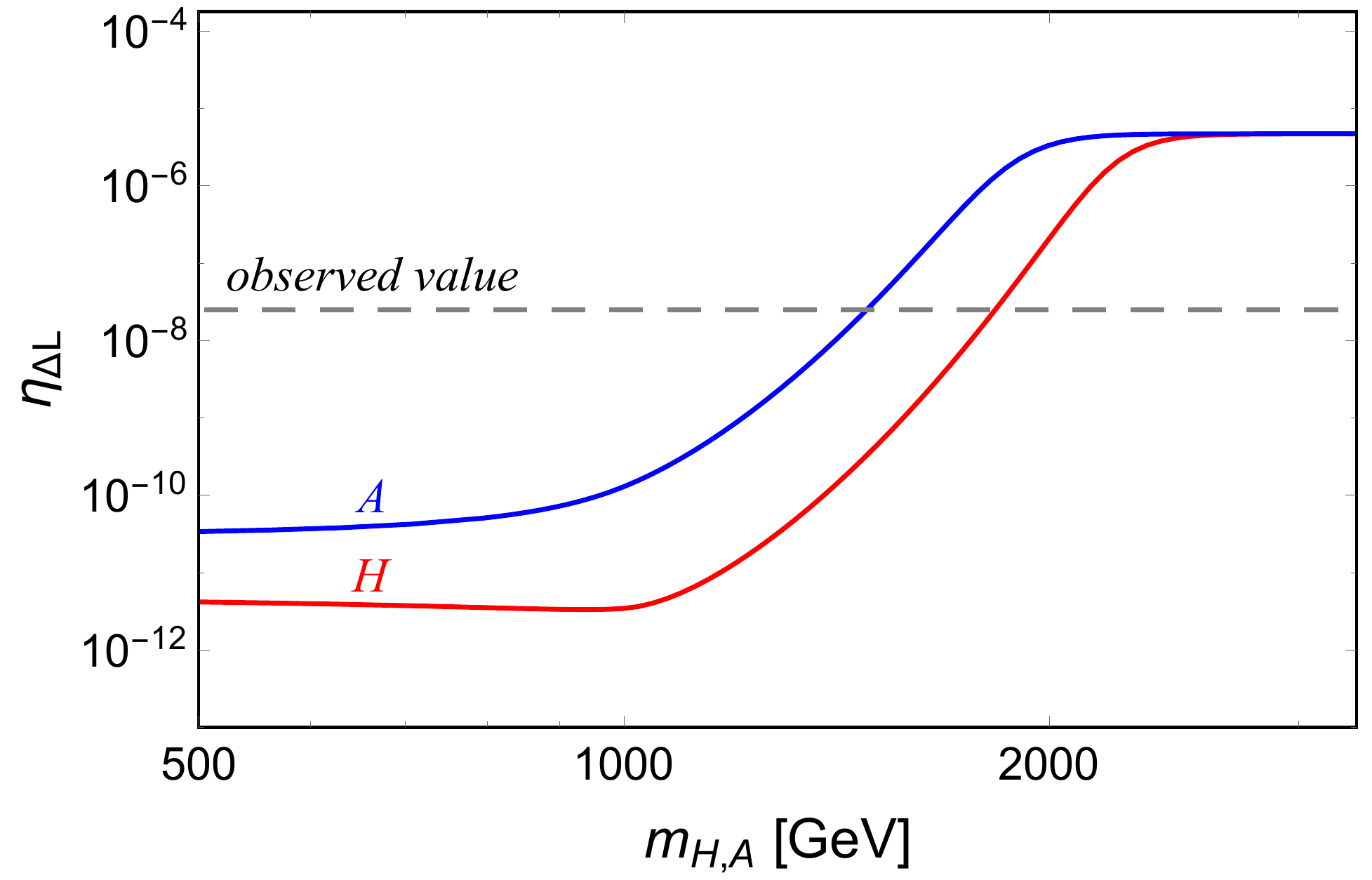}
  \caption{Dependence of lepton asymmetry on the scalar mass $m_{H,\,A}$, with $\widetilde{m} = 50$ meV, $m_N = 1$ TeV, $f = 1$ and $\varepsilon_{\rm CP} = 10^{-3}$. The horizontal dashed line corresponds to the value of lepton asymmetry for which the observed baryon asymmetry can be reproduced, after taking into account the sphaleron and entropy dilution effects.}
  \label{fig:example2}
\end{figure}

This is further illustrated in Fig.~\ref{fig:example2}, where we show the dependence of the lepton asymmetry as a function of the scalar mass, while keeping the other parameter values same as in Fig.~\ref{fig:example}.  We find that for $m_{H,\,A} \lesssim m_N$, the RHN density and lepton asymmetry are strongly diluted,  whereas for $m_{H,\,A} \gg m_N$, the dilution effect becomes negligible, as expected from the kinematics.

In the strong washout regime, the RHN is typically close to the thermal equilibrium, i.e. $|\eta_N / \eta_N^{\rm eq} - 1| \ll 1$, as can be seen from its number density in Fig.~\ref{fig:example}. In this case, the RHS of Eq.~(\ref{eqn:Boltzmann}) can be simplified as
\begin{eqnarray}
- \left( \frac{\eta_N}{\eta_N^{\rm eq}} - 1 \right) \left[ \gamma_D + \gamma_{s} + 2\gamma_{t} + 4 \gamma_{HH,\,AA} \right] \,.
\end{eqnarray}
The leptogenesis constraints on the new scalar can thus be qualitatively obtained by requiring that $4 \gamma_{HH,\,AA} < \gamma_D + \gamma_{s} + 2\gamma_{t}$. In most of the parameter space of interest, one can solve analytically the Boltzmann equations (\ref{eqn:Boltzmann}) and (\ref{eqn:Boltzmann2}) by taking the equilibrium limit, which is a good approximation in presence of the additional Yukawa (and gauge) interactions. For instance, following Ref.~\cite{Dev:2014iva}, the final lepton asymmetry around $z\sim z_c$ can be factorized as
\begin{eqnarray}
\label{eqn:asymmetry}
\eta_{\Delta L} \ \simeq \ \frac{3 \, \varepsilon_{\rm CP}}{2 z K^{\rm eff}}
\frac{\gamma_D}{\gamma_D + 2 \gamma_s + 4 \gamma_t + 4 \gamma_{HH,\,AA}} \,,
\end{eqnarray}
with the effective $K$-factor
\begin{eqnarray}
K^{\rm eff} \ = \ \frac{\Gamma_N}{H_N} \frac{\gamma_D + 2 \gamma_s + 4 \gamma_t}{\gamma_D} \,.
\end{eqnarray}
It is clear from Eq.~(\ref{eqn:asymmetry}) that the lepton asymmetry would be highly suppressed when the $\gamma_{HH,\,AA}$ term is large. Throughout this paper, we follow this analytic approximation to derive the bounds and switch to the numerical solutions whenever necessary.

\begin{figure}[t!]
  \centering
  \includegraphics[height=0.31\textwidth]{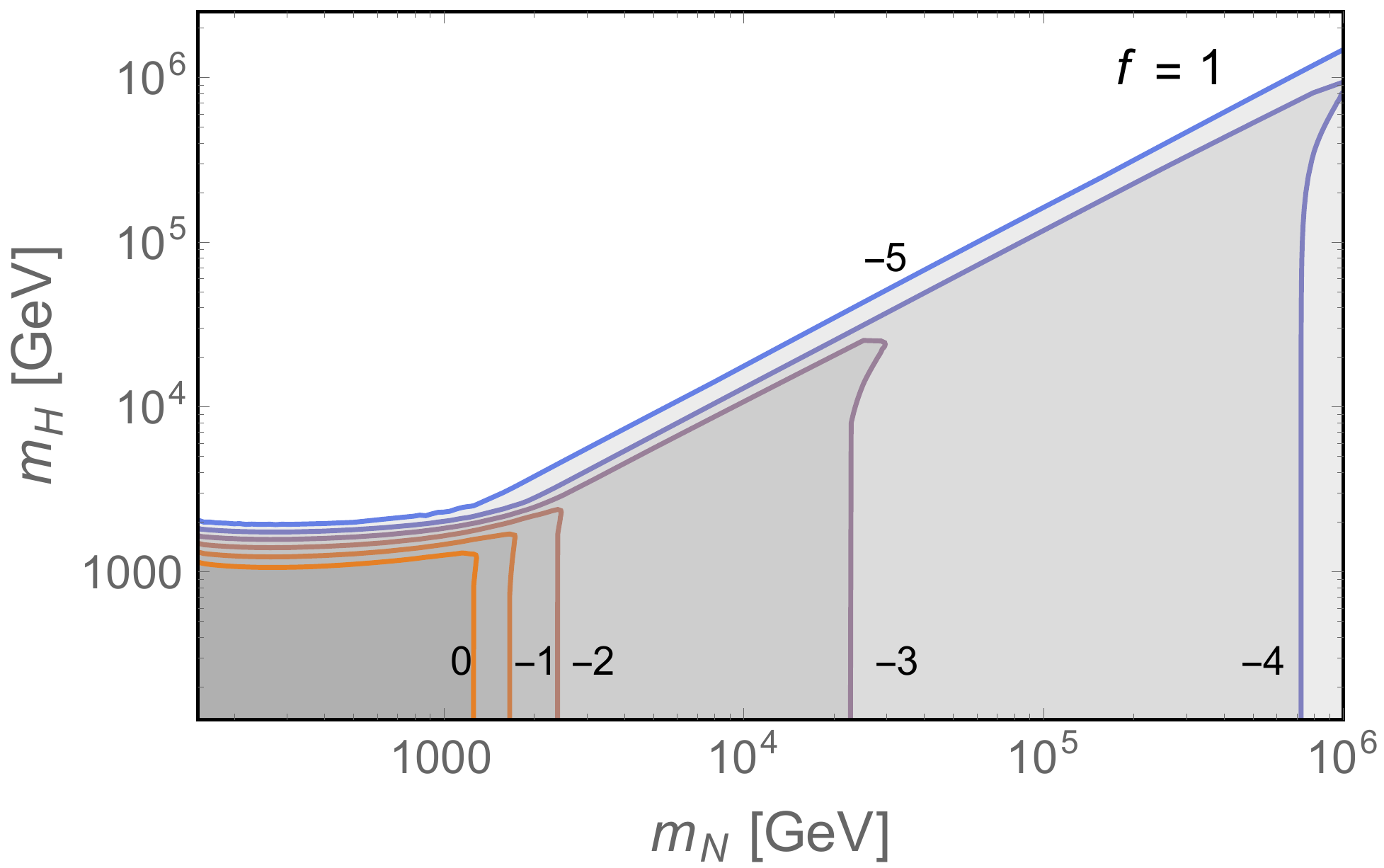}
  \includegraphics[height=0.31\textwidth]{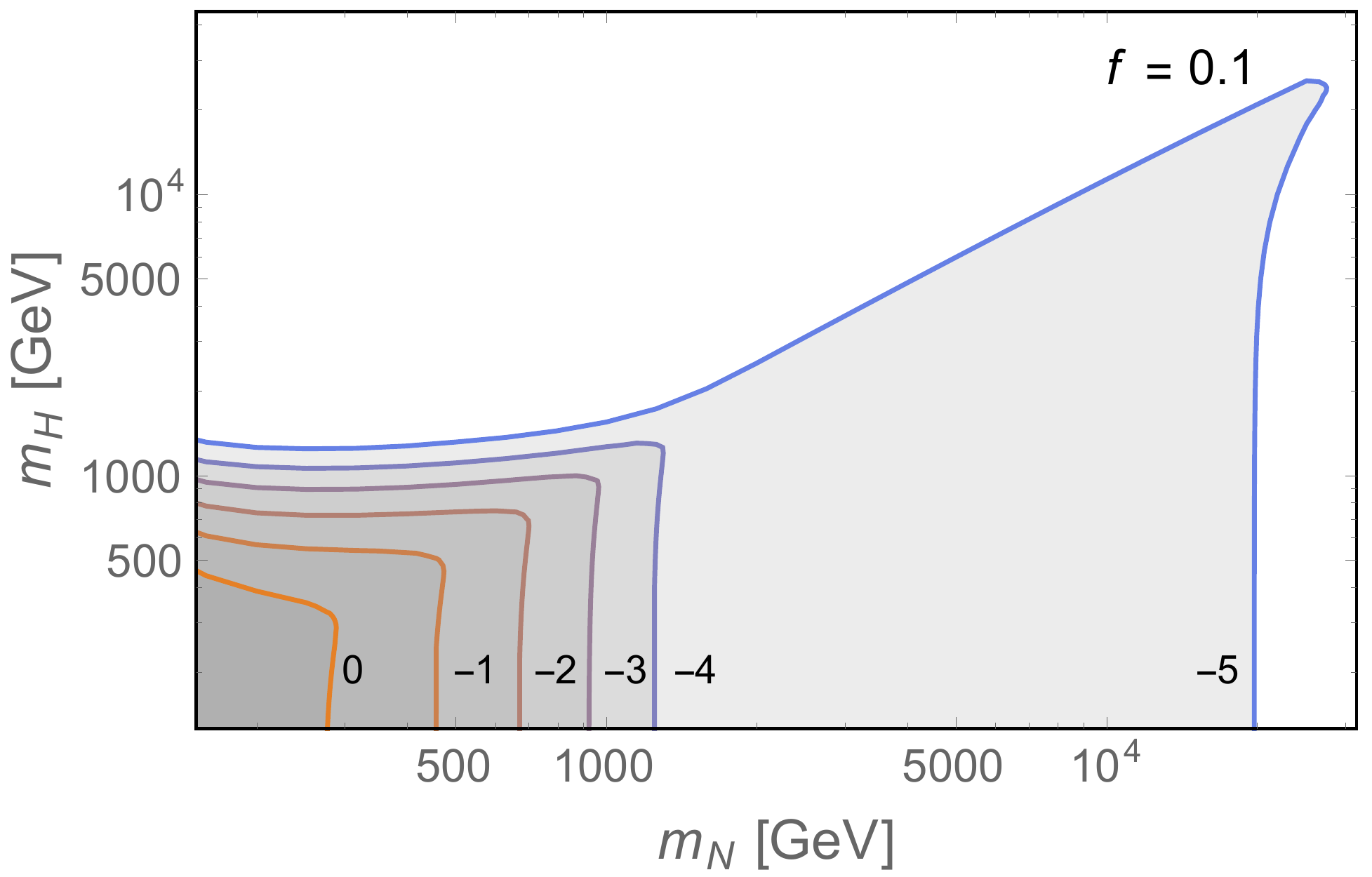}
  \caption{Contours of the asymmetry $\log_{10} \varepsilon_{\rm CP}$, as functions of the RHN mass $m_N$ and the CP-even scalar mass $m_H$, with $\widetilde{m} = 50$ meV and $f_H = 1$ (left) and $0.1$ (right). Within the shaded regions the lepton asymmetry is strongly washed out by the process $NN \to HH$ (cf. Fig.~\ref{fig:diagrams}), the type-I seesaw leptogenesis is falsified. }
  \label{fig:EFT1}
\end{figure}

\begin{figure}[t!]
  \centering
  \includegraphics[height=0.31\textwidth]{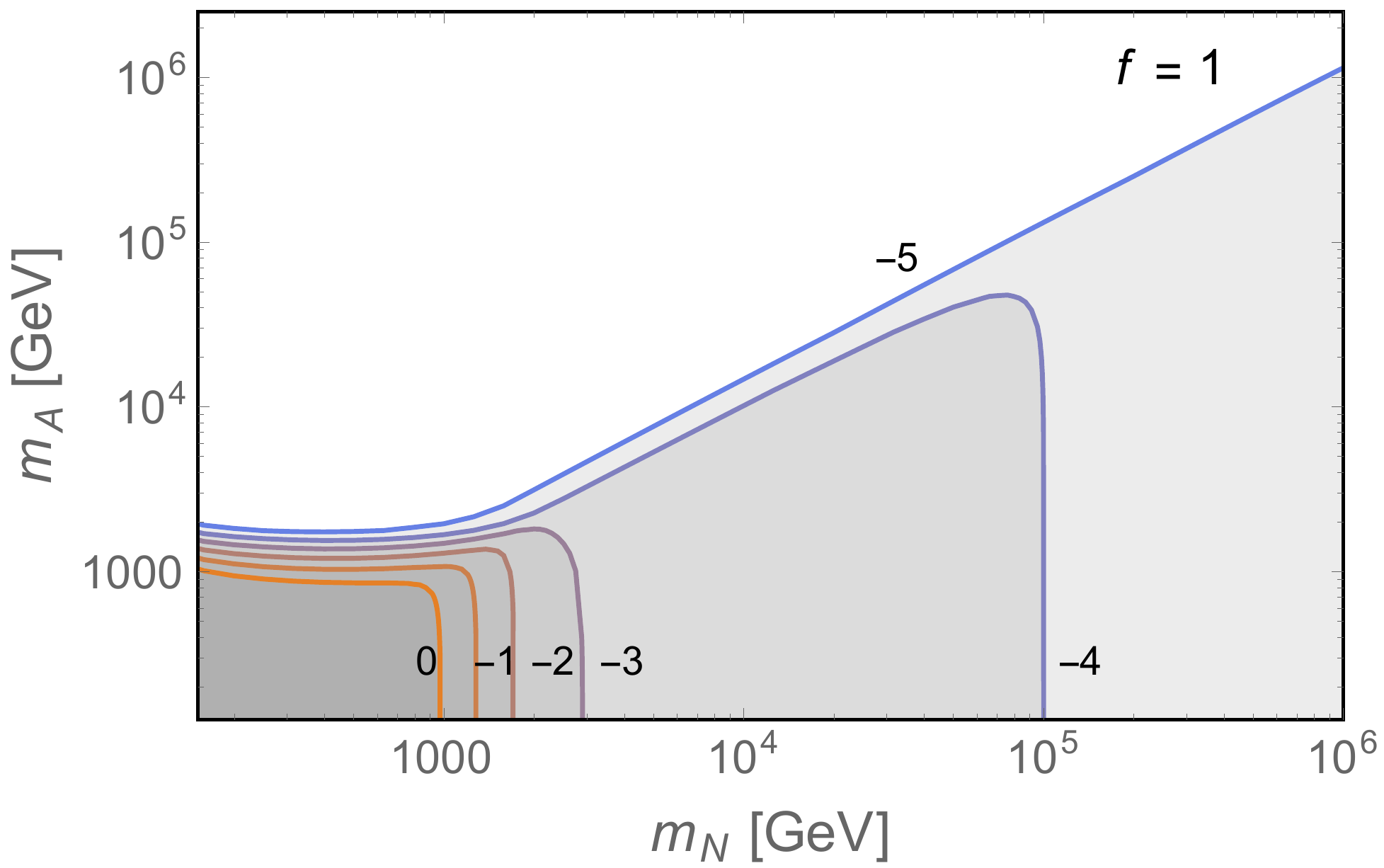}
  \includegraphics[height=0.31\textwidth]{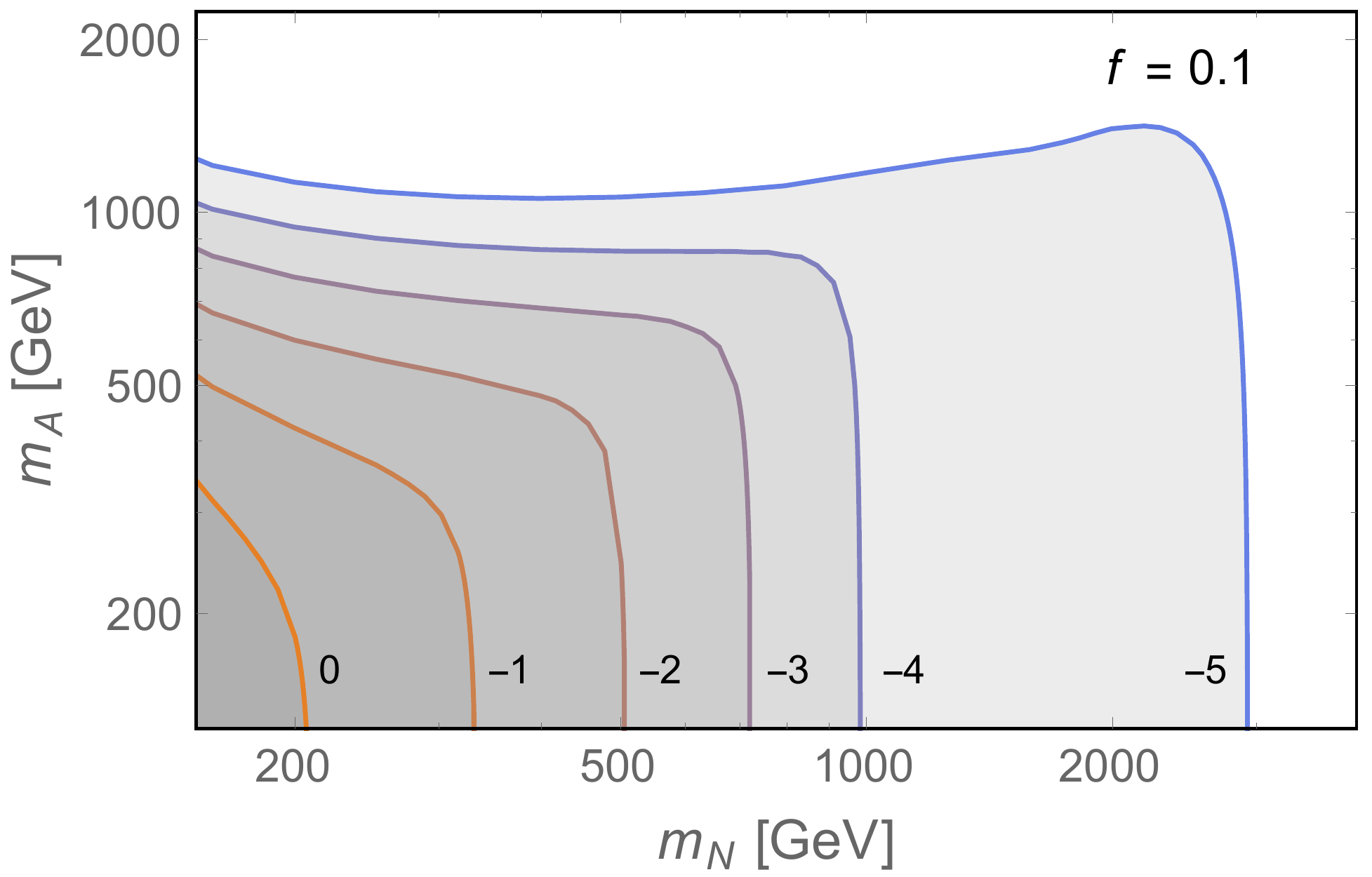}
  \caption{The same as in Fig.~\ref{fig:EFT1}, for the CP-odd scalar $A$, with respectively the Yukawa coupling $f_A = 1$ (left) and $0.1$ (right).}
  \label{fig:EFT2}
\end{figure}

In Fig.~\ref{fig:EFT1}, we show the required values of $\varepsilon_{\rm CP}$  to produce the correct lepton asymmetry $\eta_{\rm obs}^{\Delta L} \simeq 2.5 \times 10^{-8}$~\cite{Dev:2014laa}, which reproduces the observed baryon asymmetry $\eta_B^{\rm obs}\simeq 6\times 10^{-10}$~\cite{Ade:2015xua}, as functions of $m_N$ and the scalar mass $m_{H}$ {for two different values of $f=1$ (left panel) and 0.1 (right panel).}   The shaded regions below these contours are excluded {for the corresponding values of $\varepsilon_{\rm CP}$} due to the dilution effect of the new scalar. The same is shown in Fig.~\ref{fig:EFT2} for the CP-odd scalar case. As clearly seen in Figs.~\ref{fig:EFT1} and \ref{fig:EFT2}, successful leptogenesis demands that $m_{H,\, A} \gtrsim m_N$ in many regions of interest. {Since $\varepsilon_{\rm CP}$ cannot exceed one, a light $H/A$ thus requires a {\rm lower} bound on the RHN mass, as shown by the $\log_{10}\varepsilon_{\rm CP}=0$ contours. Depending on the actual value of $\varepsilon_{\rm CP}$ in a given model, a larger parameter space could be excluded, as illustrated by the other $\log_{10}\varepsilon_{\rm CP}$ contours.}

\begin{figure}[t!]
  \centering
  \includegraphics[height=0.31\textwidth]{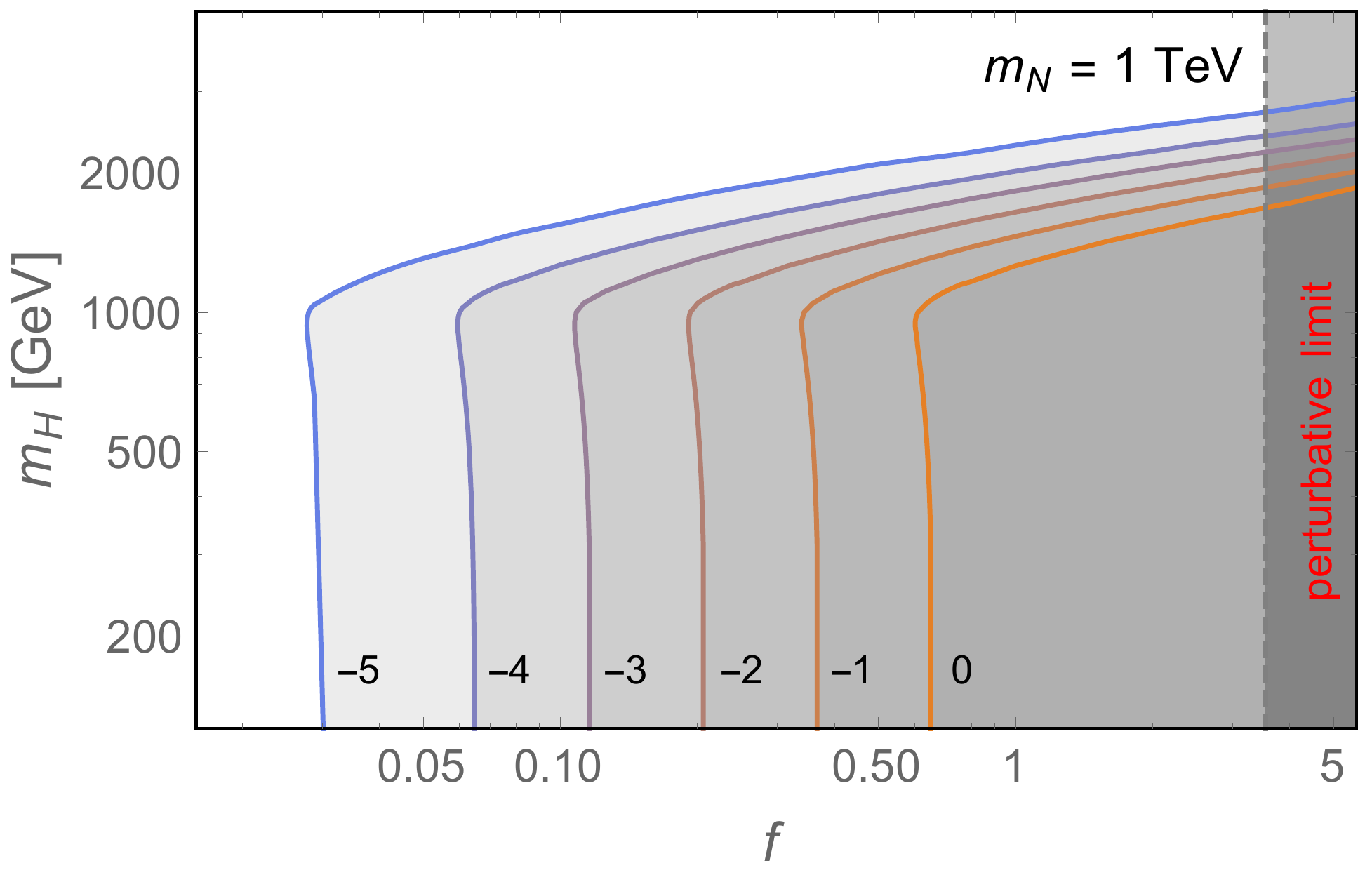}
  \includegraphics[height=0.31\textwidth]{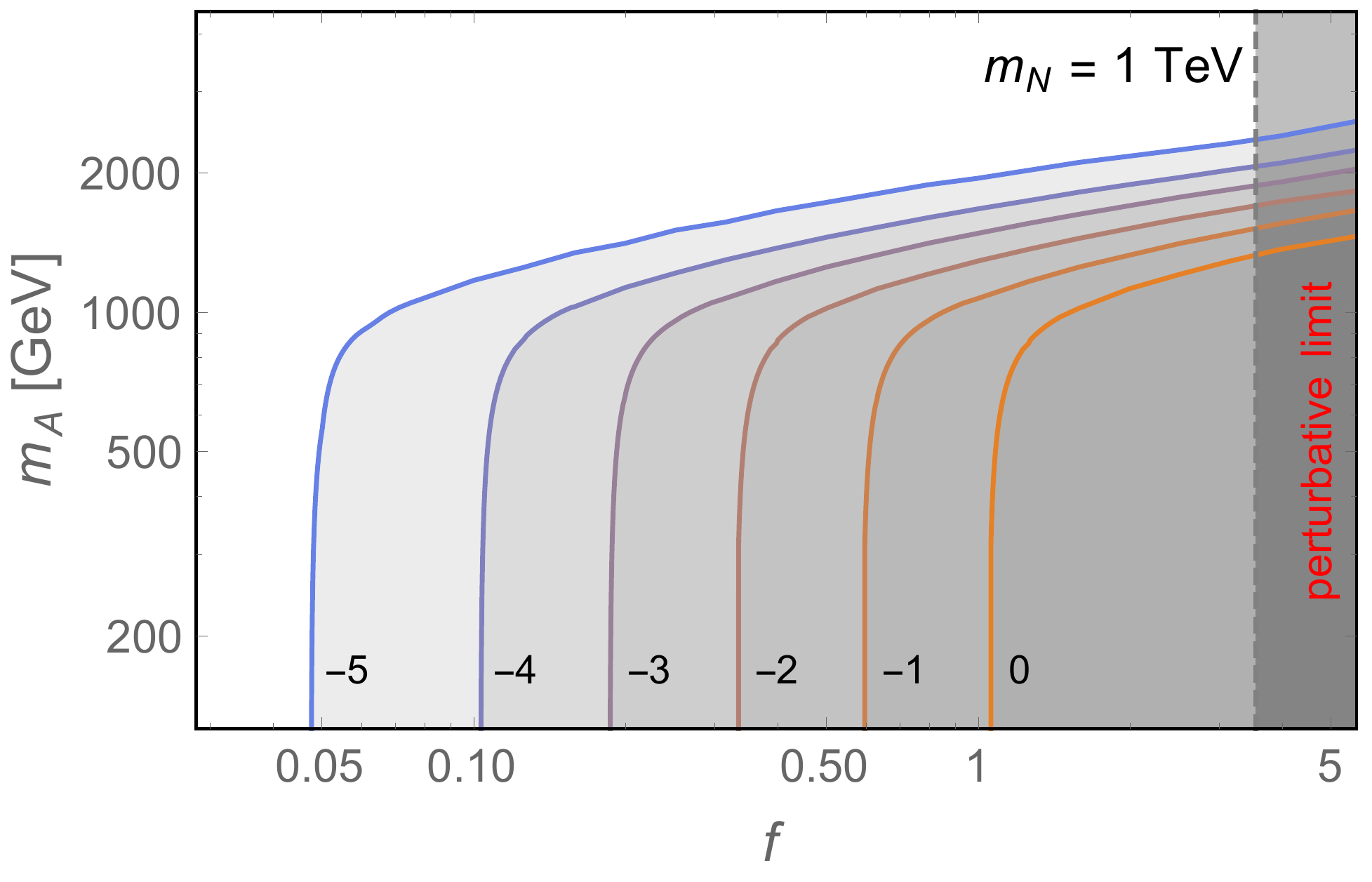}
  \caption{Contours of the asymmetry $\log_{10} \varepsilon_{\rm CP}$, as functions of the effective Yukawa coupling $f$ and scalar mass $m_H$ (left) or $m_A$ (right). Within the shaded regions the process $NN \to HH/AA$ in Fig.~\ref{fig:diagrams} makes type-I seesaw leptogenesis falsified, for $\widetilde{m} = 50$ meV and $m_N = 1$ TeV. The gray shaded region to the right of the vertical dashed line shows the non-perturbative regime $f > \sqrt{4\pi}$.}
  \label{fig:EFT3}
\end{figure}

By comparing the left and right panels of Figs.~\ref{fig:EFT1} and \ref{fig:EFT2} for different effective Yukawa couplings $f = 1$ and $0.1$, we find that a larger $f$ excludes larger regions of $m_N$ and $m_{H,\,A}$, as expected, since the scattering cross section $\sigma (NN \to HH/AA) \propto f^4$, and therefore, the dilution effect is stronger for larger values of $f$. This can be used to set limits on the coupling $f$, given the masses $m_N$ and $m_{H,\,A}$. As a explicit example, we show the contours of $\varepsilon_{\rm CP}$ in Fig.~\ref{fig:EFT3} as functions of $f$ and $m_{H,\,A}$ for $m_N = 1$ TeV. The shaded regions below these contours are excluded from successful leptogenesis requirement.
The gray bands on the right side of the vertical dashed line in Fig.~\ref{fig:EFT3} show the non-perturbative regime $f > \sqrt{4\pi}$. 
We find that for a TeV scale RHN, the effective Yukawa coupling is required to be smaller than 0.2 for a CP-even scalar $H$ below the 1 TeV scale if $\varepsilon_{\rm CP} = 10^{-2}$ and smaller than $0.03$ for $\varepsilon_{\rm CP} = 10^{-5}$, just to give some examples. Comparing the analytic cross sections in Eqs.~(\ref{eqn:sigmaNNHH}) and (\ref{eqn:sigmaNNAA}), one finds that the wash-out effect of the CP-odd scalar $A$ is somewhat weaker than that of the CP-even scalar $H$; therefore, in Fig.~\ref{fig:EFT2} the excluded regions are to some extent smaller than in Fig.~\ref{fig:EFT1}, and in Fig.~\ref{fig:EFT3} the limits on the effective coupling $f$ for the CP-odd scalar $A$ is comparatively weaker: $f \gtrsim 0.3$ if $\varepsilon_{\rm CP} = 10^{-2}$ and $f \gtrsim 0.05$ for $\varepsilon_{\rm CP} = 10^{-5}$.

If there are more than one physical scalars that couple to the RHN simultaneously, then roughly speaking the dilution effect will be strengthened by the extra degrees of freedom, unless there is a destructive interference due to some specific (fine-tuned) choice of the phases. For instance, assuming the interaction $f_i S_i N N$ with roughly a universal coupling $f_i$, the cross section $\sigma(NN \to S_i S_i) \propto \textsf{N}_S \, \sigma(NN \to SS)$ ($SS$ could be either $HH$ or $AA$ in our case), with $\textsf{N}_S$ the numbers of scalars that are lighter than the RHN. Then more stringent limits could be imposed on the scalar masses and couplings $f_i$. This gets more involved when the scalar self-interactions are also taken into account, e.g. in the case of the global $U(1)_{B-L}$ model in Sec.~\ref{sec:global} (see below)  where we have both a CP-even and odd scalars.

\section{Global $U(1)_{B-L}$ model}
\label{sec:global}

We now extend the `effective theory' results of Sec.~\ref{sec:EFT} to two realistic models, which are based on the $U(1)_{B-L}$ symmetry. In this section we deal with a global $B-L$ symmetry (the local $B-L$ symmetry case will be considered in the Sec.~\ref{sec:local}).  This is a well-motivated simplified model to generate the RHN masses and implement the seesaw mechanism, and could originate from a more fundamental UV-theory at high scale~\cite{Chikashige:1980ui,Schechter:1981cv}. In this class of models, we have both the CP-even and odd scalars as well as new interaction in the scalar sector beyond the diagram in Fig.~\ref{fig:diagrams}, that could lead to resonant enhancement of the dilution effect. {See also Ref.~\cite{Sierra:2014sta} for related discussion.}

The physical scalars in this scenario arise from a complex scalar singlet $\Delta_R$ that carries two units of lepton number and couples to the RHNs through the Yukawa Lagrangian
\begin{eqnarray}
\label{eqn:LYukawa2}
{\cal L}_Y \ = \ - f \Delta_R \overline{N}^c N + {\rm H.c.} \,.
\end{eqnarray}
After symmetry breaking with the VEV $\langle \Delta_R \rangle = v_R$, we obtain the RHN Majorana mass $m_N = 2 f v_R$.
The scalar potential of $\Delta_R$ reads (omitting its possible interactions with the SM Higgs boson)
\begin{eqnarray}
\label{eqn:potential}
V_\Delta \ = \ - \mu^2 (\Delta_R^\dagger \Delta_R) + \rho (\Delta_R^\dagger \Delta_R)^2 \,.
\end{eqnarray}
Expanding the $\Delta_R$ field about its VEV, $\Delta_R = v_R + \frac{1}{\sqrt2} H_3 + \frac{i}{\sqrt2} J$, we can obtain mass $m_{H_3}^2 = 4 \rho v_R^2$ for the physical CP-even scalar $H_3$,\footnote{We follow the same notation as in Ref.~\cite{Dev:2016dja}, although the model considered  here is different.} while the accompanying Goldstone component is the massless Majoron particle $J$. In principle, the Majoron could have a non-vanishing mass through the gravitational interactions~\cite{Akhmedov:1992hi, Rothstein:1992rh} or by interacting with other particles beyond the type-I seesaw scheme. Even if it is the case, as long as its mass $m_J \ll m_N$, that does not essentially make any difference to our following discussion. For a light Majoron, e.g. with mass $\lesssim {\rm GeV}$, in the type-I seesaw its couplings to the active neutrinos are proportional to the heavy-light neutrino mixing angle $V_{\nu N}^2$. For a TeV scale $N$, the mixing $V_{\nu N} \sim 10^{-6}$, and we are then safe from the Majoron limits from supernova observations~\cite{Heurtier:2016otg} and other astrophysical and terrestrial constraints~\cite{Pilaftsis:1993af, Arnold:2006sd}. Here we assume this to be the case. {For simplicity, we assume that the Majoron decouples from the SM particles much above the QCD phase transition temperature ($\sim 200$ MeV) as is the case for the singlet Majoron model~\cite{Chikashige:1980ui}, so that its effect on $N_{\rm eff}$ is very small due to entropy dilution.}



\begin{figure}[t!]
  \centering
  \begin{tikzpicture}[]
  \draw[neutrino,thick] (-1.4,1)node[left]{{\footnotesize$N$}} -- (-0.6,0);
  \draw[neutrino] (-1.4,-1)node[left]{{\footnotesize$N$}} -- (-0.6,0);
  \draw[dashed,thick](-0.6,0)--(0,0)node[above]{{\footnotesize$H_3$}}--(0.6,0);
  \draw[dashed,thick](1.4,1)node[right]{{\footnotesize$H_3$}}--(0.6,0);
  \draw[dashed,thick](1.4,-1)node[right]{{\footnotesize$H_3$}}--(0.6,0);
  \end{tikzpicture}
\hspace{0.5cm}
  \begin{tikzpicture}[]
  \draw[neutrino,thick] (-1.4,1)node[left]{{\footnotesize$N$}} -- (-0.6,0);
  \draw[neutrino] (-1.4,-1)node[left]{{\footnotesize$N$}} -- (-0.6,0);
  \draw[dashed,thick](-0.6,0)--(0,0)node[above]{{\footnotesize$J$}}--(0.6,0);
  \draw[dashed,thick](1.4,1)node[right]{{\footnotesize$H_3$}}--(0.6,0);
  \draw[dashed,thick](1.4,-1)node[right]{{\footnotesize$J$}}--(0.6,0);
  \end{tikzpicture}
\hspace{0.5cm}
  \begin{tikzpicture}[]
  \draw[neutrino,thick] (-1.4,1)node[left]{{\footnotesize$N$}} -- (-0.6,0);
  \draw[neutrino] (-1.4,-1)node[left]{{\footnotesize$N$}} -- (-0.6,0);
  \draw[dashed,thick](-0.6,0)--(0,0)node[above]{{\footnotesize$H_3$}}--(0.6,0);
  \draw[dashed,thick](1.4,1)node[right]{{\footnotesize$J$}}--(0.6,0);
  \draw[dashed,thick](1.4,-1)node[right]{{\footnotesize$J$}}--(0.6,0);
  \end{tikzpicture}
  \caption{Feynman diagrams for the scattering $N N \to H_3H_3,\, H_3 J,\, JJ$. }
  \label{fig:diagrams2}
\end{figure}
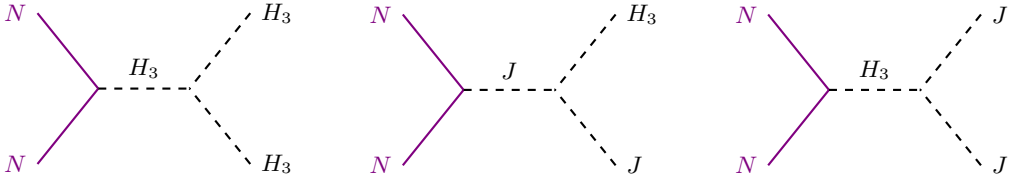

From the potential~(\ref{eqn:potential}) we could get the triple scalar couplings, which can be parameterized as functions of the CP-even scalar mass $m_{H_3}$ and the VEV $v_R$~\cite{Dev:2016dja}
\begin{eqnarray}
\label{eqn:triple}
\lambda_{H_3 H_3 H_3} & \ = \ & - 6 \times \frac{m_{H_3}^2}{2\sqrt2 v_R} \,, \qquad \qquad
\lambda_{H_3 JJ}  \ = \  - 2 \times \frac{m_{H_3}^2}{2\sqrt2 v_R}
\end{eqnarray}
with $6$ and $2$ the symmetric factors for identical particles. These couplings induce the extra diagrams in Fig.~\ref{fig:diagrams2}, that contributes to the processes $N N \to H_3 H_3$, $H_3 J$, $J J$ and interfere with those from the pure Yukawa interactions in Fig.~\ref{fig:diagrams}. The partial decay widths of $H_3$ into heavy RHNs and massless Majorons are respectively
\begin{eqnarray}
\label{eqn:width1}
\Gamma (H_3 \to N N) & \ = \ &
\frac{m_{H_3} m_N^2}{64\pi v_R^2}
\left( 1 - \frac{4m_N^2}{m_{H_3}^2} \right)^{3/2}
\Theta (m_{H_3} - 2 m_N) \,, \\
\label{eqn:width2}
\Gamma (H_3 \to JJ) & \ = \ &
\frac{m_{H_3}^3}{64\pi v_R^2} \,,
\end{eqnarray}
where we have traded the Yukawa and scalar coupling into the particle masses via $f = m_N / 2 v_R$ and $\lambda_{H_3 JJ} = - m_{H_3}^2 / \sqrt2 v_R$. The reduced cross sections $\hat{\sigma} (NN \to H_3 H_3 ,\, H_3 J, \, JJ)$ are collected in Appendix~\ref{sec:appendixA} (see also~\cite{Pilaftsis:2008qt, Sierra:2014sta}). It should be noted that in the third diagram in Fig.~\ref{fig:diagrams2}, i.e. $NN \to H_3^{(\ast)} \to JJ$, if the $H_3$ mediator is on-shell, this corresponds to the inverse decay process $N N \to H_3$ with the scalar $H_3$ decays further into two massless Majorons $H_3 \to JJ$. Compared to the two-to-two process $NN \to JJ$ (combining both the $s$ and $t$ channels in Fig.~\ref{fig:diagrams} and \ref{fig:diagrams2}, with the $H_3$ propagator off-shell), the inverse decay could enhance largely the dilution effect, as a result of the resonance structure.

\begin{figure}[t!]
  \centering
  \includegraphics[height=0.315\textwidth]{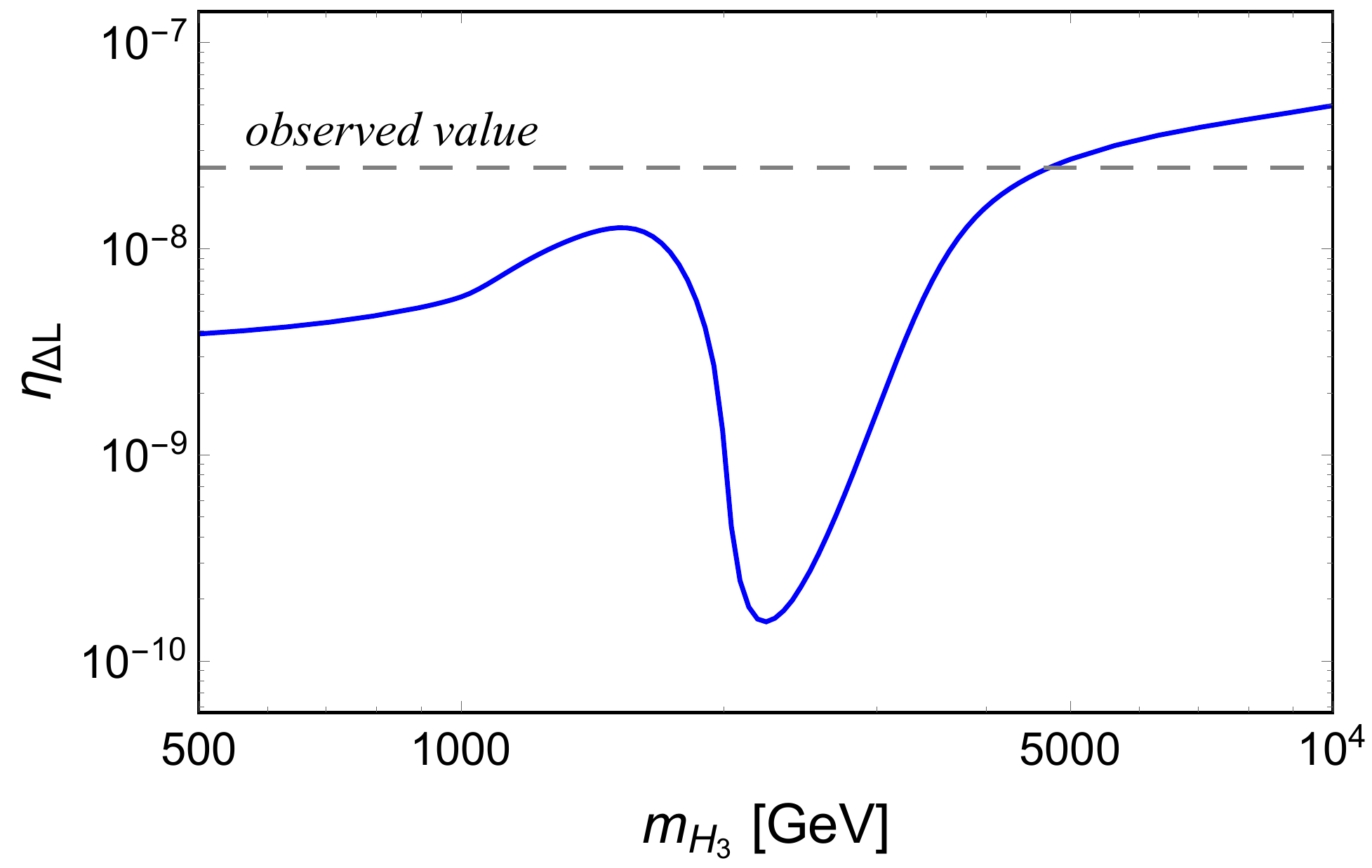}
  \caption{Dependence of lepton asymmetry on the scalar mass $m_{H_3}$ in the global $U(1)_{B-L}$ model, with $\widetilde{m} = 50$ meV, $m_N = 1$ TeV, $v_R = 1$ TeV and $\varepsilon_{\rm CP} = 1$.
}
  \label{fig:example3}
\end{figure}

\begin{figure}[t!]
  \centering
  \includegraphics[height=0.34\textwidth]{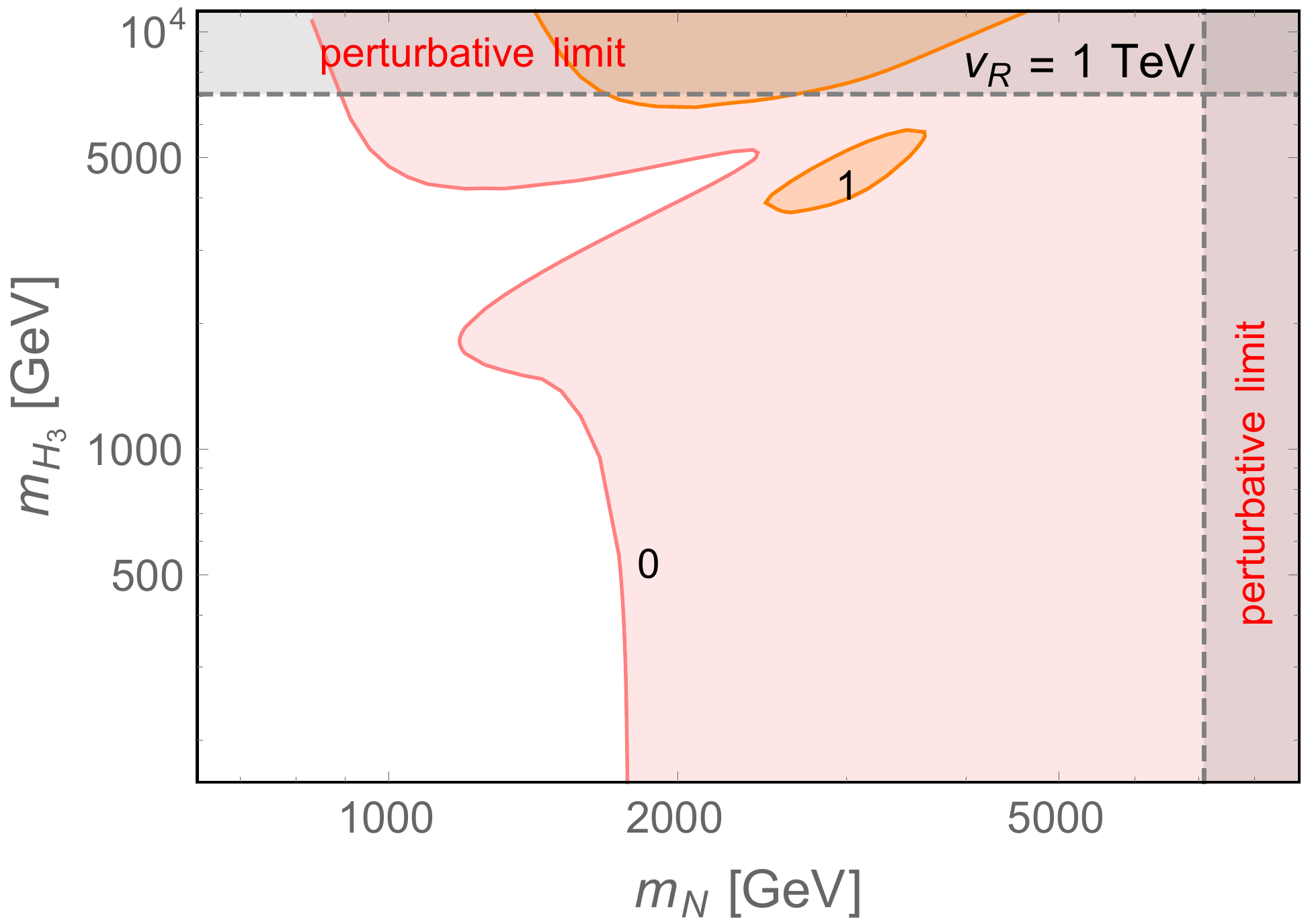}
  \includegraphics[height=0.345\textwidth]{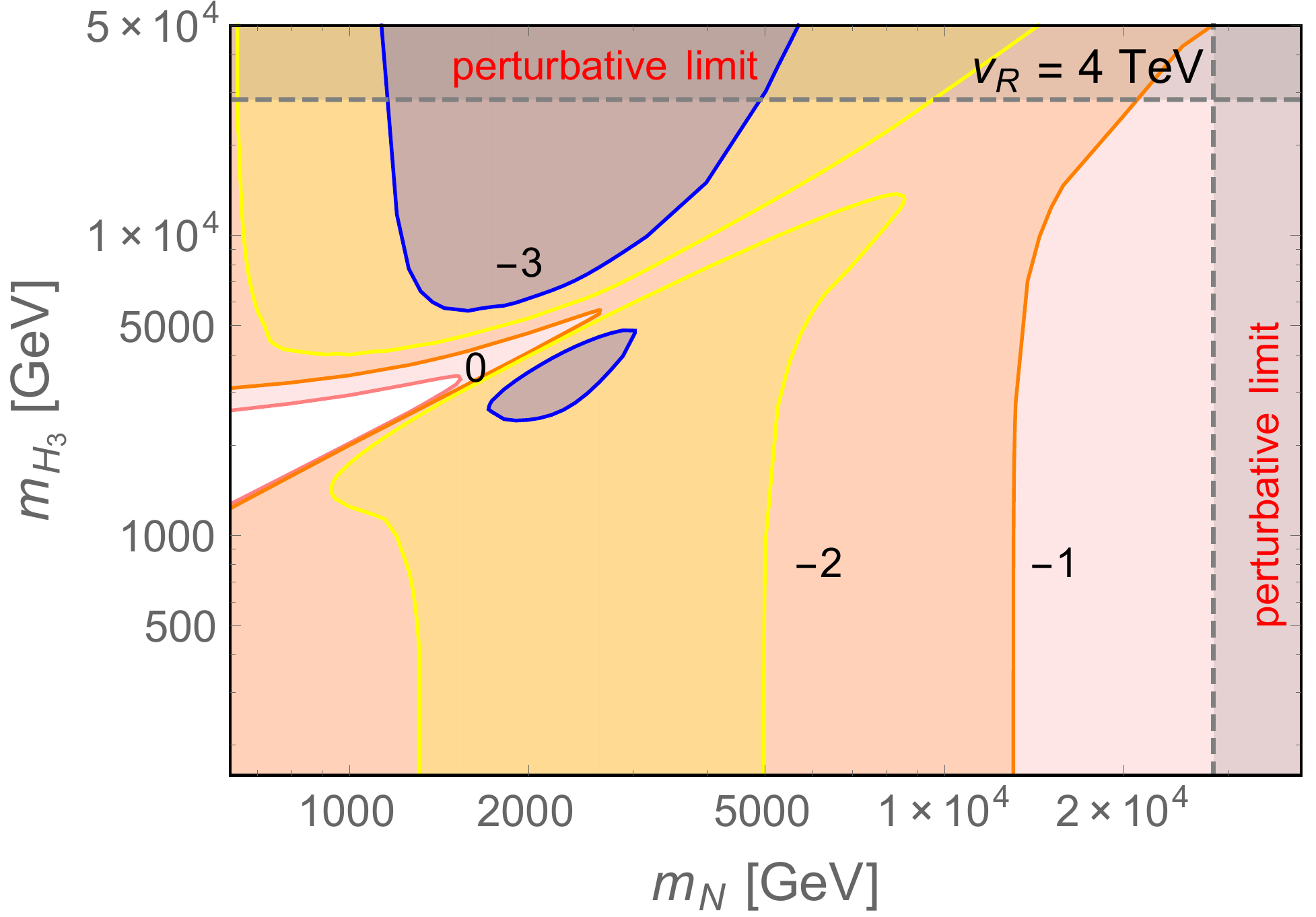}
  \caption{Contours of the asymmetry $\log_{10} \varepsilon_{\rm CP}$, as functions of the RHN mass $m_N$ and the scalar mass $m_H$ in the global $U(1)_{B-L}$ model, with $\widetilde{m} = 50$ meV, $v_R = 1$ TeV (left) and $4$ TeV (right). Here the colored regions could generate the observed lepton asymmetry (and not falsified) by the processes $NN \to H_3 H_3$, $H_3 J$, $JJ$. The gray vertical (horizontal) bands on the right (top) are exclusion regions corresponding to the perturbative limits of the couplings $f < \sqrt{4\pi}$ and $\rho < 4\pi$, respectively. }
  \label{fig:global}
\end{figure}

 An example is presented in Fig.~\ref{fig:example3} where one can see clearly that the resonance structure occurs at $m_{H_3} \simeq 2 m_N$. The corresponding Boltzmann equation in the global $U(1)_{B-L}$ model is easily obtained by replacing the $2\gamma_{HH,\,AA}$ term in Eq.~(\ref{eqn:Boltzmann}) by summation of all the terms which reduce the RHN number by unit of two, i.e.
\begin{eqnarray}
2 (\gamma_{H_3H_3} + \gamma_{H_3 J} + \gamma_{JJ}) \,.
\end{eqnarray}
As in Fig.~\ref{fig:EFT1} and \ref{fig:EFT2}, one can easily figure out in which regions could leptogenesis be falsified by the extra scalars, which are presented in Fig.~\ref{fig:global}, for two benchmark values of $v_R = 1$ TeV and $4$ TeV. Within the colored regions, we can generate the observed lepton asymmetry from RHN decay, with the indicated values of $\varepsilon_{\rm CP}$ and $\widetilde{m} = 50$ meV. The gap structure corresponds to the resonance $NN \to H_3 \to JJ$ [cf. Fig.~\ref{fig:example3}]. Note that for fixed value of $v_R$ in both panels, the Yukawa coupling changes as $f = m_N / 2 v_R$. Together with $m_{H_3}^2 = 4 \rho v_R^2$, we set the perturbative boundaries $f < \sqrt{4\pi}$ and $\rho < 4\pi$ in Fig.~\ref{fig:global}.\footnote{The perturbative limits are different for the Yukawa and scalar quartic couplings, because they appear with different powers in the renormalization group equations; see e.g., Ref.~\cite{Dev:2015vjd}.} By comparing the two panels, one can see that when the VEV $v_R$ is larger, the Yukawa coupling $f$ gets smaller, and more regions are allowed, as shown by the broader colorful regions in the right panel of Fig.~\ref{fig:global}.

\begin{figure}[t!]
  \centering
  \includegraphics[height=0.35\textwidth]{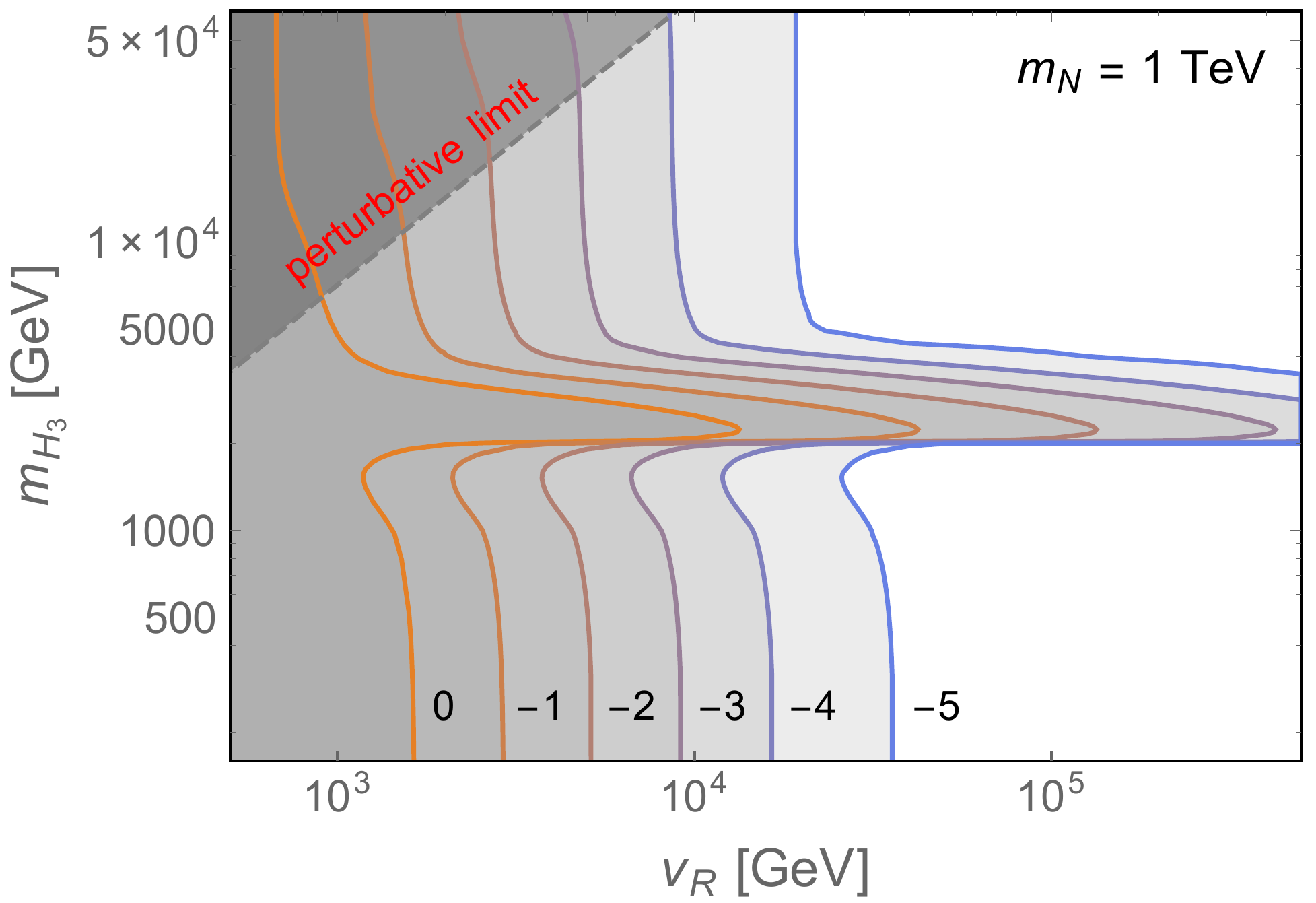}
  \caption{Contours of the asymmetry $\log_{10} \varepsilon_{\rm CP}$, as functions of $v_R$ and the scalar mass $m_H$ in the global $U(1)_{B-L}$ model, with $\widetilde{m} = 50$ meV and $m_N = 1$ TeV. The shaded regions are falsified by the processes $NN \to H_3 H_3$, $H_3 J$, $JJ$.
 The darker region at the top left corner is excluded by the perturbativity limit $\rho < 4\pi$.}
  \label{fig:global2}
\end{figure}

In analogy to Fig.~\ref{fig:EFT3} we can set limits on the $v_R$ scale and scalar mass $m_{H_3}$, which are shown in Fig.~\ref{fig:global2} with $m_N = 1$ TeV. In this plot one can see clearly the leptogenesis limit on the $v_R$ scale, and the dependence on the physical scalar mass $m_{H_3}$. For a 1 TeV RHN, the $v_R$ scale is required to be larger than 2.1 TeV for $\varepsilon_{\rm CP} = 10^{-2}$ and 19 TeV if $\varepsilon_{\rm CP} = 10^{-5}$, in the large $m_{H_3}$ limit. The limits become more stringent when $H_3$ is lighter. At the resonance $m_{H_3} \simeq 2 m_N$, the constraints on $v_R$ could even be an  order of magnitude higher, which is determined largely by the $H_3$ width. One should note that in the large $m_{H_3}$ limit the dilution is dominated by the (almost) massless final state $NN \to H_3^\ast \to JJ$, which however does not decouple as $1/m_{H_3}^4$, as naively expected, as the triple scalar coupling $\lambda_{H_3 JJ}$ in Eq.~(\ref{eqn:triple}) scales like $m_{H_3}^2$ which cancels out the scalar mass $1/m_{H_3}^2$ in the propagator. Thus the limits on $v_R$ approaches to be a constant in Fig.~\ref{fig:global2} in the large $m_{H_3}$ limit, as long as the quartic coupling $\rho$ in Eq.~(\ref{eqn:potential}) is perturbative.

\section{Local $U(1)_{B-L}$ model}
\label{sec:local}

In this section we consider a local $U(1)_{B-L}$ model, with the gauge group $SU(2)_L \times U(1)_{I_{3R}} \times U(1)_{B-L}$ before symmetry breaking, which can be viewed
in some sense as the ``effective'' theory of LR model~\cite{Pati:1974yy, Mohapatra:1974gc, Senjanovic:1975rk} at the TeV scale with the $SU(2)_R$ breaking scale and the mass of the heavy $W_R$ bosons much higher.\footnote{For $m_{W_R}< m_{Z_R}$, as expected in generic LR models, the $W_R$ boson will play more important role than the $Z_R$ boson in diluting the lepton asymmetry. For instance, it could mediate $\Delta L=1$ processes like $NL \to W_R^\ast \to Qu_R$ and the three-body decay $N \to L W_R^\ast \to L q \bar{q}$ if $m_{W_R} > m_N$, which would dilute the RHN number density before the temperature $T_c$, thus imposing very stringent limits on the $W_R$ boson mass~\cite{Frere:2008ct, Dev:2015vra}. If $m_{W_R} < m_N$, then the situation becomes worse, as the two-body decay $N \to L W_R$ is open, which does not generate any lepton asymmetry, as it is dictated by the gauge interaction, but only contributes to dilution of the asymmetry. 
} The SM fermion doublets $Q$, $L$ and singlets $u_R$, $d_R$, $e_R$ have the following quantum number assignments under the gauge group:
\begin{align}
& Q=(u_L, \: d_L)^{\sf T}: \left({\bf 2}, 0, \frac13\right);  \qquad
L=(\nu, \:  e_L)^{\sf T}: \left({\bf 2}, 0, -1 \right); \nonumber \\
& u_R: \left({\bf 1}, \frac12, \frac13 \right); \quad
d_R: \left({\bf 1}, -\frac12, \frac13 \right); \quad
e_R: \left({\bf 1},-\frac12, -1 \right).
\end{align}
Anomaly freedom necessitates the introduction of three RHNs $N_i: ({\bf 1}, 1/2, -1)$. In the minimal scalar sector, we have the SM Higgs doublet $\phi({\bf 2},-1/2, 0)$ and a singlet $\Delta_R ({\bf 1},-1,2)$. The singlet VEV $\langle\Delta_R\rangle=v_R$ breaks the gauge symmetry down to the SM gauge group $SU(2)_L \times U(1)_Y$ which is further broken by  the doublet VEV $\langle \phi^0\rangle=v_{\rm EW}$ to $U(1)_{\rm em}$, leading to the type I seesaw formula for neutrino masses. In this model, $H_3 = {\rm Re}(\Delta_R)$ is a physical CP-even scalar that couples directly to the RHNs as in Eq.~(\ref{eqn:LYukawa2}), and the Goldstone mode is eaten by the heavy $Z_R$ boson, which acquires a mass
\begin{eqnarray}
\label{eqn:MZR}
m_{Z_R}^2 \ = \ 2 (g_R^2 + g_{BL}^2) v_R \,,
\end{eqnarray}
where $g_R$ and $g_{BL}$ are the gauge couplings associated with the $U(1)_{I_{3R}}$ and $U(1)_{B-L}$ gauge groups.
The $Z_R$ couplings to the chiral fermions $f_{L,R}$ with electric charge $Q_f$ and third-component of isospin $I_{3,f}$ are given by
\begin{eqnarray}
\label{eqn:ZR1}
g_{Z_R f_L f_L} & \ = \ & \frac{e}{\cos\theta_w} \,
(I_{3,f} - Q_f) \,
\frac{\sin\phi}{\cos\phi} \,, \\
\label{eqn:ZR2}
g_{Z_R f_R f_R} & \ = \ & \frac{e}{\cos\theta_w} \,
(I_{3,f} - Q_f \sin^2\phi) \,
\frac{1}{\sin\phi \cos\phi} \,
\end{eqnarray}
with $\tan\phi \equiv g_{BL} / g_R$ the right-handed gauge mixing angle, $\theta_w$ the weak mixing angle and $e$ the electric charge. One can find more details on how to obtain these gauge couplings in Appendix~\ref{sec:appendixB}.

In presence of the scalar $H_3$ and the $Z_R$ boson, we can have the processes $NN \to Z_R^{(\ast)} \to f \bar{f}$ ($f$ being the SM fermions) as wall as $NN \to H_3 H_3$, $H_3 Z_R$, $Z_R Z_R$, which all reduce the RHN number by two units~\cite{Heeck:2016oda}. Our model in this section is however different from Ref.~\cite{Heeck:2016oda} in the following aspects: (i) The gauge structure in our model is different from that in Ref.~\cite{Heeck:2016oda}, although both are claimed to be $U(1)_{B-L}$ models. The $U(1)_{B-L}$ in~\cite{Heeck:2016oda} does not contribute to electric charge and their $Z'$ is a pure $B-L$ boson, while in our case the $Z_R$ is a mixture of the ${I_{3R}}$ and $B-L$; (ii) Ref.~\cite{Heeck:2016oda} concentrates on the gauge boson, while in this paper we focus on the impact of $B-L$ breaking scalars on leptogenesis, which are largely complementary to each other.

Before proceeding to the leptogenesis constraints on the scalar and gauge bosons, we would like to mention that the $Z_R$ mass is tightly constrained by the LHC dilepton data $pp \to Z_R \to \ell^+\ell^-$ with $\ell = e,\,\mu$~\cite{ATLAS:2016cyf, CMS:2016abv, Patra:2015bga, Klasen:2016qux}. The current limits can be cast onto our model by rescaling the couplings and production cross sections of the sequential $Z'$ boson adopted in the data analysis~\cite{ATLAS:2016cyf, CMS:2016abv}. In our model, the gauge coupling $g_R$ is a free parameter, which is related to the $U(1)_Y$ coupling $g_Y$ via $g_Y^{-2} = g_R^{-2} + g_{BL}^{-2}$. To keep $g_{BL}$ {\it real}, there is a lower limit on $g_R$, i.e. $g_R > g_L \tan\theta_w \simeq 0.55 g_L$~\cite{Dev:2016dja}.\footnote{If the coupling $g_{BL}$ is required to be perturbative, i.e. $< \sqrt{4\pi}$, the lower limit on $g_R$ is slightly more stringent~\cite{Dev:2017dui}.} For simplicity, we take $g_R = g_L$ for which the dilepton limit is $m_{Z_R} > 3.72$ TeV~\cite{Dev:2016xcp}.

\begin{figure}[t!]
  \centering
  \includegraphics[height=0.315\textwidth]{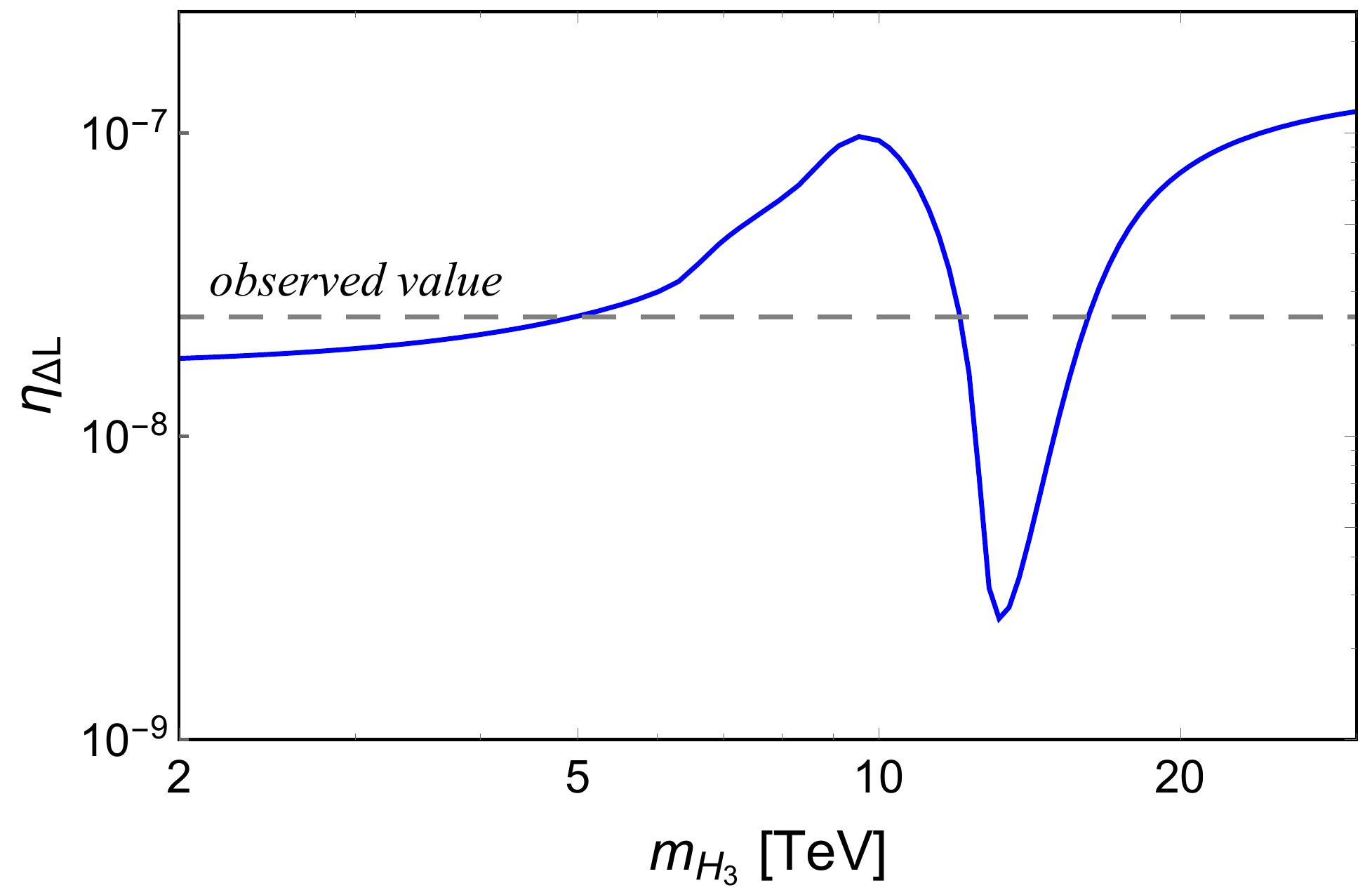}
  \caption{Dependence of lepton asymmetry on the scalar mass $m_{H_3}$ in the local $U(1)_{B-L}$ model, with $\widetilde{m} = 50$ meV, $\varepsilon_{\rm CP}=1$, $v_R$ = 4 TeV, $m_N = 6.3$ TeV, $g_R = g_L$.
}
  \label{fig:example3b}
\end{figure}

\begin{figure}[t!]
  \centering
  \includegraphics[width=0.48\textwidth]{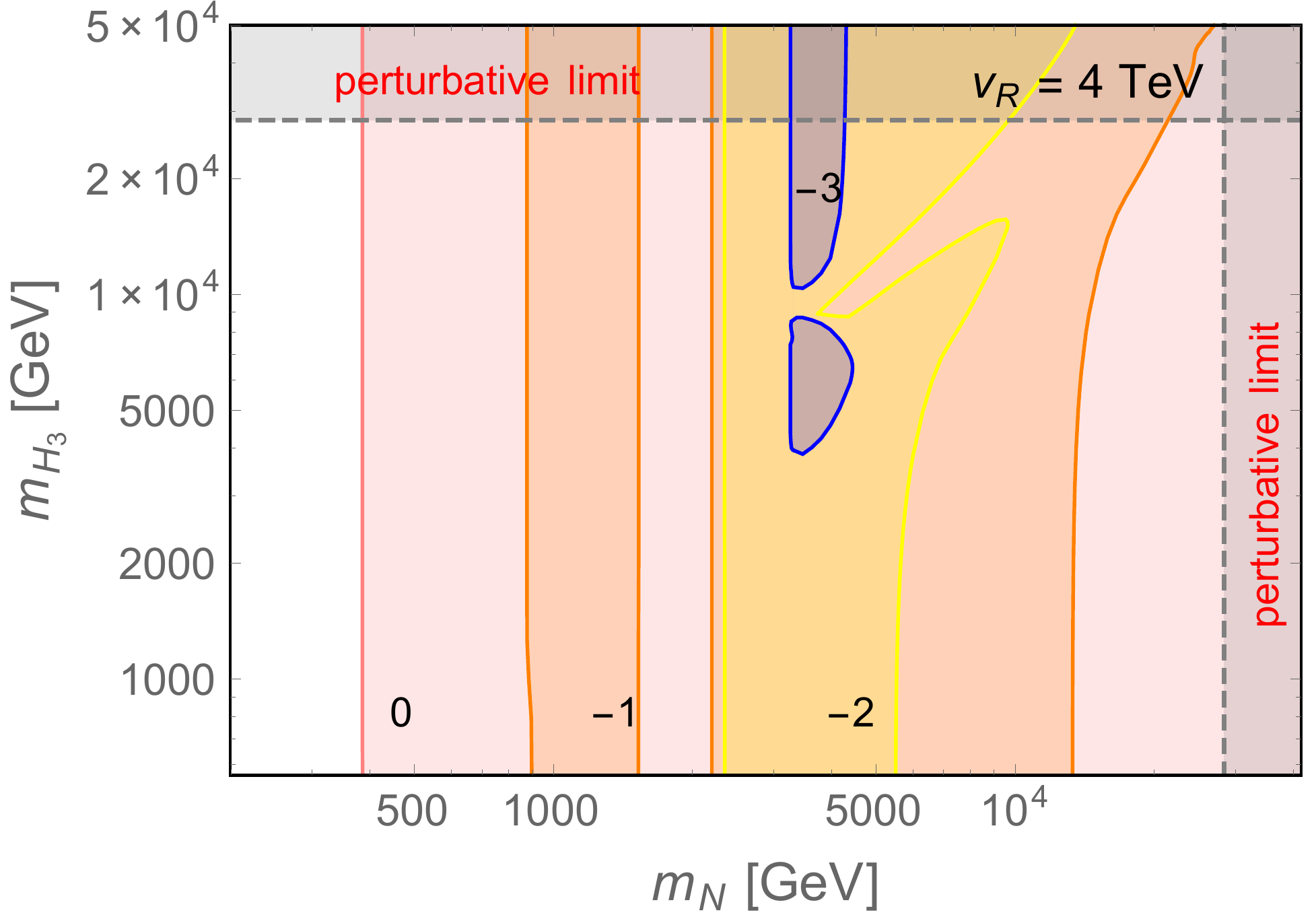}
  \includegraphics[width=0.484\textwidth]{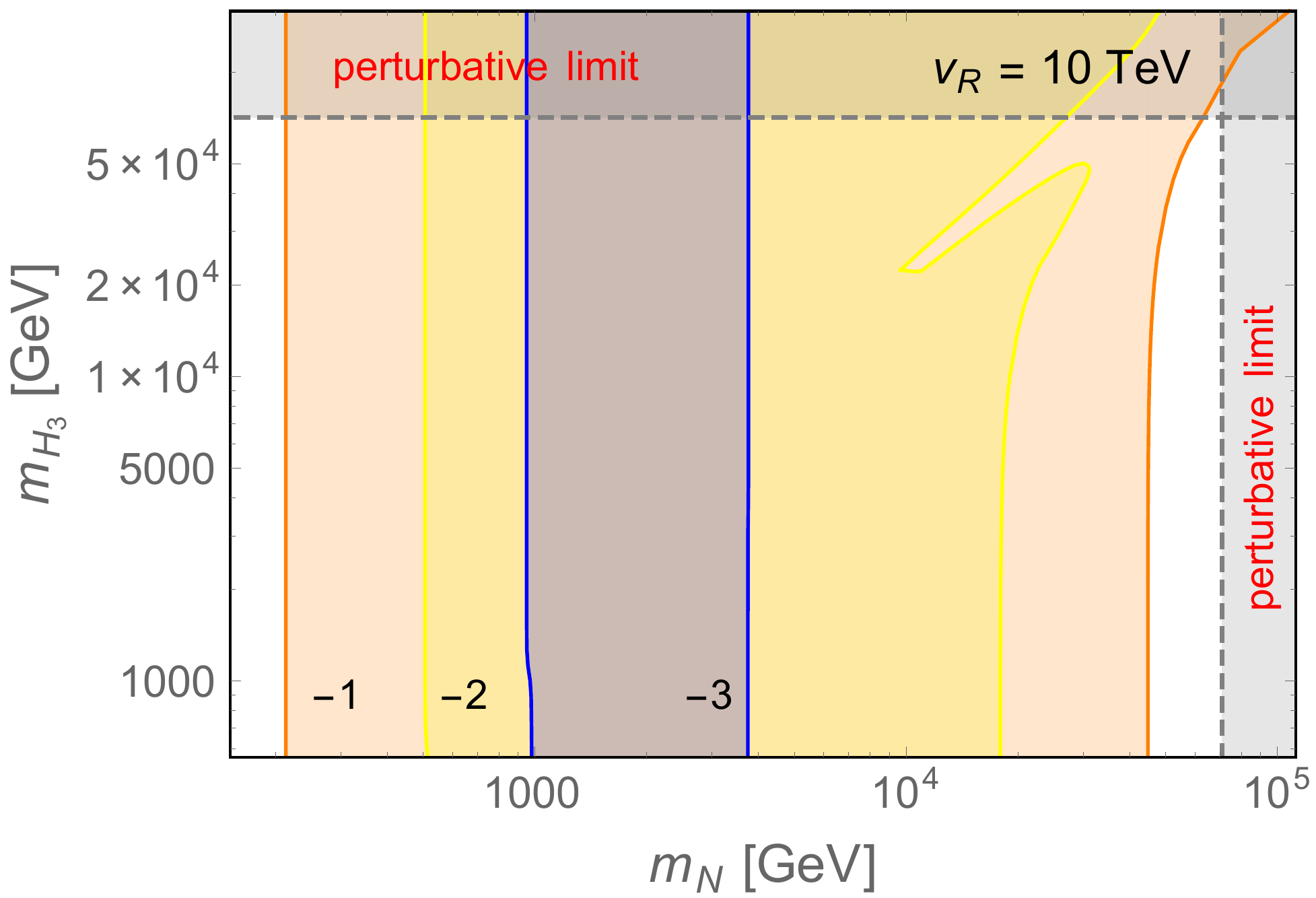}
  \caption{Contours of the CP-asymmetry $\log_{10} \varepsilon_{\rm CP}$, as functions of the RHN mass $m_N$ and the scalar mass $m_H$ in gauged $U(1)_{B-L}$ model, with $\widetilde{m} = 50$ meV and $v_R = 4$ TeV (left) and $10$ TeV (right). The colored regions could generate the observed lepton asymmetry and not falsified by the processes $NN \to H_3 H_3$, $H_3 Z_R$, $Z_R Z_R$ and $NN \to f \bar{f}$. The gray bands on the right and at the top are excluded, respectively, by the perturbativity constraints $f < \sqrt{4\pi}$ and $\rho < 4\pi$. }
  \label{fig:local}
\end{figure}

All the reduced cross sections are collected in Appendix~\ref{sec:appendixA}, with the widths of $H_3$ and $Z_R$ bosons properly included, respectively, in the channels $NN \to H_3 \to Z_R Z_R$ and $NN \to Z_R \to f \bar{f}$. As in the global $U(1)_{B-L}$ model where we have $NN \to H_3 \to JJ$, the processes with on-shell $s$-channel $H_3/Z_R$ bosons correspond to the inverse decay $NN \to H_3 / Z_R$ which significantly enhances the dilution cross sections, as shown in Fig.~\ref{fig:example3b}. The leptogenesis predictions in two benchmark scenarios are shown in Fig.~\ref{fig:local}, where for the sake of comparison, two different values of the $v_R$ scale are adopted, i.e. $v_R = 4$ TeV and $10$ TeV, which are both above the current dilepton limits on $Z_R$ mass. As in Fig.~\ref{fig:global}, only within the colorful regions, the observed lepton asymmetry could be generated for the indicated $\varepsilon_{\rm CP}$, with $\widetilde{m} = 50$ meV. It is transparent in both panels that when $m_N \lesssim m_{Z_R}$, there is almost no limit on the scalar mass $m_{H_3}$, as in this case the dilution is dominated by the $Z_R$ mediated process $NN \to f \bar{f}$, benefiting from the (almost) massless fermions in the final states and the large number of degrees of freedom. When the RHN is relatively light, the Yukawa coupling is to some extent suppressed via $f = m_N / 2v_R$; when $H_3$ is light, the triple scalar coupling could also be suppressed by $m_{H_3}^2$, as shown in Eq.~(\ref{eqn:triple}), thus the process $NN \to H_3 H_3$ could not compete against $NN \to f \bar{f}$.

When the RHNs are sufficiently heavy, e.g. heavier than the current dilepton limits on $Z_R$ boson mass of around $3.7$ TeV for $g_R = g_L$, such that they could annihilate into two $Z_R$ bosons, the scalar $H_3$ returns to play an important role, in particular when it is close to the resonance $m_{H_3} \simeq 2 m_N \gtrsim 2 m_{Z_R}$. One can see the clear resonance structure around $4$ TeV in the left panel of Fig.~\ref{fig:local}.\footnote{There exists another resonance effect that occurs at $m_N \simeq m_{Z_R}/2 \simeq 2$ TeV where the process $NN \to Z_R \to f\bar{f}$ is largely enhanced, and thus we have the gap at around $2$ TeV in the left panel of Fig.~\ref{fig:local}.} When the $v_R$ scale is higher, as exemplified in the right panel of Fig.~\ref{fig:local}, the $Z_R$ gets heavier, and the leptogenesis limits on the scalar mass become less stringent, and making it more challenging for the LHC and future collider tests (see Sec.~\ref{sec:collider} below).

\begin{figure}[t!]
  \centering
  \includegraphics[width=0.52\textwidth]{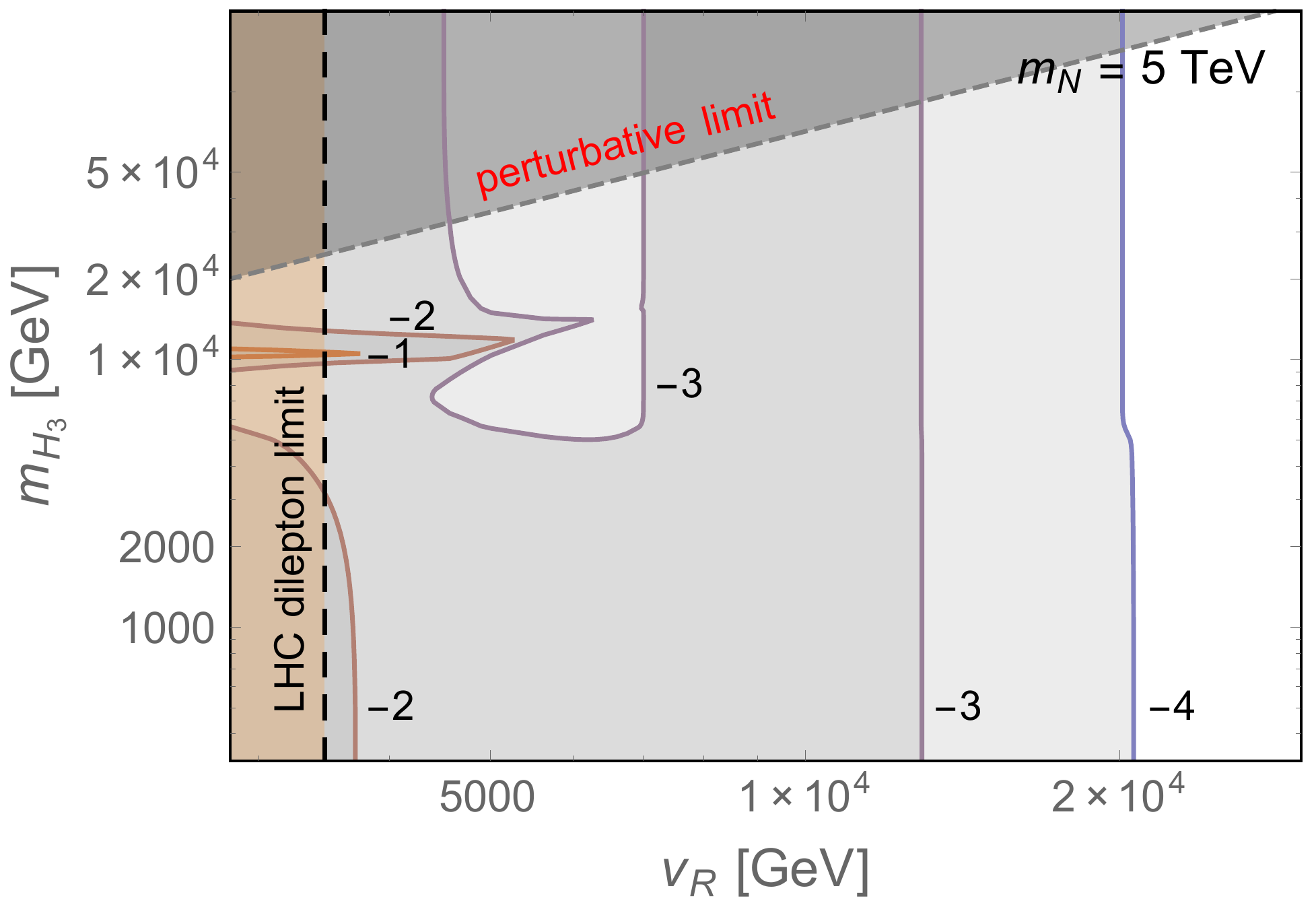}
  \caption{Contours of the CP-asymmetry $\log_{10} \varepsilon_{\rm CP}$, as functions of the scale $v_R$ and the scalar mass $m_H$ in gauged $U(1)_{B-L}$ model, with $\widetilde{m} = 50$ meV and $m_N = 5$ TeV. In the shaded regions, leptogenesis is ruled out by the processes $NN \to H_3 H_3$, $H_3 Z_R$, $Z_R Z_R$ and $NN \to f \bar{f}$. The darker shaded region on the top left corner is excluded by $\rho < 4\pi$, and the brown band is the current LHC dilepton limits on the $Z_R$ boson~\cite{ATLAS:2016cyf,CMS:2016abv}. }
  \label{fig:local2}
\end{figure}

Analogous to Figs.~\ref{fig:EFT3} and \ref{fig:global2}, we present in Fig.~\ref{fig:local2} the leptogenesis limits on $m_{H_3}$, as functions of the scale $v_R$, with fixed $m_N = 5$ TeV, $\widetilde{m} = 50$ meV and the indicated values of $\varepsilon_{\rm CP}$. All the shaded regions below (or on the left of) these curves are excluded, which correspond to the lighter scalar mass. The peak structure at around $m_{H_3} \simeq 10$ TeV corresponds to the resonance $m_{H_3} \simeq 2m_N$, which helps to exclude large regions in the $m_N - m_{H_3}$ plane. Note that here we do not consider any flavor structure; if the couplings of the $Z_R$ boson is flavor-dependent, e.g. dominantly to the tau-flavor, then it might be lighter, because the limits from $Z_R \to \tau^+ \tau^-$ searches at the LHC~\cite{CMS:2016zxk, ATLAS:2017mpg} are not so stringent as the dilepton limits (for $\ell=e,\mu$). With lighter $Z_R$ and RHN, the leptogenesis limits on $H_3$ for a given $g_R$ might change accordingly.


\section{Collider prospects}
\label{sec:collider} In this section, we focus on collider prospects for the seesaw scalar.
Since we are considering the role of beyond SM scalar in leptogenesis via the processes like $NN \to HH$, 
we will assume the scalar to be lighter than the RHNs involved, e.g. scalar mass $m_{H,\,A}$ (or $m_{H_3}$) down from $m_N$ to about 1 GeV.\footnote{Scalar masses below the GeV scale are tightly constrained  from low-energy (e.g. $K$-meson decay) and cosmological (e.g. BBN) data~\cite{Dev:2017dui}, so we do not consider that region in this analysis.} Since the scalar primarily couples to the RHNs, it is largely free of existing collider constraints. 
We will be interested in two mass ranges: (a) masses heavier than 5-10 GeV and (b) masses below 5 GeV and try to delineate possible ways to test leptogenesis in both cases.

In the latter case (b),
 the {\it tree-level} scalar quartic coupling $\rho = m_{H_3}^2 / 4 v_R^2$ is required to be very small, in order to generate small or vanishing scalar masses at the tree level in both the global and local $U(1)_{B-L}$ models, as in Refs.~\cite{Dev:2016vle,Dev:2017dui}. The small scalar mass can be assumed to be generated at 1-loop level via the Coleman-Weinberg mechanism, inspired by a conformal theory set-up ~\cite{Holthausen:2009uc}. The searches for displaced vertices and long-lived particles at the high energy frontier, i.e. the LHC and proposed dedicated experiments such as MATHUSLA~\cite{Chou:2016lxi,Evans:2017lvd}, and the high intensity experiments like SHiP~\cite{Alekhin:2015byh} and DUNE~\cite{Adams:2013qkq} provide excellent motivations to study the light scalars.

 In case (a) on the other hand, 
 the scalar mass has an upper limit $m_{H_3} < \sqrt{16\pi} v_R$, from the perturbativity of $\rho < 4\pi$, since it breaks the $B-L$ symmetry and is responsible for neutrino mass generation via the seesaw mechanism. 

 As we have seen above, successful leptogenesis forbids certain ranges for the seesaw scalar mass depending on the RHN masses, and therefore, any evidence for scalar masses in these forbidden mass regions would rule out leptogenesis as a mechanism for origin of matter.

\subsection{Constraints}
\begin{figure}[t!]
  \centering
  \includegraphics[width=0.48\textwidth]{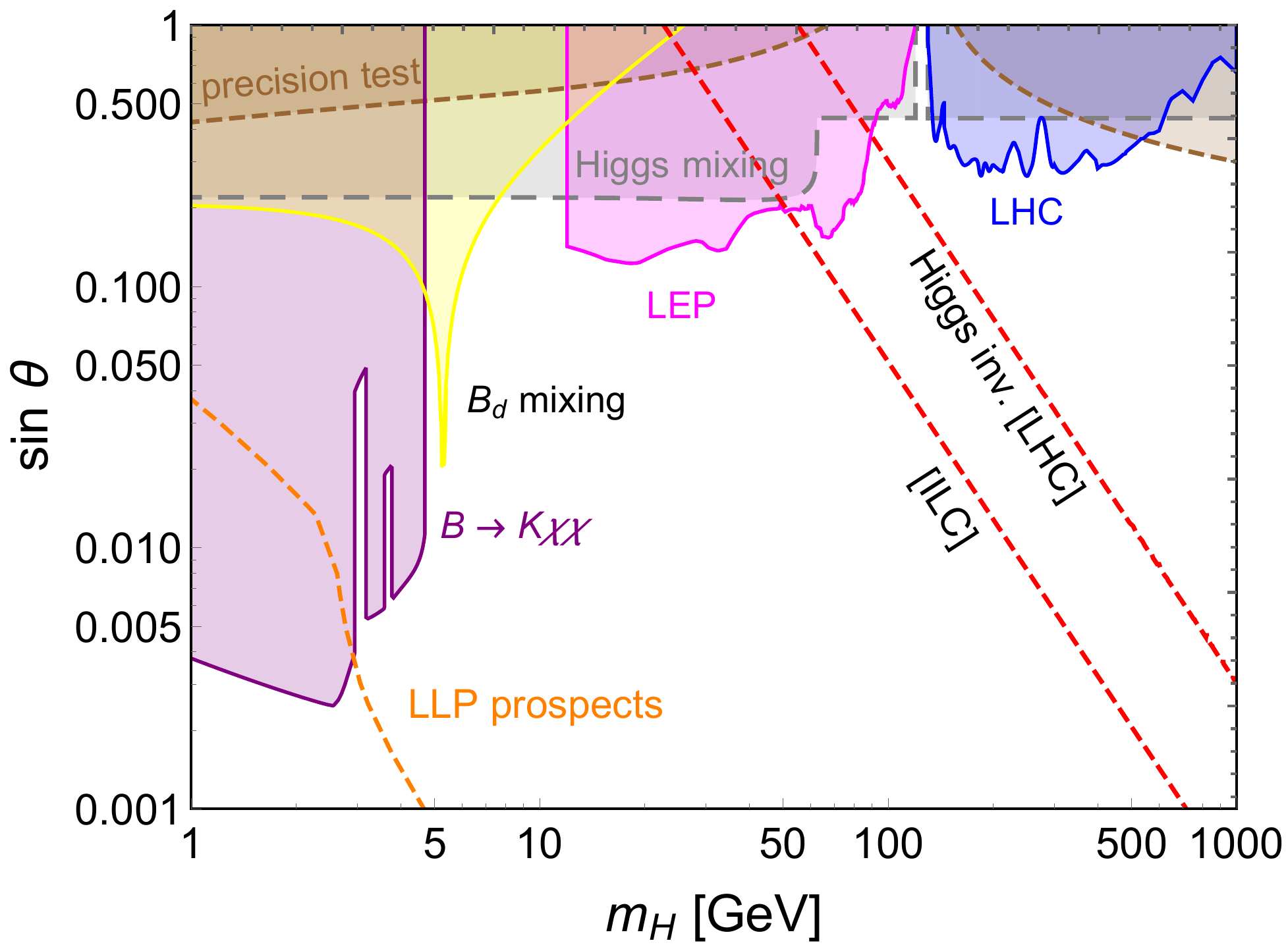}
  \caption{Limits on the CP-even scalar mass $m_H$ and its mixing with the SM Higgs $\sin\theta$ in the effective theory framework. The limits from precision test (brown) and SM Higgs coupling measurements (gray) are from Ref.~\cite{Falkowski:2015iwa}. The limits from direct searches at LEP (magenta), LHC (blue), and the flavor limits from $B \to K \chi\chi$ (purple) and $B_d - \bar{B}_d$ mixing (yellow) in the low mass regime are from Ref.~\cite{Dev:2017dui}. Also shown are the LLP prospects at LHC~\cite{Dev:2017dui}. The future sensitivity from invisible decay of the SM Higgs (red dashed) apply to the CP-even scalar in the global $U(1)_{B-L}$ model with $v_R = 1$ TeV. See text for more details. }
  \label{fig:limit}
\end{figure}

In both regimes (a) and (b), constraints and predictions for signals of $H, A$ come from the fact that
a beyond SM scalar could in principle mix with the SM Higgs after EW symmetry breaking. This scalar mixing however does not play any role in freeze-out leptogenesis since the latter takes place prior to EW symmetry breaking. 
First of all, the scalar mixing would  universally rescale the couplings of SM Higgs and contribute to the oblique parameters. It could  also induce flavor-changing rare meson decays to the extra scalar when the latter is light, leading to stringent constraints in that mass range~\cite{Dev:2016vle, Dev:2017dui}. There are also direct searches of these extra scalars at LEP and LHC, e.g. when they decay into two SM fermions~\cite{ATLAS:2016pyq}, gauge bosons~\cite{Chatrchyan:2013mxa, Khachatryan:2015cwa, Aad:2015kna, Aaboud:2016okv, Aad:2014ioa, ATLAS:2016eeo, Khachatryan:2016yec}, or di-Higgs~\cite{Khachatryan:2015yea, Khachatryan:2016sey, ATLAS:2016ixk}. All these data could be used to set limits on generic beyond SM scalars, i.e. on its mass and mixing angle $\sin\theta$ with the SM Higgs, as shown in Fig.~\ref{fig:limit}. The purpose of this section is to connect the rich phenomenology of heavy or light scalars to leptogenesis: if a (light) scalar could be found at the LHC or in the future higher energy colliders, that would have strong implications on the type-I seesaw leptogenesis and the heavy RHNs, thus providing a complementary probe of the seesaw mechanism.

For a CP-even scalar, the constraints from SM Higgs data and EW precision tests in the effective theory framework of Sec.~\ref{sec:EFT} can be found e.g. in Ref.~\cite{Falkowski:2015iwa}, depicted as the gray and brown regions in Fig.~\ref{fig:limit}, respectively. The direct searches have been performed at the LHC in the final states of $H \to WW/ZZ$~\cite{Chatrchyan:2013mxa, Khachatryan:2015cwa, Aad:2015kna, Aaboud:2016okv}, induced by mixing with the SM Higgs. All these data are combined, following the procedure in Ref.~\cite{Bian:2017jpt}, and are shown as the blue shaded region in Fig.~\ref{fig:limit}. There are also searches for heavy CP-even or odd scalar resonances in the diphoton spectra~\cite{Aad:2014ioa, ATLAS:2016eeo, Khachatryan:2016yec}. However, in these searches, the interference between the continuum background $gg \to \gamma\gamma$ and the signal $gg \to H/A \to \gamma\gamma$ is in general very important, even dominating over the pure resonance contribution~\cite{Bian:2017jpt}. The interference effect is not taken into account properly in the data analysis, thus these exclusion data can not be na\"ively interpreted as limits on the effective mixing angle $\sin\theta$ in our model. When kinematically allowed (i.e. for $m_H > 2 m_h$), the heavy scalar could decay also into the SM Higgs pair, $H \to hh$; however, the limits from Refs.~\cite{Khachatryan:2015yea, Khachatryan:2016sey, ATLAS:2016ixk} can not set any limits on the mixing angle $\sin\theta$, suppressed by the small branching ratio.\footnote{Here for simplicity we have assumed the cubic scalar coupling stems only from the term $\lambda_{hhh}^{\rm SM}$ in the SM via the mixing $\sin\theta$. In principle, one could also have contributions from the beyond SM cubic terms like $\lambda_{HHH}$.} The limits from Higgs searches at LEP are shown as the magenta region in Fig.~\ref{fig:limit}~\cite{Barate:2003sz}, excluding  the scalar mass from $\sim 12$ GeV up to 120 GeV and setting an upper limit on the mixing angle of order ${\cal O} (0.1)$.

For a light scalar $H$, the most stringent limits are from the flavor sector, as $H$ could have loop-level flavor-changing couplings to quarks, such as $H \bar{s} b$, thus contributing significantly to neutral meson oscillations like $B_d - \bar{B}_d$ and rare meson decays like $B \to K \ell^+ \ell^-$. A detailed discussion can be found, e.g. in Ref.~\cite{Dev:2017dui}. In Fig.~\ref{fig:limit} we only show the most stringent limits in the low mass regime, i.e. those from $B \to K \chi\chi$ (with $\chi$ SM charge leptons) combining the data from Refs.~\cite{Aubert:2003cm, Wei:2009zv, Aaij:2012vr} and from $B_d - \bar{B}_d$ oscillation~\cite{Dev:2017dui}. Enhanced by the small scalar mass $m_H$, the limits on the mixing could go even up to the order of $10^{-3}$. For such small mixing values, when $H$ is light enough, the decay length at the LHC could be sizable compared to the detector radius, thus motivating a long lived particle (LLP) search. The region below the orange curve in Fig.~\ref{fig:limit} can be probed at LHC with a decay length $1 \, {\rm cm} < b \tau_0 < 1.5 \, {\rm m}$ for an integrated luminosity of 3000 fb$^{-1}$ at $\sqrt{s}  = 14$ TeV~\cite{Dev:2017dui}.

For the CP-odd scalar $A$, if it mixes with the SM Higgs, then the two physical scalars from $h - A$ mixing are no longer CP eigenstates but rather admixtures of the CP-even and odd components, with one at 125 GeV and the other one heavier or lighter depending on the mass $m_A$. A mixing angle of $\simeq 0.2$ is still compatible with current LHC data~\cite{Aad:2015tna, Aad:2016nal, Khachatryan:2016tnr} at 95\% CL. The limits on the scalar $H$ presented in Fig.~\ref{fig:limit} apply also to the CP-odd $A$.

The limits on the physical scalar $H_3$ in the global and local $U(1)_{B-L}$ model (cf. Sec.~\ref{sec:global} and \ref{sec:local} respectively) are roughly similar to the CP-even scalar $H$ in the EFT framework (cf. Fig.~\ref{fig:limit}) with one extra decay mode  $H_3 \to JJ$ open for the global case. The branching ratio ${\rm BR} (H_3 \to JJ)$ depends on the $v_R$ scale and the scalar mass $m_{H_3}$ [cf. the couplings in Eq.~(\ref{eqn:triple})], with the corresponding partial decay width given by Eq.~\eqref{eqn:width2}, with the RHS multiplied by $\cos^2\theta$.
So all the flavor and direct search constraints in Fig.~\ref{fig:limit} are weakened by the factor of $\big[ 1- {\rm BR} (H_3 \to JJ) \big]$.

Through the mixing with $H_3$, the SM Higgs could also decay into two (almost) massless Majorons, with the partial width
\begin{eqnarray}
\Gamma (h \to JJ) \ = \
\frac{\sin^2\theta \, m_{H_3}^4}{64 \pi \, m_h \, v_R^2} \,.
\label{eq:5.2}
\end{eqnarray}
The Majorons are expected to escape the detectors without leaving any tracks even if they can decay into active neutrinos when they are massive, thus contributing to the invisible branching ratio of the SM Higgs ${\rm BR} (h \to {\rm inv.})$. At the $\sqrt{s} = 14$ TeV LHC, with an integrated luminosity of 3000 fb$^{-1}$, the Higgs invisible BR can be constrained to be smaller than 9\% at the 95\% C.L~\cite{Peskin:2012we}, while at $\sqrt{s} = 1$ TeV ILC with a luminosity of 1000 fb$^{-1}$, the BR limit can reach up to 0.26\%~\cite{Baer:2013cma}. These prospects can be used to set future limits on the scalar mass $m_{H_3}$ and mixing angle $\sin\theta$ for a given $v_R$ scale. A representative example is shown in Fig.~\ref{fig:limit} as the red dashed lines (the region above which can be excluded), where we have set $v_R = 1$ TeV. This is largely complementary to the direct search of heavy scalars, in particular when the mixing angle is relatively small and the scalar mass relatively large, until it hits the perturbativity limit (roughly at 3 TeV for $v_R=1$ TeV in Fig.~\ref{fig:limit}).

\begin{figure}[t!]
  \centering
  \begin{tikzpicture}[]
  \draw[gluon,thick] (-1.4,1)node[left]{{\footnotesize$g$}} -- (-0.6,0);
  \draw[gluon,thick] (-1.4,-1)node[left]{{\footnotesize$g$}} -- (-0.6,0);
  \shade[top color=violet, bottom color=white] (-0.6,0)circle(2.5pt);
  \draw[dashed,thick](-0.6,0)--(0,0)node[above]{{\footnotesize$h/H_3$}}--(0.6,0);
  \draw[dashed,thick](1.4,1)node[right]{{\footnotesize$J$}}--(0.6,0);
  \draw[dashed,thick](1.4,-1)node[right]{{\footnotesize$J$}}--(0.6,0);
  \draw[gluon,thick] (-1,0.5) -- (0.3,1)node[right]{{\footnotesize$g$}};
  \end{tikzpicture}
  \caption{A representative Feynman diagram for the Majoron pair-production process at the LHC in the global $U(1)_{B-L}$ model. }
  \label{fig:diagrams3}
\end{figure}
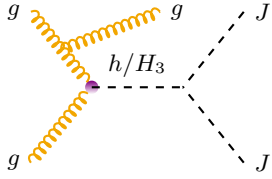

\begin{figure}[t!]
  \centering
  \includegraphics[width=0.52\textwidth]{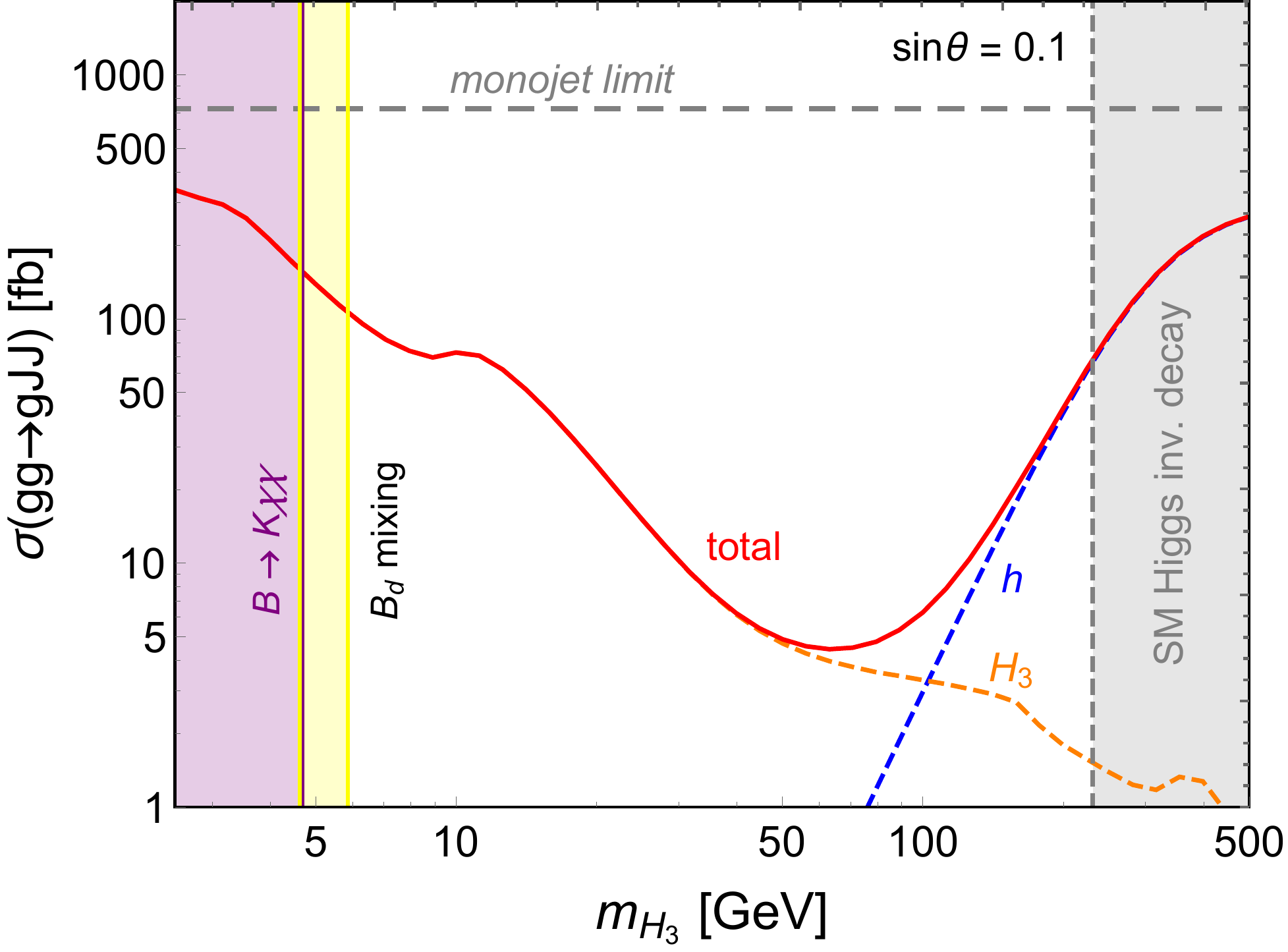}
  \caption{Production cross section $\sigma (gg \to gJJ)$ from both the $h$ portal (dashed blue) and the $H_3$ portal (dashed orange) and the total one (solid red), with $\sin\theta = 0.1$ and $v_R = 1$ TeV. Also shown are the limits from flavor-changing decay $B \to K \chi\chi$ and $B_d - \bar{B}_d$ oscillation data, as well as from the invisible decay of the SM Higgs. The horizontal dashed line is the current limit from the LHC monojet search~\cite{Aad:2015zva}.}
  \label{fig:monojet}
\end{figure}

In the global $U(1)_{B-L}$ model, the Majorons can be pair produced from its couplings to the SM Higgs $h$ and $H_3$ in Eq.~(\ref{eqn:triple}) and the $h - H_3$ scalar mixing, via the gluon fusion processes $gg \to h/H_3$. If one gluon is emitted from one of the initial-state gluons, as shown in Fig.~\ref{fig:diagrams3}, this resembles the mono-jet process at the LHC, with the two Majorons acting like (long-lived) dark matter particles leaving only missing transverse energy at hadron colliders.\footnote{With a non-zero but small mass, the Majoron could decay into the active neutrinos through the heavy-light neutrino mixing~\cite{Chikashige:1980ui} and other SM fermions at loop level~\cite{Garcia-Cely:2017oco}, which renders it a good long-lived decaying dark matter candidate. } Thus the production cross section for $gg \to g JJ$ in Fig.~\ref{fig:diagrams3} is constrained by the monojet searches at the LHC~\cite{Aad:2015zva, Khachatryan:2014rra, Aaboud:2016tnv}. The parton level production cross section is estimated by implementing the simulations $gg \to g h$ and $gg \to g H_3$ in {\tt CalcHEP}~\cite{Belyaev:2012qa} and summing up the cross sections from both the two portals via
\begin{eqnarray}
\sigma (gg \to g JJ) & \ = \ &
\sigma (gg \to g h) \times {\rm BR} (h \to JJ) 
+ \ \sigma (gg \to g H_3) \times {\rm BR} (H_3 \to JJ) \,,
\end{eqnarray}
where the BRs can be evaluated using Eqs.~(\ref{eqn:width2}) and \eqref{eq:5.2}.
Then the monojet limits on the $H_3$ mass and mixing angle $\sin\theta$ can be obtained by comparing with the experimental data, after imposing appropriate cuts on the gluon jet. It turns out that when the scalar mass $m_{H_3}$ is small, $H_3$ decays mostly into the (almost massless) Majorons, while the SM Higgs decay $h \to JJ$ is suppressed by both the mixing angle $\sin\theta$ and the small coupling $m_{H_3}^2/v_R$ in Eq.~(\ref{eqn:triple}); then the production is dominated by the $H_3$ portal, as shown in Fig.~\ref{fig:monojet} where we have set explicitly $\sin\theta = 0.1$ and $v_R = 1$ TeV. The low $m_{H_3} \lesssim 5$ GeV range is however excluded by the $B$ meson data for a sizable $\sin\theta$ (cf. the $B \to K \chi\chi$ decay and $B_d - \bar{B}_d$ oscillation limits in Fig.~\ref{fig:limit}). When the scalar $H_3$ gets heavier, the production rate $gg \to g H_3$ decreases largely, which diminishes the contribution from the $H_3$ portal; at the same time, the triple scalar coupling $m_{H_3}^2/ v_R$ becomes larger and larger which enhance the branching ratio of $h \to JJ$, and eventually the SM Higgs takes over as the dominant channel, at $m_{H_3} \simeq 100$ GeV, as shown in Fig.~\ref{fig:monojet}. For $m_{H_3} \gtrsim 230$ GeV, the ${\rm BR} (h \to JJ)$ exceeds the current limit of $28\%$ for the invisible decay of the SM Higgs~\cite{PDG}, and hence, excluded as shown in Fig.~\ref{fig:monojet}. With both the limits from $B$ meson and the invisible decay of the SM Higgs taking into consideration, the scalar mass is restricted to be within a range of $5 \, {\rm GeV} \lesssim m_{H_3} \lesssim 230$ GeV; within this range, the current monojet data~\cite{Aad:2015zva, Khachatryan:2014rra, Aaboud:2016tnv} does not provide any limits on the mass $m_{H_3}$ and the mixing angle $\sin\theta$, if the scale $v_R = 1$ TeV.

\subsection{Testing leptogenesis at the LHC}

In the EFT framework, the RHNs have only the scalar portal to couple to the SM particles,\footnote{The RHNs could also be produced through their Dirac Yukawa coupling~\cite{Dev:2012zg, Cely:2012bz, Hessler:2014ssa, Das:2017zjc, Das:2017rsu, Ruiz:2017yyf} to the SM Higgs [cf.~Eq.~\eqref{eqn:LYukawa}]; however, in the TeV-scale type-I seesaw with Majorana RHNs, the Yukawa coupling is $\lesssim 10^{-6}$, thereby suppressing this production channel.} i.e. through the $h - H$ scalar mixing $\sin\theta$; then the RHNs are produced predominantly from the gluon fusion process via the SM quark loops: $gg \to h^\ast/H^{(\ast)} \to NN$. After production, due to their Majorana nature, the RHNs could decay into both positively and negatively charged leptons, emitting a SM $W$ boson: $N \to \ell^\pm W^\mp$. One expects the signals with same-sign charged leptons in the final state, with roughly a branching fraction
\begin{eqnarray}
\label{eqn:BR}
\frac{\sigma (NN \to \ell^\pm \ell^\pm W^\mp W^\mp)}{\sigma(NN \to {\rm anything})} \ \simeq \ \frac18 \,,
\end{eqnarray}
in the limit of large $m_N\gg m_W$. The factor of 8 takes into account other RHN-pair decay modes, such as $\ell^\pm \ell^\mp W^\mp W^\pm$, $\nu \nu \phi^0 \phi^0$ and $\ell^\pm \nu W^\mp \phi^0$ with $\phi^0 = h,\,Z$ (and $\nu$ standing for both $\nu$ and $\bar{\nu}$), all of which are proportional to the heavy-light neutrino mixing angle $V_{\nu N}^2$.
Assuming the hadronic decay of $W$ bosons, we have the signal of
\begin{eqnarray}
gg \to h / H \to N N \to \ell^\pm \ell^\pm jjjj \,,
\end{eqnarray}
which is absent in the SM, barring lepton charge mis-identification in the $\ell^\pm \ell^\mp+{\rm jets}$ final state. One should note that here the $h$ and $H$ mediated diagrams interfere destructively with each other, as the couplings of $h$ and $H$ to the gluons and the RHNs are respectively proportional to the combinations of factors of
\begin{eqnarray}
\label{eqn:gXgg}
\text{couplings of $h$}:&& g_{hgg} \, (-f \sin\theta) \,, \nonumber \\
\text{couplings of $H$}:&& ( \sin\theta \, g_{Hgg}) \, f \,,
\end{eqnarray}
where
\begin{eqnarray}
g_{Xgg} \ \simeq \ \frac{\alpha_s}{16\pi \, v_{\rm EW}} A_{1/2} (\tau_X)
\end{eqnarray}
(with $X = h,\,H$) is the effective scalar coupling to the gluons through the SM quarks (dominated by the top quark loop), $\tau_X = m_X^2 / 4m_q^2$ ($m_q$ being the SM quark masses), and the loop function~\cite{Djouadi:2005gi}
\begin{eqnarray}
A_{1/2} (\tau) & \ \equiv \ & 2 \left[ \tau + (\tau-1) f(\tau) \right] \tau^{-2} \,,
\label{eq:Ahalf}
\end{eqnarray}
with
\begin{align}
f(\tau) \ \equiv \ \left\{ \begin{array}{cc}
{\rm \arcsin}^2\sqrt{\tau} & ({\rm for}~\tau\leq 1) \\
-{\displaystyle \frac{1}{4}}\left[\log \left( \frac{1+\sqrt{1-1/\tau}}{1-\sqrt{1-1/\tau}}\right)-i\pi  \right]^2 & ({\rm for}~\tau>1) \;.
\end{array}\right.
\label{fx}
\end{align}
The relative minus sign in Eq.~(\ref{eqn:gXgg}) comes from the orthogonal scalar mixing matrix
\begin{eqnarray}
\left( \begin{matrix} h \\ H \end{matrix} \right) \ = \
\left( \begin{matrix} \cos\theta & - \sin\theta \\ \sin\theta & \cos\theta \end{matrix} \right)
\left( \begin{matrix} \phi^{0} \\ \Delta_R \end{matrix} \right) \,.
\end{eqnarray}
The resultant production amplitude is
\begin{eqnarray}
\label{eqn:amplitude}
{\cal A}(pp\to NN) \ \sim \ f \sin\theta \left( \frac{g_{Hgg}}{q^2 - m_H^2} - \frac{g_{hgg}}{q^2 - m_h^2} \right)
\end{eqnarray}
with $q = \sqrt{\hat{s}}$ the center-of-mass energy at the parton level. When $H$ is lighter than the SM Higgs, it dominates the RHN production, while when $H$ is heavier, the SM Higgs takes over to be the dominant term. When the two scalars are almost degenerate, which is mildly disfavored by the precision Higgs data~\cite{PDG} (barring fine-tuned cancellations), the two terms in Eq.~(\ref{eqn:amplitude}) almost cancel out with each other, diminishing largely the production cross section of RHNs.\footnote{In the limit of $m_H \to m_h$, the effective loop couplings $g_{H gg} \to g_{hgg}$. In the case of the CP-odd scalar $A$, as it does not couple directly to the SM quarks but through mixing with the SM Higgs, we again have the cancellation of the $h$ and $A$ mediated diagrams in the limit of $m_A \to m_h$. }

\begin{figure}[t!]
  \centering
  \includegraphics[width=0.48\textwidth]{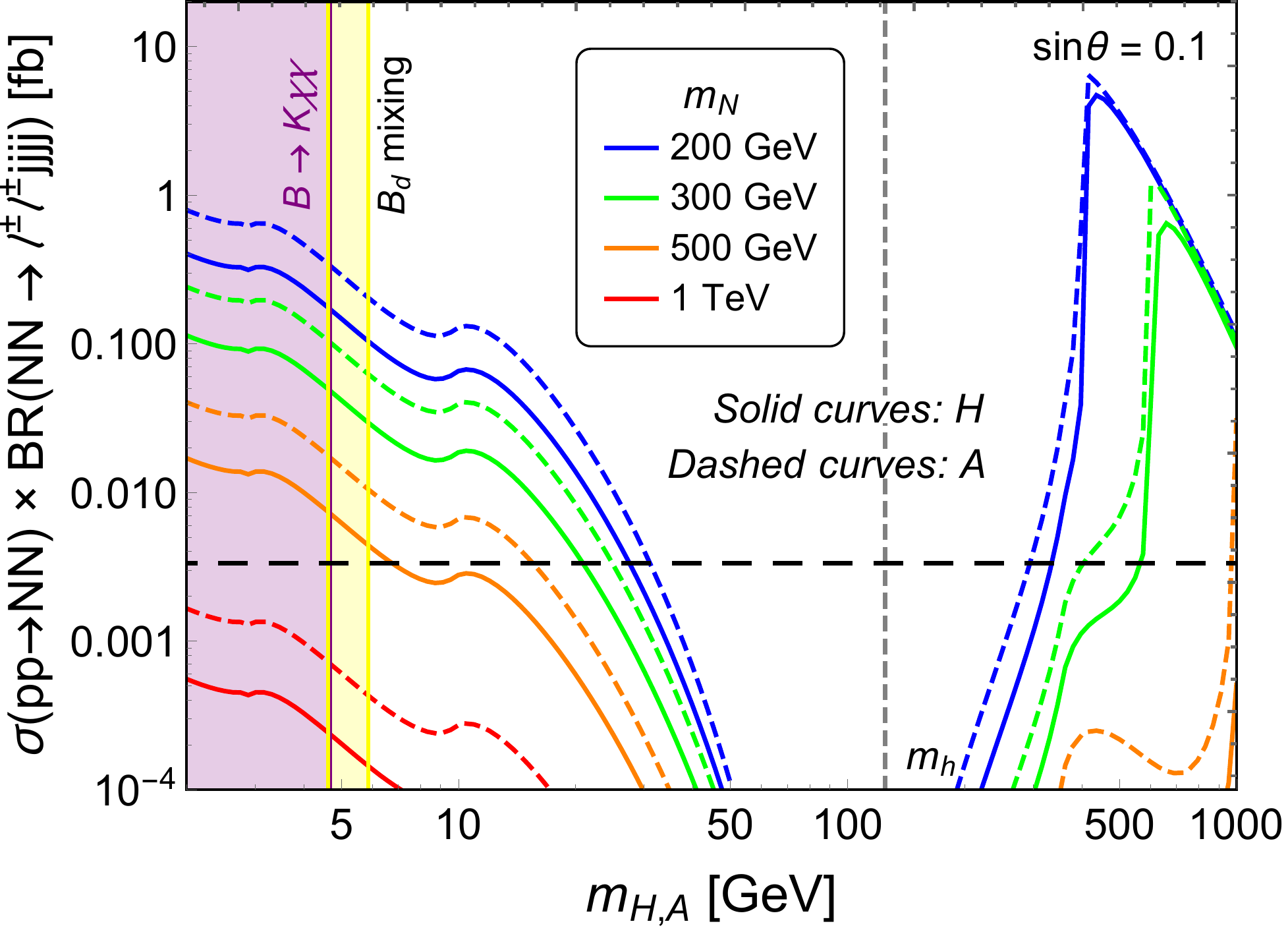}
  \includegraphics[width=0.48\textwidth]{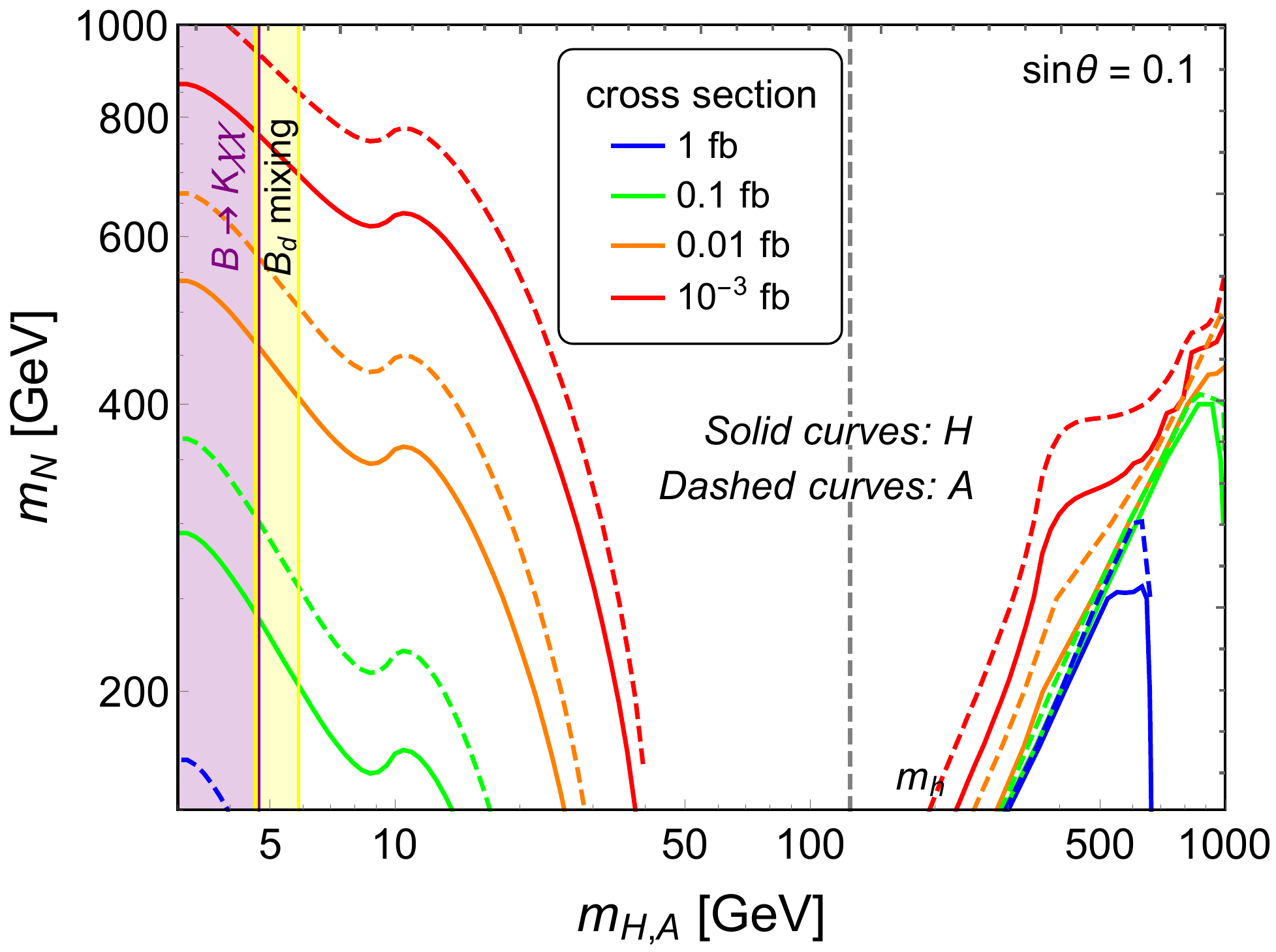}
  \caption{{\it Left}: Examples of production cross section of RHNs in the scalar portal from gluon fusion $gg \to h/H (A) \to NN$ times the branching ratio of the RHN decays into same-sign charged leptons with four jets ${\rm BR} ( NN \to \ell^\pm \ell^\pm jjjj )$, as functions of the heavy scalar mass $m_{H,\,A}$ in the EFT framework, for $m_N = 200$, 300, 500 GeV and 1 TeV, $f = 1$ and $\sin\theta = 0.1$. The solid (dashed) curves are for the CP-even (odd) scalar. The purple and yellow bands are excluded respectively by the $B \to K \ell^+ \ell^-$ and $B_d - \bar{B}_d$ oscillation data. The vertical dashed line corresponds to the SM Higgs mass $m_h = 125$ GeV. The horizontal long dashed line indicates the cross section with at least 10 signal events at the high-luminosity LHC with an integrated luminosity of 3000 fb$^{-1}$ at $\sqrt{s} = 14$ TeV. {\it Right}: the contours of production cross section $\sigma (gg \to h/H (A) \to NN \to \ell^\pm \ell^\pm jjjj) = 1$, 0.1, 0.01 and $10^{-3}$ fb, as functions of $m_{N}$ and $m_{H,\,A}$, with the other parameters the same as in the left panel.}
  \label{fig:production}
\end{figure}

The production cross sections for four different benchmark values of $m_N = 200$ GeV, 300 GeV, 500 GeV and 1 TeV are shown in the left panel of Fig.~\ref{fig:production}, respectively, as the solid blue, green, orange and red curves, where the Yukawa coupling $f = 1$, the scalar mixing angle $\sin\theta = 0.1$ and the cross sections are obtained by rescaling the next-to-next-to-leading order production of a SM Higgs in the gluon fusion process~\cite{Heinemeyer:2013tqa} by appropriate couplings given by Eq.~\eqref{eqn:gXgg}. The same-sign dilepton branching ratio in Eq.~(\ref{eqn:BR}) and ${\rm BR} (W \to {\rm hadrons})\simeq 0.67$ have been taken into consideration. If the CP-even scalar $H$ is replaced by the CP-odd $A$, the production cross section would be raised by a factor of $\simeq 2$, as shown in Fig.~\ref{fig:production} by the dashed curves. This is mainly due to the structure of the effective Yukawa couplings in Eq.~\eqref{eqn:LYukawa}. The destructive interference of the $h$ and $H$ ($A$) diagrams are taken into consideration, which is particularly important when $m_{H,A} \simeq m_h$. The shape for $m_{H,\,A} < m_h$ is determined mainly by the SM fermion loop $\sum_f A_{1/2} (\tau_f)$ [cf.~Eq.~\eqref{eq:Ahalf}] for the effective $Hgg$ ($Agg$) coupling; for $m_{H,\,A} > m_h$, the cross sections could be orders of magnitude larger when the RHNs are produced from on-shell $H$ ($A$) decay, i.e. $m_{H,\,A} > 2m_N$. When the scalar gets much heavier, $m_{H,A} \gtrsim$ TeV, the production is then kinematically suppressed by the heavy scalar mass $m_{H,\,A}$.

As seen on the left panel of Fig.~\ref{fig:limit}, the mixing angle $\sin\theta = 0.1$ is excluded by the the rare $B$ meson decay $B \to K \ell^+ \ell^-$ data and $B_d - \bar{B}_d$ oscillation data when the light scalar $H$ is lighter than $\sim 5$ GeV. It is transparent in Fig.~\ref{fig:production} that a few hundred GeV scale RHN $N$ could be probable in the Higgs portal, through its couplings to the seesaw scalar $H,\,A$ and the SM Higgs, at the high-luminosity LHC. Assuming the luminosity of 3000 fb$^{-1}$ at $\sqrt{s} = 14$ TeV, we can have at least 10 signal events for $m_N \lesssim 500$ GeV, if the scalar mass $m_{H,\,A} \lesssim 30$ GeV or $200\, {\rm GeV} \lesssim m_{H,\,A} \lesssim 1$ TeV, for $f = 1$ and scalar mixing angle $\sin\theta = 0.1$. For other values of $f$ and $\sin\theta$, the production cross section can be just simply rescaled by $f^2 (\sin\theta/0.1)^2$.

The dependence of production cross section on the RHN mass $m_{N}$ and the scalar mass $m_{H,\,A}$ is more clearly seen in the right panel of Fig.~\ref{fig:production}, where we show the contours for some benchmark values of 1, 0.1, 0.01 and $10^{-3}$ fb. With highly suppressed SM background of same-sign charged lepton events, it is promising to test leptogenesis at the LHC for a large range of $m_{H,A} \lesssim 30$ GeV and $\gtrsim 200$ GeV with $m_N \lesssim$ TeV. Recall from Sec.~\ref{sec:EFT} that when the RHN mass $m_N \lesssim$ TeV, even if the scalar $H,\,A$ is heavier than $N$, there is large parameter space of inefficient leptogenesis, see e.g. Fig.~\ref{fig:EFT1} and \ref{fig:EFT2}. At higher energy colliders such as FCC-hh~\cite{Arkani-Hamed:2015vfh, Golling:2016gvc, Contino:2016spe} and SPPC~\cite{CEPC-SPPCStudyGroup:2015csa}, with larger production cross section $\sigma (pp \to NN)$, a broader parameter region of leptogenesis could be tested.

\begin{figure}[t!]
  \centering
  \includegraphics[width=0.48\textwidth]{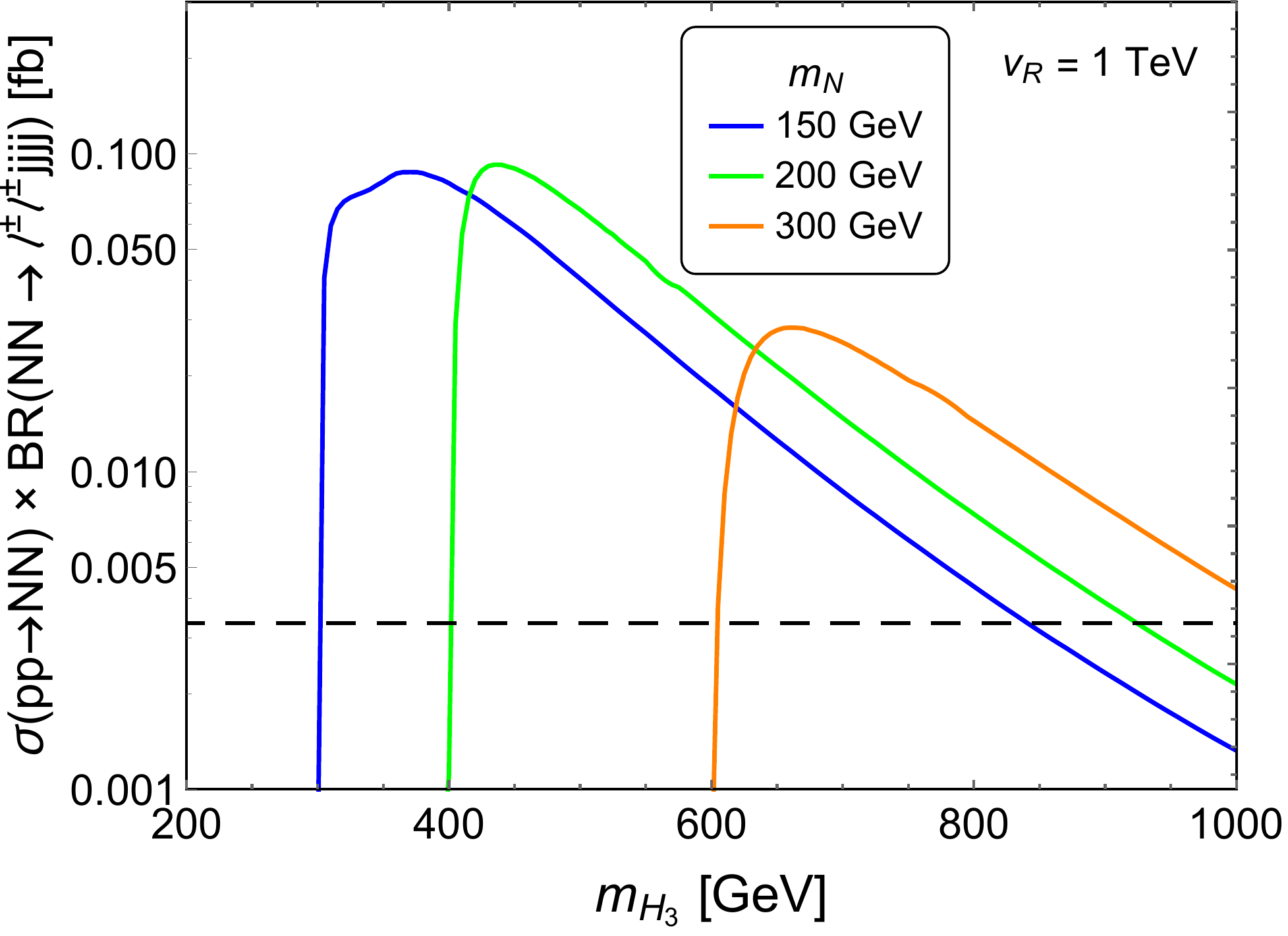}
  \includegraphics[width=0.47\textwidth]{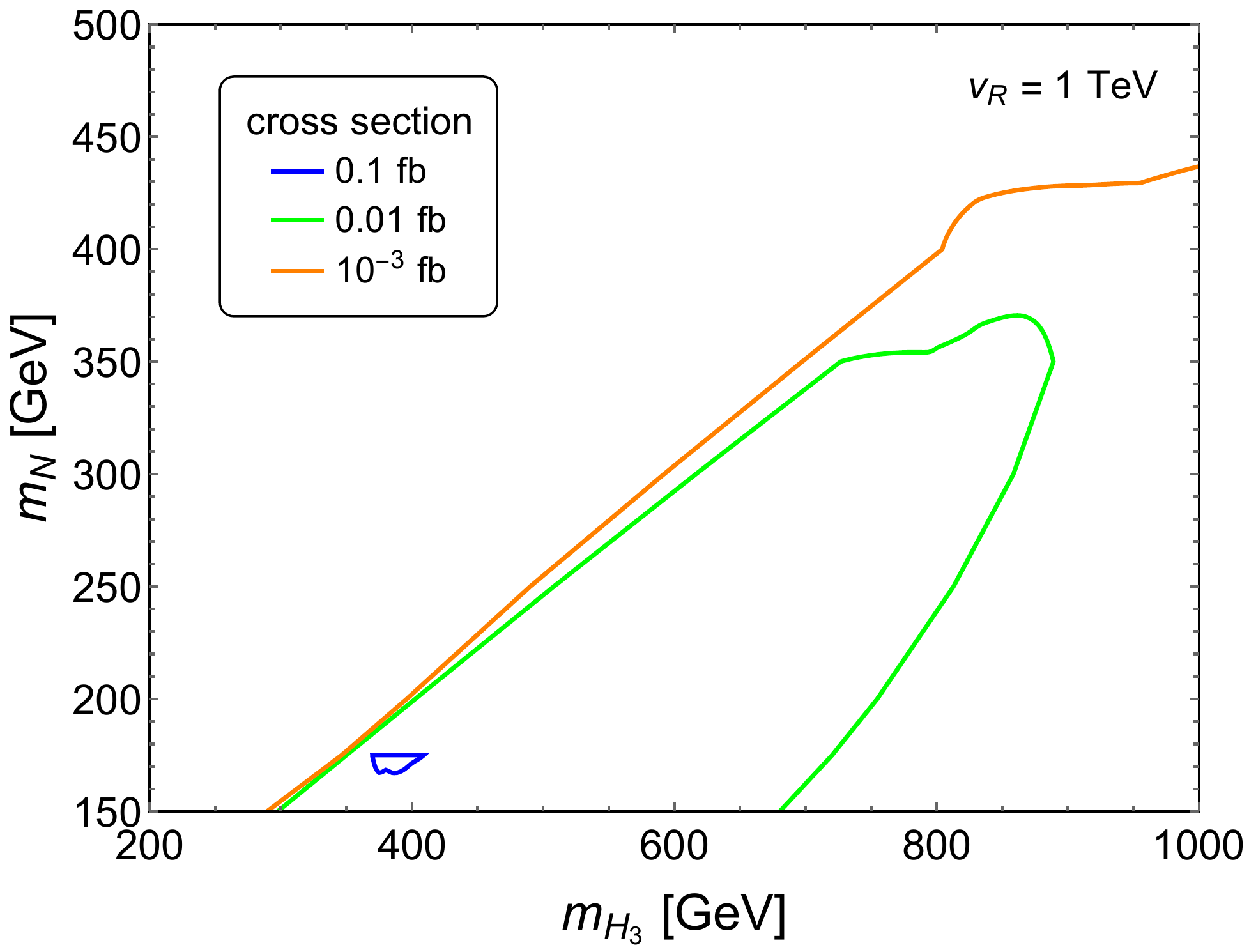}
  \caption{The same as in Fig.~\ref{fig:production}, for the global $U(1)_{B-L}$ model, with $v_R = 1$ TeV and $\sin\theta = 0.1$.}
  \label{fig:production2}
\end{figure}

Regarding the global $U(1)_{B-L}$ model, in the simplest scenario without any CP violation in the scalar sector (i.e. assuming all the quartic couplings in the scalar potential to be real and no spontaneous CP violations in the VEV $v_R$), the SM Higgs mixes with only the CP-even scalar $H_3$. Without the CP violating mixing $h - J$, the collider prospects for the scalar $H_3$ is the same as that for $H$, with the effective Yukawa coupling of $f_{\rm eff} = m_N / \sqrt2 v_R$. In other words, for a fixed value of $v_R$, the solid contours in Fig.~\ref{fig:production} have to be just rescaled by the factor of $(m_N / \sqrt2 v_R)^2$. So for lighter $N$, the production cross section is relatively suppressed by the RHN mass $m_N$. An explicit example with $v_R = 1$ TeV and $\sin\theta = 0.1$ is shown in Fig.~\ref{fig:production2}. For larger mixing angles which is still allowed by current Higgs precision data and direct searches at LHC, as well as for different $v_R$ values, the same-sign dilepton cross section in Fig.~\ref{fig:production2} should be rescaled by a factor of $\left( {\sin\theta}/{0.1} \right)^2 \times
\left( {v_R}/{1\,{\rm TeV}} \right)^{-2}$.
It is clear that for a small scalar mixing $\sin\theta = 0.1$, if the RHNs is pair produced through an on-shell heavy seesaw scalar, i.e. $m_{H_3} > 2 m_N$, then the production cross section $\sigma (pp \to H_3 \to NN)>0.01 \, {\rm fb}$
in a large region of parameter space of the global $U(1)_{B-L}$ model. So it seems promising that leptogenesis in this model could be directly tested at the high luminosity LHC, if both the seesaw scalar and RHNs are in the few hundred GeV range.

In the gauged $U(1)_{B-L}$ model, the RHNs could also be pair produced through the gauge portal~\cite{Basso:2008iv, Deppisch:2013cya, Kang:2015uoc, Cox:2017eme, Das:2017pvt}, i.e. through both the $Z$ and $Z_R$ bosons, as the SM $Z$ boson mixes with the $Z_R$ boson~\cite{Dev:2016dja,Dev:2016xcp} while $Z_R$ couples directly to the SM quarks and leptons [cf. Eqs.~(\ref{eqn:ZR1}) and (\ref{eqn:ZR2})]. However, in the heavy $Z_R$ limits, both the $Z$ and $Z_R$ mediators are effectively suppressed by the large mass  $m_{Z_R}^2$ (note that the $Z - Z_R$ mixing angle is suppressed by $\sim m_Z^2 / m_{Z_R}^2$). As in the global model, the collider prospects of the RHNs can be easily obtained by rescaling the solid contours in Fig.~\ref{fig:production2} by the factor of $\left( {\sin\theta}/{0.1} \right)^2 \times
\left( {v_R}/{1\,{\rm TeV}} \right)^{-2}$. However, in light of the stringent dilepton limits on the $Z_R$ mass~\cite{ATLAS:2016cyf,CMS:2016abv}, the $v_R$ scale in this case has to be larger than about $3.4$ TeV [cf. Eq.~(\ref{eqn:MZR})], which implies that the production cross sections will be suppressed for both gauge and scalar portals.\footnote{This could be partly ameliorated in $U(1)_X$ models with different $Z_R$ couplings to charged leptons and RHNs, where the $pp\to Z_R\to NN$ production cross section can be enhanced by up to a factor of 5~\cite{Das:2017flq}.}

\section{Conclusion}
\label{sec:conclusion}
In this paper, we have studied the impact of the neutral scalar field that breaks $B-L$ symmetry in a TeV scale seesaw on leptogenesis. The Yukawa coupling of the new CP-even (odd) scalar $H$ ($A$) to RHNs induces dilution of the RHN number density through $2\leftrightarrow 2$ scattering processes like $NN\leftrightarrow HH (AA)$, which renders leptogenesis ineffective as a mechanism for understanding matter-antimatter asymmetry of the universe for a wide range of the scalar mass and coupling parameters. We have reported the leptogenesis constraints in the context of three different scenarios: in an effective low-energy theory [cf. Figs.~\ref{fig:EFT1}, \ref{fig:EFT2}, \ref{fig:EFT3}] and with global [cf. Figs.~\ref{fig:global}, \ref{fig:global2}] and local [cf.~Figs.~\ref{fig:local}, \ref{fig:local2}] $U(1)_{B-L}$ symmetry. We also comment on ways to probe these mass ranges for the seesaw scalar in collider settings and point out how this could provide a new window to probe the origin of neutrino masses and the baryon asymmetry.

\section*{Acknowledgments}

We would like to thank Julian Heeck, Daniele Teresi and Ye-ling Zhou for very useful discussions and communication. The work of R.N.M. is supported by the US National Science Foundation under Grant No. PHY1620074. Y.Z. would like to thank the IISN and Belgian Science Policy (IAP VII/37) for partial support. For some of the numerical calculations in this paper, computational resources have been provided by the Shared ICT Services Centre, Universit\'{e} Libre de Bruxelles.

\appendix
\section{The reduced cross sections}
\label{sec:appendixA}
In this appendix, we give the explicit analytic formulas for the reduced cross sections for various $2\leftrightarrow 2$ scatterings involving the RHNs used in our leptogenesis calculations in Sec.~\ref{sec:EFT}, \ref{sec:global} and \ref{sec:local}.

\subsection{Effective theory}
In the EFT framework of Sec.~\ref{sec:EFT}, the reduced cross sections read
\begin{eqnarray}
\label{eqn:sigmaNNHH}
&& \hat{\sigma} (NN \to H H) \ = \ \frac{f^4}{16 \pi }
\bigg[ -2\beta_H +
\frac{4 \beta _H \left(r_H-4\right){}^2}{x^2 \beta _H^2-\left(x-2
   r_H\right){}^2} \nonumber \\
&& \qquad - \frac{x^2 -4 \left(r_H-4\right)x +2 \left(r_H-4\right) \left(3 r_H+4\right)}{x(x-2r_H)}
\log \left( \frac{(1-\beta_H)x -2r_H}{(1+\beta_H)x -2r_H} \right)\bigg] \,, \\
\label{eqn:sigmaNNAA}
&& \hat{\sigma} (NN \to AA) \ = \ \frac{f^4}{16 \pi }
\bigg[ -2\beta_A +
\frac{4 \beta_A r_A^2 }{x^2 \beta_A^2-\left(x-2
   r_A\right){}^2} \nonumber \\
&& \qquad \qquad \qquad \qquad \qquad \qquad - \frac{x^2-4 r_Ax +6 r_A^2}{x(x-2r_A)}
\log \left( \frac{(1-\beta_A)x -2r_A}{(1+\beta_A)x -2r_A} \right)\bigg] \,,
\end{eqnarray}
where $x = s/m_N^2$, $r_X = m_{X}^2 / m_N^2$ (with $X = H,\,A$), and the functions
\begin{eqnarray}
\beta_X & \ \equiv \ & \sqrt{ ( 1 - 4x^{-1} ) ( 1 - 4r_X x^{-1} ) } \,.
\end{eqnarray}
Note the difference in the two cross sections~\eqref{eqn:sigmaNNHH} and \eqref{eqn:sigmaNNAA} for the CP-even and odd scalar, respectively, which is due to the $\gamma^5$ structure of the CP-odd scalar coupling to the Majorana neutrinos in Eq.~\eqref{eqn:LYukawa}.

\subsection{Global $U(1)_{B-L}$ model}
In the global $U(1)_{B-L}$ model of Sec.~\ref{sec:global}, the reduced cross sections are
\begin{eqnarray}
\label{eqn:H3H3}
\hat{\sigma} (NN \to H_3 H_3) & \ = \ & \frac{f^4}{8 \pi }
\left( {\cal A}_{SS}^{(1)} + {\cal A}_{SS}^{(2)} + {\cal A}_{SS}^{(3)} \right) \,, \\
\hat{\sigma} (NN \to H_3 J) & \ = \ & \frac{f^4}{4 \pi x^2 }
\bigg[ \beta_{H_3 \, 0} r^2 x - 2 (x-r)(x-2r)
\log \left( \frac{(1-\beta_{H_3\,0})x -r}{(1+\beta_{H_3\,0})x -r} \right) \bigg] \,, \\
\hat{\sigma} (NN \to JJ) & \ = \ & \frac{f^4}{8 \pi |x-r|^2 }
\bigg[ \beta_0 x \left(r^2-4 x\right) - 2 \left(x^2-r^2\right)
\log \left( \frac{1-\beta_0}{1+\beta_0} \right) \bigg] \,,
\end{eqnarray}
with $r = m_{H_3}^2 / m_N^2$, $\beta_{H_3\,0} = \beta_{XY} (X = r_{H_3}, Y = 0)$ and $\beta_0 = \beta_{XY} (X = Y = 0)$, where
\begin{eqnarray}
\beta_{XY} & \ \equiv \ & x^{-1}
\sqrt{ ( 1 - 4x^{-1} ) ( x^2 + r_X^2 + r_Y^2 - 2xr_X -2xr_Y -2r_X r_Y ) } \,.
\end{eqnarray}
The ${\cal A}_{SS}$ terms are defined as
\begin{eqnarray}
{\cal A}_{SS}^{(1)} & \ \equiv \ &
\frac{9 \beta _1 (x-4) r^2}{|x-r|^2} \,, \\
{\cal A}_{SS}^{(2)} & \ \equiv \ &
- \frac{12 r}{x |x-r|}
\bigg[ 2 \beta _1 x -
\Big(x+2 (r-4)\Big)
\log \left( \frac{(1-\beta_1)x - 2r}{(1+\beta_1)x - 2r} \right) \bigg] \,, \\
{\cal A}_{SS}^{(3)} & \ \equiv \ &
- 4 \beta _1 \left(1+\frac{2 (r-4)^2}{(x-2 r)^2 - \beta _1^2 x^2}\right) 
- \frac{2}{x (x-2 r)}
\Big( x^2 -4 (r-4)x + 2 (r-4) (3 r+4) \Big) \nonumber \\
&& \qquad \qquad \qquad \qquad \qquad \qquad \qquad \qquad \times \log \left( \frac{(1-\beta_1)x - 2r}{(1+\beta_1)x - 2r} \right) \,.
\end{eqnarray}
At the $H_3$ resonance, the propagator $1/|x-r|$ should be modified accordingly to include the $H_3$ width.

\subsection{Local $U(1)_{B-L}$ model}
In the gauged $U(1)_{B-L}$ model of Sec.~\ref{sec:local}, the reduced cross sections are
\begin{eqnarray}
\label{eqn:sigmaNNff}
\hat{\sigma} (NN \to f\bar{f}) &=& \frac{ c_{N}^2 g_L^4}{12 \pi \cos^4\theta_w }
\bigg( \sum_{f} S_f N_C^f \left( c_{f,\,L}^2 + c_{f,\,R}^2 \right) \bigg)
\frac{\sqrt{x} (x-4)^{3/2}}{|x-w|^2} \,, \\
\hat{\sigma} (NN \to Z_R Z_R) &=& \frac{g_L^4}{32\pi w^2}
\left( {\cal A}_{VV}^{(1)} + {\cal A}_{VV}^{(2)} + {\cal A}_{VV}^{(3)} \right) \,, \\
\hat{\sigma} (NN \to Z_R Z_R) &=& \frac{c_N^2 g_L^2}{32 \pi  \cos^2\theta_w \, w^2}
\left( {\cal A}_{VS}^{(1)} + {\cal A}_{VS}^{(2)} + {\cal A}_{VS}^{(3)} \right) \,,
\end{eqnarray}
where $w = m_{Z_R}^2 / m_N^2$, $c_{f,\,L} = g_{Z_R f_L f_L}\left( \frac{e}{\sin\theta_w \cos\theta_w}\right)^{-1}$ and $c_{f,\,R} = g_{Z_R f_R f_R}\left(\frac{e}{\sin\theta_w \cos\theta_w}\right)^{-1}$ the effective couplings of $Z_R$ boson to the fermions, and $c_N = \frac{\sin\theta_w}{\sin{2\phi}}$ to the RHNs (with the RH gauge mixing $\tan\phi = g_{BL} / g_R$). In Eq.~(\ref{eqn:sigmaNNff}) $f$ runs over all the flavors of quarks, charged leptons and active neutrinos, with $N_C^f$ the color degrees of freedom (3 for quarks and 1 otherwise) and the symmetry factor $S_f=1$ for the charged fermions and $1/2$ for neutrinos. The ${\cal A}_{VV}$ and ${\cal A}_{VS}$ terms are
\begin{eqnarray}
{\cal A}_{VV}^{(1)} & \ = \ &
\frac{\beta_{Z_R} (x-4) \left(x^2 -4 w x + 12 w^2 \right)}{\cos^4\phi \, |x-r|^2} \,, \\
{\cal A}_{VV}^{(2)} & \ = \ &
-\frac{8 c_N^2}{\cos^2\theta_w \cos^2\phi \, x (x-r)} \bigg[
\beta_{Z_R} x \left( x^2 -2 w x +4w^2\right)  \nonumber \\
&& \qquad  +2 \left(x^2-4 w x-2w^2(w-4) \right)
\log \left( \frac{(1-\beta_{Z_R})x - 2w}{(1+\beta_{Z_R})x - 2w} \right) \bigg] \,, \\
{\cal A}_{VV}^{(3)} & \ = \ &
\frac{8 c_N^4}{\cos^4\theta_w \, x}
\bigg[ \beta_{Z_R} x (2 x-w (w+8))
-\frac{4 \beta _{Z_R} x w^2(w-4)^2}{(x-2 w)^2-\beta _{Z_R}^2 x^2} \nonumber \\
&& \qquad - \frac{1}{x-2w} \Big( (w (w+4)-4) x^2+4 (4-3 w) w x + 4 (w-4) w^3 \Big) \nonumber \\
&& \qquad \qquad \qquad \qquad \qquad \qquad \times \log \left( \frac{(1-\beta_{Z_R})x - 2w}{(1+\beta_{Z_R})x - 2w} \right) \bigg] \,, \\
{\cal A}_{VS}^{(1)} & \ = \ &
\frac{\beta _{3} g_R^4 / f^2}{\cos^4\phi \, w (x-w)^2}
\bigg[
4 x^3 + ((w-16) w-8 r) x^2  \nonumber \\
&& \qquad + 2(2 r^2- r (w-4) w+ w^2 (3 w+10)) x \nonumber  \\
&& \qquad  + w \left(r^2 (w-8)-2 r (w-8) w+(w-40) w^2\right)
-\frac{1}{3} \beta _3^2 w^2 x^2 \bigg] \,, \\
{\cal A}_{VS}^{(2)} & \ = \ &
-\frac{16 g_R^2}{\cos^2\phi \, x (x-w)} \bigg[
\beta _3 x \left(x^2 -(r+w)x +4 w^2\right)  \nonumber \\
&& \qquad + 2 \Big(
x^2 + (r (w-2)-w (w+2))x +  r^2-r w (w+2)-(w-9) w^2
\Big) \nonumber \\
&& \qquad \qquad \qquad \qquad \qquad \qquad \times \log \left(\frac{(1-\beta_3)x - (r+w)}{(1+\beta_3)x - (r+w)}\right) \bigg] \,, \\
{\cal A}_{VS}^{(3)} & \ = \ &
16 f^2 w \bigg[
\beta _3 \left(x-2 w-\frac{4 (4-r) (4-w) w}{(x-r-w)^2-\beta _3^2 x^2}\right) \nonumber \\
&& \qquad -\frac{1}{x(x-r-w)} \Big( (w-2) x^2 -2(2 r (w-1)+ (w-10) w)x  \nonumber \\
&& \qquad + r^2 (w-2)+4 rw(w-1) +w ((w-10) w-32) \Big) \nonumber \\
&& \qquad \qquad \qquad \qquad \qquad \qquad \times \log \left(\frac{(1-\beta_3)x - (r+w)}{(1+\beta_3)x - (r+w)}\right) \bigg] \,,
\end{eqnarray}
with $\beta_{Z_R} = \beta_X (X = w = m_{Z_R}^2/m_N^2)$ and $\beta_3 = \beta_{XY} (X = r, \, Y = w)$.

\section{Couplings of the $Z_R$ boson to fermions}
\label{sec:appendixB}

In the gauged $SU(2)_L \times U (1)_{I_{3R}} \times U(1)_{B-L}$ model of Sec.~\ref{sec:local}, neglecting the mixing with the SM $Z$ boson, the $Z_R$ boson is given by the linear combination~\cite{Dev:2016dja}
\begin{eqnarray}
Z_{R,\,\mu} \ = \ \cos\phi \, U_{I_{3R},\, \mu} - \sin\phi \, U_{BL,\, \mu} \,.
\end{eqnarray}
Since the left-handed fermions do not carry any right-handed isospin (i.e. $I_{3R}=0$), the couplings to left-handed fermions are
\begin{eqnarray}
g_{Z_R f_L f_L} \ = \ - \frac12 \sin\phi \, g_{BL} (B-L) \,.
\end{eqnarray}
Using the electric charge formula
\begin{align}
Q \ = \ I_{3L} + I_{3R} \frac12 (B-L) \, ,
\label{eqn:charge}
\end{align}
and $\sin\phi \, g_{BL} = \frac{e}{\sin\theta_w \cos\theta_w} \frac{\sin\theta_w \, \sin\phi}{\cos\phi}$, we get
\begin{eqnarray}
\label{eqn:fL}
g_{Z_R f_L f_L}  &\ = \ & \sin\phi \, g_{BL} (I_{3L}-Q) \ = \  \frac{e}{\cos\theta_w} (I_{3L}-Q) \frac{\sin\phi}{\cos\phi} \, .
\end{eqnarray}

For the couplings to right-handed fermions,
\begin{eqnarray}
g_{Z_R f_R f_R} &\ = \ & \cos\phi \, g_{R} I_{3R} - \frac12 \sin\phi \, g_{BL} (B-L) \nonumber \\
&\ = \ & \cos\phi \, g_{R} I_{3R} - \sin\phi \, g_{BL} (Q-I_{3R}) \, ,
\label{eq:B5}
\end{eqnarray}
where in the second step, we have used Eq.~\eqref{eqn:charge} with $I_{3L}=0$. With
\begin{eqnarray}
&& \cos\phi \, g_R = \frac{e}{\sin\theta_w \cos\theta_w} \frac{\sin\theta_w \, \cos\phi}{\sin\phi} \,, 
\qquad
\sin\phi \, g_{BL} = \frac{e}{\sin\theta_w \cos\theta_w} \frac{\sin\theta_w \, \sin\phi}{\cos\phi} \,,
\end{eqnarray}
we can write Eq.~\eqref{eq:B5} as
\begin{eqnarray}
g_{Z_R f_R f_R}
&\ = \ & \frac{e}{\sin\theta_w \cos\theta_w}
\left[
\frac{\sin\theta_w \, \cos\phi}{\sin\phi} I_{3R} -
\frac{\sin\theta_w \, \sin\phi}{\cos\phi} (Q-I_{3R})
\right] \nonumber \\
%
& \ = \ & \frac{e}{\cos\theta_w}
( I_{3R} - Q \sin^2\phi)
\frac{1}{\sin\phi \, \cos\phi} \,.
\label{eqn:fR}
\end{eqnarray}
In Eqs.~(\ref{eqn:fL}) and Eq.~(\ref{eqn:fR}), we can write both $I_{3L}$ and $I_{3R}$ generically as $I_{3,f}$ as the third component of isospin for both left and right-handed fermions, thus producing the couplings of $Z_R$ bosons given in Eqs.~(\ref{eqn:ZR1}) and (\ref{eqn:ZR2}), respectively.

Using these couplings, we can write down the partial widths of $Z_R$ into SM fermions:
\begin{eqnarray}
\Gamma (Z_R \to f \bar{f}) & \ = \ & \frac{S_f N_C^f \, e^2 \, m_{Z_R} \left( c_{f,\,L}^2 + c_{f,\,R}^2 \right)}{24\pi \sin^2\theta_w \cos^2\theta_w} \, ,
\end{eqnarray}
Similarly, for the $Z_R$ decay into RHNs, we have
\begin{eqnarray}
\Gamma (Z_R \to NN) & \ = \ & \frac{e^2 m_{Z_R} c_N^2}{48\pi \sin^2\theta_w \cos^2\theta_w}
\left( 1 - \frac{4m_N^2}{m_{Z_R}^2} \right)^{3/2}
\Theta (m_{Z_R} - 2 m_N) \,.
\end{eqnarray}

The scalar $H_3$ could decay into two RHNs with the width given in Eq.~(\ref{eqn:width1}). If kinematically allowed, it also decays into the heavy gauge bosons, with the partial width
\begin{eqnarray}
\Gamma (H_3 \to Z_R Z_R) & \ = \ &
\frac{m_{H_3}^3}{64\pi v_R^2}
\left( 1 - \frac{4m_{Z_R}^2}{m_{H_3}^2} \right)^{1/2}
\left( 1 - \frac{4m_{Z_R}^2}{m_{H_3}^2} + \frac{12 m_{Z_R}^4}{m_{H_3}^4} \right)
\Theta (m_{H_3} - 2 m_{Z_R}) \,. \nonumber \\ &&
\end{eqnarray}


\begin{thebibliography}{99}

\bibitem{seesaw1} P. Minkowski, Phys. Lett. B {\bf 67}, 421 (1977).

\bibitem{seesaw2} R. N. Mohapatra and G. Senjanovi\'{c}, Phys. Rev. Lett. {\bf 44}, 912 (1980).

\bibitem{seesaw3} T. Yanagida, Conf.  Proc.  C {\bf 7902131},  95  (1979).

\bibitem{seesaw4} M. Gell-Mann, P. Ramond and R. Slansky, Con
f. Proc. C {\bf 790927}, 315 (1979) [arXiv:1306.4669 [hep-th]].

\bibitem{seesaw5} S.~L.~Glashow, NATO Sci. Ser. B {\bf 61}, 687 (1980).

\bibitem{Mohapatra:2005wg}
  R.~N.~Mohapatra {\it et al.},
  Rept.\ Prog.\ Phys.\  {\bf 70}, 1757 (2007)
  [hep-ph/0510213].

\bibitem{Drewes:2013gca}
  M.~Drewes,
  Int.\ J.\ Mod.\ Phys.\ E {\bf 22}, 1330019 (2013)
  [arXiv:1303.6912 [hep-ph]].

\bibitem{Deppisch:2015qwa}
  F.~F.~Deppisch, P.~S.~B.~Dev and A.~Pilaftsis,
  New J.\ Phys.\  {\bf 17}, no. 7, 075019 (2015)
  [arXiv:1502.06541 [hep-ph]].

\bibitem{Cai:2017mow}
  Y.~Cai, T.~Han, T.~Li and R.~Ruiz,
  arXiv:1711.02180 [hep-ph].

\bibitem{deGouvea:2013zba}
  A.~de Gouvea and P.~Vogel,
  Prog.\ Part.\ Nucl.\ Phys.\  {\bf 71}, 75 (2013)
  [arXiv:1303.4097 [hep-ph]].

\bibitem{Alekhin:2015byh}
  S.~Alekhin {\it et al.},
  Rept.\ Prog.\ Phys.\  {\bf 79}, no. 12, 124201 (2016)
  [arXiv:1504.04855 [hep-ph]].

\bibitem{Vissani:1997ys}
  F.~Vissani,
  Phys.\ Rev.\ D {\bf 57}, 7027 (1998)
  [hep-ph/9709409].

\bibitem{Clarke:2015gwa}
  J.~D.~Clarke, R.~Foot and R.~R.~Volkas,
  Phys.\ Rev.\ D {\bf 91}, no. 7, 073009 (2015)
  [arXiv:1502.01352 [hep-ph]].



\bibitem{Bambhaniya:2016rbb}
  G.~Bambhaniya, P.~S.~B.~Dev, S.~Goswami, S.~Khan and W.~Rodejohann,
  Phys.\ Rev.\ D {\bf 95}, no. 9, 095016 (2017)
  [arXiv:1611.03827 [hep-ph]].

\bibitem{Marshak:1979fm}
  R.~E.~Marshak and R.~N.~Mohapatra,
  Phys.\ Lett.\  {\bf 91B}, 222 (1980).

  \bibitem{Mohapatra:1980qe}
  R.~N.~Mohapatra and R.~E.~Marshak,
  Phys.\ Rev.\ Lett.\  {\bf 44}, 1316 (1980)
  Erratum: [Phys.\ Rev.\ Lett.\  {\bf 44}, 1643 (1980)].

\bibitem{Maiezza:2015lza}
  A.~Maiezza, M.~Nemev\v{s}ek and F.~Nesti,
  Phys.\ Rev.\ Lett.\  {\bf 115}, 081802 (2015)
  [arXiv:1503.06834 [hep-ph]].

\bibitem{Nemevsek:2016enw}
  M.~Nemev\v{s}ek, F.~Nesti and J.~C.~Vasquez,
  JHEP {\bf 1704}, 114 (2017)
  [arXiv:1612.06840 [hep-ph]].

\bibitem{Dev:2016vle}
  P.~S.~B.~Dev, R.~N.~Mohapatra and Y.~Zhang,
  Phys.\ Rev.\ D {\bf 95}, no. 11, 115001 (2017)
  [arXiv:1612.09587 [hep-ph]].


\bibitem{Dev:2017dui}
  P.~S.~B.~Dev, R.~N.~Mohapatra and Y.~Zhang,
  Nucl.\ Phys.\ B {\bf 923}, 179 (2017)
  [arXiv:1703.02471 [hep-ph]].

\bibitem{Dev:2017ozg}
  P.~S.~B.~Dev, R.~N.~Mohapatra and Y.~Zhang,
  Acta Phys.\ Polon.\ B {\bf 48}, 969 (2017).

\bibitem{Fukugita:1986hr}
  M.~Fukugita and T.~Yanagida,
  Phys.\ Lett.\ B {\bf 174}, 45 (1986).

\bibitem{Davidson:2008bu}
  S.~Davidson, E.~Nardi and Y.~Nir,
  Phys.\ Rept.\  {\bf 466}, 105 (2008)
  [arXiv:0802.2962 [hep-ph]].


\bibitem{Blanchet:2012bk}
  S.~Blanchet and P.~Di Bari,
  New J.\ Phys.\  {\bf 14}, 125012 (2012)
  [arXiv:1211.0512 [hep-ph]].


\bibitem{Fong:2013wr}
  C.~S.~Fong, E.~Nardi and A.~Riotto,
  Adv.\ High Energy Phys.\  {\bf 2012}, 158303 (2012)
  [arXiv:1301.3062 [hep-ph]].


\bibitem{Pilaftsis:1997dr}
  A.~Pilaftsis,
  Nucl.\ Phys.\ B {\bf 504}, 61 (1997)
  [hep-ph/9702393].

\bibitem{Pilaftsis:1997jf}
  A.~Pilaftsis,
  Phys.\ Rev.\ D {\bf 56}, 5431 (1997)
  [hep-ph/9707235].

\bibitem{Pilaftsis:2003gt}
  A.~Pilaftsis and T.~E.~J.~Underwood,
  Nucl.\ Phys.\ B {\bf 692}, 303 (2004)
  [hep-ph/0309342].

\bibitem{Dev:2017wwc}
  P.~S.~B.~Dev, M.~Garny, J.~Klaric, P.~Millington and D.~Teresi,
  arXiv:1711.02863 [hep-ph].

\bibitem{Davidson:2002qv}
  S.~Davidson and A.~Ibarra,
  Phys.\ Lett.\ B {\bf 535}, 25 (2002)
  [hep-ph/0202239].



\bibitem{Deppisch:2013jxa}
  F.~F.~Deppisch, J.~Harz and M.~Hirsch,
  Phys.\ Rev.\ Lett.\  {\bf 112}, 221601 (2014)
  [arXiv:1312.4447 [hep-ph]].

\bibitem{Deppisch:2015yqa}
  F.~F.~Deppisch, J.~Harz, M.~Hirsch, W.~C.~Huang and H.~Päs,
  Phys.\ Rev.\ D {\bf 92}, no. 3, 036005 (2015)
  [arXiv:1503.04825 [hep-ph]].

\bibitem{Chun:2017spz}
  E.~J.~Chun {\it et al.},
  arXiv:1711.02865 [hep-ph].

\bibitem{Pati:1974yy}
  J.~C.~Pati and A.~Salam,
  Phys.\ Rev.\ D {\bf 10}, 275 (1974)
  Erratum: [Phys.\ Rev.\ D {\bf 11}, 703 (1975)].

\bibitem{Mohapatra:1974gc}
  R.~N.~Mohapatra and J.~C.~Pati,
  Phys.\ Rev.\ D {\bf 11}, 2558 (1975).

\bibitem{Senjanovic:1975rk}
  G.~Senjanovic and R.~N.~Mohapatra,
  Phys.\ Rev.\ D {\bf 12}, 1502 (1975).

\bibitem{Frere:2008ct}
  J.~M.~Frere, T.~Hambye and G.~Vertongen,
  JHEP {\bf 0901}, 051 (2009)
  [arXiv:0806.0841 [hep-ph]].

\bibitem{Dev:2014iva}
  P.~S.~B.~Dev, C.~H.~Lee and R.~N.~Mohapatra,
  Phys.\ Rev.\ D {\bf 90}, no. 9, 095012 (2014)
  [arXiv:1408.2820 [hep-ph]].

\bibitem{Dev:2015vra}
  P.~S.~B.~Dev, C.~H.~Lee and R.~N.~Mohapatra,
  J.\ Phys.\ Conf.\ Ser.\  {\bf 631}, no. 1, 012007 (2015)
  [arXiv:1503.04970 [hep-ph]].

\bibitem{Dhuria:2015cfa}
  M.~Dhuria, C.~Hati, R.~Rangarajan and U.~Sarkar,
  Phys.\ Rev.\ D {\bf 92}, no. 3, 031701 (2015)
  [arXiv:1503.07198 [hep-ph]].


\bibitem{Blanchet:2009bu}
  S.~Blanchet, Z.~Chacko, S.~S.~Granor and R.~N.~Mohapatra,
  Phys.\ Rev.\ D {\bf 82}, 076008 (2010)
  [arXiv:0904.2174 [hep-ph]].

\bibitem{Blanchet:2010kw}
  S.~Blanchet, P.~S.~B.~Dev and R.~N.~Mohapatra,
  Phys.\ Rev.\ D {\bf 82}, 115025 (2010)
  [arXiv:1010.1471 [hep-ph]].

\bibitem{Iso:2010mv}
  S.~Iso, N.~Okada and Y.~Orikasa,
  Phys.\ Rev.\ D {\bf 83}, 093011 (2011)
  [arXiv:1011.4769 [hep-ph]].


\bibitem{Okada:2012fs}
  N.~Okada, Y.~Orikasa and T.~Yamada,
  Phys.\ Rev.\ D {\bf 86}, 076003 (2012)
  [arXiv:1207.1510 [hep-ph]].

\bibitem{Heeck:2016oda}
  J.~Heeck and D.~Teresi,
  Phys.\ Rev.\ D {\bf 94}, no. 9, 095024 (2016)
  [arXiv:1609.03594 [hep-ph]].




\bibitem{Shuve:2017jgj}
  B.~Shuve and C.~Tamarit,
  JHEP {\bf 1710}, 122 (2017)
  [arXiv:1704.01979 [hep-ph]].






\bibitem{Chikashige:1980ui}
  Y.~Chikashige, R.~N.~Mohapatra and R.~D.~Peccei,
  Phys.\ Lett.\  {\bf 98B}, 265 (1981).

\bibitem{Aad:2015tna}
  G.~Aad {\it et al.} [ATLAS Collaboration],
  Phys.\ Lett.\ B {\bf 753}, 69 (2016)
  [arXiv:1508.02507 [hep-ex]].

\bibitem{Aad:2016nal}
  G.~Aad {\it et al.} [ATLAS Collaboration],
  Eur.\ Phys.\ J.\ C {\bf 76}, no. 12, 658 (2016)
  [arXiv:1602.04516 [hep-ex]].

\bibitem{Khachatryan:2016tnr}
  V.~Khachatryan {\it et al.} [CMS Collaboration],
  Phys.\ Lett.\ B {\bf 759}, 672 (2016)
  [arXiv:1602.04305 [hep-ex]].


\bibitem{Giudice:2003jh}
  G.~F.~Giudice, A.~Notari, M.~Raidal, A.~Riotto and A.~Strumia,
  Nucl.\ Phys.\ B {\bf 685}, 89 (2004)
  [hep-ph/0310123].

\bibitem{Dev:2014laa}
  P.~S.~B.~Dev, P.~Millington, A.~Pilaftsis and D.~Teresi,
  Nucl.\ Phys.\ B {\bf 886}, 569 (2014)
  [arXiv:1404.1003 [hep-ph]].

\bibitem{DOnofrio:2014rug}
  M.~D'Onofrio, K.~Rummukainen and A.~Tranberg,
  Phys.\ Rev.\ Lett.\  {\bf 113}, no. 14, 141602 (2014)
  [arXiv:1404.3565 [hep-ph]].

\bibitem{Lazarides:1991wu}
  G.~Lazarides and Q.~Shafi,
  Phys.\ Lett.\ B {\bf 258}, 305 (1991).


\bibitem{Giudice:1999fb}
  G.~F.~Giudice, M.~Peloso, A.~Riotto and I.~Tkachev,
  JHEP {\bf 9908}, 014 (1999)
  [hep-ph/9905242].

\bibitem{Asaka:1999yd}
  T.~Asaka, K.~Hamaguchi, M.~Kawasaki and T.~Yanagida,
  Phys.\ Lett.\ B {\bf 464}, 12 (1999)
  [hep-ph/9906366].



\bibitem{Akhmedov:1998qx}
  E.~K.~Akhmedov, V.~A.~Rubakov and A.~Y.~Smirnov,
  Phys.\ Rev.\ Lett.\  {\bf 81}, 1359 (1998)
  [hep-ph/9803255].

\bibitem{Asaka:2005pn}
  T.~Asaka and M.~Shaposhnikov,
  Phys.\ Lett.\ B {\bf 620}, 17 (2005)
  [hep-ph/0505013].


\bibitem{Drewes:2017zyw}
  M.~Drewes {\it et al.},
  arXiv:1711.02862 [hep-ph].


\bibitem{Ade:2015xua}
  P.~A.~R.~Ade {\it et al.} [Planck Collaboration],
  Astron.\ Astrophys.\  {\bf 594}, A13 (2016)
  [arXiv:1502.01589 [astro-ph.CO]].

\bibitem{Schechter:1981cv}
  J.~Schechter and J.~W.~F.~Valle,
  Phys.\ Rev.\ D {\bf 25}, 774 (1982).

\bibitem{Sierra:2014sta}
  D.~Aristizabal Sierra, M.~Tortola, J.~W.~F.~Valle and A.~Vicente,
  JCAP {\bf 1407}, 052 (2014)
  [arXiv:1405.4706 [hep-ph]].

\bibitem{Akhmedov:1992hi}
  E.~K.~Akhmedov, Z.~G.~Berezhiani, R.~N.~Mohapatra and G.~Senjanovic,
  Phys.\ Lett.\ B {\bf 299}, 90 (1993)
  [hep-ph/9209285].

\bibitem{Rothstein:1992rh}
  I.~Z.~Rothstein, K.~S.~Babu and D.~Seckel,
  Nucl.\ Phys.\ B {\bf 403}, 725 (1993)
  [hep-ph/9301213].

\bibitem{Heurtier:2016otg}
  L.~Heurtier and Y.~Zhang,
  JCAP {\bf 1702}, no. 02, 042 (2017)
  [arXiv:1609.05882 [hep-ph]].


\bibitem{Pilaftsis:1993af}
  A.~Pilaftsis,
  Phys.\ Rev.\ D {\bf 49}, 2398 (1994)
  [hep-ph/9308258].

\bibitem{Arnold:2006sd}
  R.~Arnold {\it et al.} [NEMO Collaboration],
  Nucl.\ Phys.\ A {\bf 765}, 483 (2006)
  [hep-ex/0601021].

\bibitem{Dev:2016dja}
  P.~S.~B.~Dev, R.~N.~Mohapatra and Y.~Zhang,
  JHEP {\bf 1605}, 174 (2016)
  [arXiv:1602.05947 [hep-ph]].

\bibitem{Pilaftsis:2008qt}
  A.~Pilaftsis,
  Phys.\ Rev.\ D {\bf 78}, 013008 (2008)
  [arXiv:0805.1677 [hep-ph]].




\bibitem{Dev:2015vjd}
  P.~S.~B.~Dev, R.~N.~Mohapatra and Y.~Zhang,
  JHEP {\bf 1602}, 186 (2016)
  [arXiv:1512.08507 [hep-ph]].


\bibitem{ATLAS:2016cyf}
  The ATLAS collaboration [ATLAS Collaboration],
  ATLAS-CONF-2016-045.

\bibitem{CMS:2016abv}
  CMS Collaboration [CMS Collaboration],
  CMS-PAS-EXO-16-031.

\bibitem{Patra:2015bga}
  S.~Patra, F.~S.~Queiroz and W.~Rodejohann,
  Phys.\ Lett.\ B {\bf 752}, 186 (2016)
  [arXiv:1506.03456 [hep-ph]].


\bibitem{Klasen:2016qux}
  M.~Klasen, F.~Lyonnet and F.~S.~Queiroz,
  Eur.\ Phys.\ J.\ C {\bf 77}, no. 5, 348 (2017)
  [arXiv:1607.06468 [hep-ph]].

\bibitem{Dev:2016xcp}
  P.~S.~B.~Dev, R.~N.~Mohapatra and Y.~Zhang,
  JHEP {\bf 1611}, 077 (2016)
  [arXiv:1608.06266 [hep-ph]].

\bibitem{CMS:2016zxk}
  CMS Collaboration [CMS Collaboration],
  CMS-PAS-EXO-16-008.

\bibitem{ATLAS:2017mpg}
  The ATLAS collaboration [ATLAS Collaboration],
  ATLAS-CONF-2017-050.



\bibitem{Holthausen:2009uc}
  M.~Holthausen, M.~Lindner and M.~A.~Schmidt,
  Phys.\ Rev.\ D {\bf 82}, 055002 (2010)
  [arXiv:0911.0710 [hep-ph]].

\bibitem{Chou:2016lxi}
  J.~P.~Chou, D.~Curtin and H.~J.~Lubatti,
  Phys.\ Lett.\ B {\bf 767}, 29 (2017)
  [arXiv:1606.06298 [hep-ph]].

\bibitem{Evans:2017lvd}
  J.~A.~Evans,
  arXiv:1708.08503 [hep-ph].


\bibitem{Adams:2013qkq}
  C.~Adams {\it et al.} [LBNE Collaboration],
  arXiv:1307.7335 [hep-ex].

\bibitem{ATLAS:2016pyq}
  The ATLAS collaboration [ATLAS Collaboration],
  ATLAS-CONF-2016-073.

\bibitem{Chatrchyan:2013mxa}
  S.~Chatrchyan {\it et al.} [CMS Collaboration],
  Phys.\ Rev.\ D {\bf 89}, no. 9, 092007 (2014)
  [arXiv:1312.5353 [hep-ex]].

\bibitem{Khachatryan:2015cwa}
  V.~Khachatryan {\it et al.} [CMS Collaboration],
  JHEP {\bf 1510}, 144 (2015)
  [arXiv:1504.00936 [hep-ex]].

\bibitem{Aad:2015kna}
  G.~Aad {\it et al.} [ATLAS Collaboration],
  Eur.\ Phys.\ J.\ C {\bf 76}, no. 1, 45 (2016)
  [arXiv:1507.05930 [hep-ex]].

\bibitem{Aaboud:2016okv}
  M.~Aaboud {\it et al.} [ATLAS Collaboration],
  JHEP {\bf 1609}, 173 (2016)
  [arXiv:1606.04833 [hep-ex]].


\bibitem{Aad:2014ioa}
  G.~Aad {\it et al.} [ATLAS Collaboration],
  Phys.\ Rev.\ Lett.\  {\bf 113}, no. 17, 171801 (2014)
  [arXiv:1407.6583 [hep-ex]].

\bibitem{ATLAS:2016eeo}
  The ATLAS collaboration [ATLAS Collaboration],
  ATLAS-CONF-2016-059.

\bibitem{Khachatryan:2016yec}
  V.~Khachatryan {\it et al.} [CMS Collaboration],
  Phys.\ Lett.\ B {\bf 767}, 147 (2017)
  [arXiv:1609.02507 [hep-ex]].


\bibitem{Khachatryan:2015yea}
  V.~Khachatryan {\it et al.} [CMS Collaboration],
  Phys.\ Lett.\ B {\bf 749}, 560 (2015)
  [arXiv:1503.04114 [hep-ex]].

\bibitem{ATLAS:2016ixk}
  The ATLAS collaboration [ATLAS Collaboration],
  ATLAS-CONF-2016-049.

\bibitem{Khachatryan:2016sey}
  V.~Khachatryan {\it et al.} [CMS Collaboration],
  Phys.\ Rev.\ D {\bf 94}, no. 5, 052012 (2016)
  [arXiv:1603.06896 [hep-ex]].


\bibitem{Falkowski:2015iwa}
  A.~Falkowski, C.~Gross and O.~Lebedev,
  JHEP {\bf 1505}, 057 (2015)
  [arXiv:1502.01361 [hep-ph]].

\bibitem{Bian:2017jpt}
  L.~Bian, N.~Chen and Y.~Zhang,
  arXiv:1706.09425 [hep-ph].

\bibitem{Barate:2003sz}
  R.~Barate {\it et al.} [ALEPH and DELPHI and L3 and OPAL Collaborations and LEP Working Group for Higgs boson searches],
  Phys.\ Lett.\ B {\bf 565}, 61 (2003)
  [hep-ex/0306033].



\bibitem{Aubert:2003cm}
  B.~Aubert {\it et al.} [BaBar Collaboration],
  Phys.\ Rev.\ Lett.\  {\bf 91}, 221802 (2003)
  [hep-ex/0308042].

\bibitem{Wei:2009zv}
  J.-T.~Wei {\it et al.} [Belle Collaboration],
  Phys.\ Rev.\ Lett.\  {\bf 103}, 171801 (2009)
  [arXiv:0904.0770 [hep-ex]].

\bibitem{Aaij:2012vr}
  R.~Aaij {\it et al.} [LHCb Collaboration],
  JHEP {\bf 1302}, 105 (2013)
  [arXiv:1209.4284 [hep-ex]].


\bibitem{Peskin:2012we}
  M.~E.~Peskin,
  arXiv:1207.2516 [hep-ph].

\bibitem{Baer:2013cma}
  H.~Baer {\it et al.},
  arXiv:1306.6352 [hep-ph].

\bibitem{Garcia-Cely:2017oco}
  C.~Garcia-Cely and J.~Heeck,
  JHEP {\bf 1705}, 102 (2017)
  [arXiv:1701.07209 [hep-ph]].

\bibitem{Aad:2015zva}
  G.~Aad {\it et al.} [ATLAS Collaboration],
  Eur.\ Phys.\ J.\ C {\bf 75}, no. 7, 299 (2015)
  Erratum: [Eur.\ Phys.\ J.\ C {\bf 75}, no. 9, 408 (2015)]
  [arXiv:1502.01518 [hep-ex]].

\bibitem{Khachatryan:2014rra}
  V.~Khachatryan {\it et al.} [CMS Collaboration],
  Eur.\ Phys.\ J.\ C {\bf 75}, no. 5, 235 (2015)
  [arXiv:1408.3583 [hep-ex]].

\bibitem{Aaboud:2016tnv}
  M.~Aaboud {\it et al.} [ATLAS Collaboration],
  Phys.\ Rev.\ D {\bf 94}, no. 3, 032005 (2016)
  [arXiv:1604.07773 [hep-ex]].

\bibitem{Belyaev:2012qa}
  A.~Belyaev, N.~D.~Christensen and A.~Pukhov,
  Comput.\ Phys.\ Commun.\  {\bf 184}, 1729 (2013)
  [arXiv:1207.6082 [hep-ph]].

\bibitem{PDG}
  C. Patrignani {\it et al}. (Particle Data Group),
  Chin. Phys. C, {\bf 40}, 100001 (2016) and 2017 update.



\bibitem{Dev:2012zg}
  P.~S.~B.~Dev, R.~Franceschini and R.~N.~Mohapatra,
  Phys.\ Rev.\ D {\bf 86}, 093010 (2012)
  [arXiv:1207.2756 [hep-ph]].

\bibitem{Cely:2012bz}
  C.~G.~Cely, A.~Ibarra, E.~Molinaro and S.~T.~Petcov,
  Phys.\ Lett.\ B {\bf 718}, 957 (2013)
  [arXiv:1208.3654 [hep-ph]].

\bibitem{Hessler:2014ssa}
  A.~G.~Hessler, A.~Ibarra, E.~Molinaro and S.~Vogl,
  Phys.\ Rev.\ D {\bf 91}, no. 11, 115004 (2015)
  [arXiv:1408.0983 [hep-ph]].

\bibitem{Das:2017zjc}
  A.~Das, P.~S.~B.~Dev and C.~S.~Kim,
  Phys.\ Rev.\ D {\bf 95}, no. 11, 115013 (2017)
  [arXiv:1704.00880 [hep-ph]].

\bibitem{Das:2017rsu}
  A.~Das, Y.~Gao and T.~Kamon,
  arXiv:1704.00881 [hep-ph].

\bibitem{Ruiz:2017yyf}
  R.~Ruiz, M.~Spannowsky and P.~Waite,
  Phys.\ Rev.\ D {\bf 96}, no. 5, 055042 (2017)
  [arXiv:1706.02298 [hep-ph]].



\bibitem{Djouadi:2005gi}
  A.~Djouadi,
  Phys.\ Rept.\  {\bf 457}, 1 (2008)
  [hep-ph/0503172].

\bibitem{Heinemeyer:2013tqa}
  S.~Heinemeyer {\it et al.} [LHC Higgs Cross Section Working Group],
  arXiv:1307.1347 [hep-ph].

\bibitem{Arkani-Hamed:2015vfh}
  N.~Arkani-Hamed, T.~Han, M.~Mangano and L.~T.~Wang,
  Phys.\ Rept.\  {\bf 652}, 1 (2016)
  [arXiv:1511.06495 [hep-ph]].

\bibitem{Golling:2016gvc}
  T.~Golling {\it et al.},
  CERN Yellow Report, no. 3, 441 (2017)
  [arXiv:1606.00947 [hep-ph]].

\bibitem{Contino:2016spe}
  R.~Contino {\it et al.},
  CERN Yellow Report, no. 3, 255 (2017)
  [arXiv:1606.09408 [hep-ph]].

\bibitem{CEPC-SPPCStudyGroup:2015csa}
  CEPC-SPPC Study Group,
  IHEP-CEPC-DR-2015-01, IHEP-TH-2015-01, IHEP-EP-2015-01.

\bibitem{Basso:2008iv}
  L.~Basso, A.~Belyaev, S.~Moretti and C.~H.~Shepherd-Themistocleous,
  Phys.\ Rev.\ D {\bf 80}, 055030 (2009)
  [arXiv:0812.4313 [hep-ph]].

\bibitem{Deppisch:2013cya}
  F.~F.~Deppisch, N.~Desai and J.~W.~F.~Valle,
  Phys.\ Rev.\ D {\bf 89}, no. 5, 051302 (2014)
  [arXiv:1308.6789 [hep-ph]].

\bibitem{Kang:2015uoc}
  Z.~Kang, P.~Ko and J.~Li,
  Phys.\ Rev.\ D {\bf 93}, no. 7, 075037 (2016)
  [arXiv:1512.08373 [hep-ph]].

\bibitem{Das:2017pvt}
  A.~Das,
  arXiv:1701.04946 [hep-ph].

\bibitem{Cox:2017eme}
  P.~Cox, C.~Han and T.~T.~Yanagida,
  arXiv:1707.04532 [hep-ph].

\bibitem{Das:2017flq}
  A.~Das, N.~Okada and D.~Raut,
  arXiv:1710.03377 [hep-ph].





\end{thebibliography}
\end{document}